\documentclass[chicago]{emulateapj}
\usepackage{amsmath}
\usepackage{graphicx}
\usepackage{multirow}
\usepackage[dvips]{color}

\slugcomment{Accepted to ApJ}
\shortauthors{}

\begin{document}
\title{Fragmentation and OB Star Formation in High--Mass Molecular Hub--Filament Systems}

\author{Hauyu Baobab Liu\altaffilmark{1,2,3}} \author{Izaskun Jim{\'e}nez-Serra \altaffilmark{3}}  \author{Paul T. P. Ho\altaffilmark{2,3}} \author{Huei--Ru Chen \altaffilmark{2,4}} \author{Qizhou Zhang\altaffilmark{3}} \author{Zhi--Yun Li \altaffilmark{5}}

\affil{$^{1}$Department of Physics, National Taiwan University, No. 1, Sec. 4, Roosevelt Road, Taipei 106, Taiwan (R.O.C.)}
\affil{$^{2}$Academia Sinica Institute of Astronomy and Astrophysics, P.O. Box 23-141, Taipei, 106 Taiwan}
\affil{$^{3}$Harvard-Smithsonian Center for Astrophysics, 60 Garden Street, Cambridge, MA 02138}
\affil{$^{4}$Institute of Astronomy and Department of Physics, National Tsing Hua University, Hsinchu, Taiwan}
\affil{$^{5}$Department of Astronomy, P.O. Box 400325, Charlottesville, VA 22904}

\altaffiltext{1}{Department of Physics, National Taiwan University}
\altaffiltext{2}{Academia Sinica Institute of Astronomy and Astrophysics}
\altaffiltext{3}{Harvard-Smithsonian Center for Astrophysics}
\altaffiltext{4}{Institute of Astronomy and Department of Physics, National Tsing Hua University}
\altaffiltext{5}{Department of Astronomy, University of Virginia}

\begin{abstract}
Filamentary structures are ubiquitously seen in the interstellar medium. 
The concentrated molecular mass in the filaments allows fragmentation to occur in a shorter timescale than the timescale of the global collapse. 
Such hierarchical fragmentation may further assist the dissipation of excessive angular momentum.  
It is crucial to resolve the morphology and the internal velocity structures of the molecular filaments observationally. 

We perform 0$''$.5--2$''$.5 angular resolution interferometric observations toward the nearly face--on OB cluster forming region G33.92+0.11.
Observations of various spectral lines as well as the millimeter dust continuum emission, consistently trace several $\sim$1 pc scale, clumpy molecular arms.
Some of the molecular arms geometrically merge to an inner 3.0$^{\mbox{\scriptsize{+2.8}}}_{\mbox{-\scriptsize{1.4}}}\cdot$10$^{3}$\,$M_{\odot}$, 0.6 pc scale central molecular clump, and may directly channel the molecular gas to the warm ($\sim$50 K) molecular gas immediately surrounding the centrally embedded OB stars. 
The NH$_{3}$ spectra suggest a medium turbulence line width of FWHM$\lesssim$2\,km\,s$^{-1}$ in the central molecular clump, implying a $\gtrsim$10 times larger molecular mass than the virial mass.
Feedbacks from shocks and the centrally embedded OB stars and localized (proto)stellar clusters, likely play a key role in the heating of molecular gas and could lead to the observed chemical stratification.
Although (proto)stellar feedbacks are already present, G33.92+0.11 chemically appears to be at an early evolutionary stage given by the low abundance limit of SO$_{2}$ observed in this region.
\end{abstract}

\keywords{ stars: formation --- ISM: evolution --- ISM: individual (G33.92+0.11)}

\clearpage
\section{Introduction }
\label{chap_introduction}
Analyses of visual and infrared extinction (Rathborne et al. 2006; Butler \& Tan 2009; Schneider et al. 2011), and the Herschel Space Telescope observations of thermal dust emission (Andr{\'e} et al. 2010; Henning et al. 2010; Men'shchikov et al.2010; Molinari et al. 2010) unveiled overall filamentary and clumpy morphology of the molecular interstellar medium (ISM).
How the densest molecular clumps/cores as well as the subsequent stellar clusters will form, are of fundamental interest in the evolution of molecular filaments (c.f. Klessen \& Burkert 2000; Klessen \& Burkert 2001; Bonnell \& Bate 2002; Bate et al. 2003; Padoan et al. 2007; V{\'a}zquez-Semadeni et al. 2007; Myers 2011). 
Observational case studies in nearby low--mass star forming regions have been reviewed by Myers (2009).
At further distances of the Orion cloud and the Cygnus--X region, which form high--mass stars (e.g. $\ge$8 $M_{\odot}$), parsec scale molecular filaments are well known (e.g. Orion: Megzer 1990; Dutrey et al. 1993; Tatematsu et al. 1993, 1998; Chini et al. 1997; Buckle et al. 2010; Shimajiri et al. 2011; Cygnus--X: Chandler et al. 1993; Bontemps et al. 2010; Schneider et al. 2010), including the NH$_{3}$ emission filaments which radiate from the Orion--KL massive star forming core (Wiseman \& Ho 1996, 1998).
As the majority of massive OB cluster forming molecular clouds are located at the typical distances of a few kpc, their filamentary structures and kinematics are only now being explored with high angular resolution observations.  

Interferometric observations of the NH$_{3}$ emission on Infrared Dark Clouds (IRDCs) G28.34+0.06 and G30.88+00.13  (Wang et al. 2008; Zhang \& Wang 2011), and the submillimeter bolometer observations on IRDC G304.74+01.32 (Miettinen \& Harju 2010), unveiled systems of parsec scale massive molecular filaments. 
High angular resolution observations of the millimeter and submillimeter thermal dust continuum emission on IRDC G28.34+0.06 (Zhang et al. 2009) and G030.88+00.13 (Swift 2009; Zhang \& Wang 2011) have demonstrated that fragmentation takes place within parsec scale molecular filaments, which leads to the formation of regularly spaced 10$^{1}$--10$^{2}$\,$M_{\odot}$ molecular cores embedded in multiple distinct condensations or clumps (Wang et al. 2011).
Stability and chemical evolution of filamentary IRDCs have been addressed in observational studies of G29.96--0.02, G35.20--1.74 (Pillai et al. 2011) and IRAS 20293+3952 (Busquet et al. 2010). 
By combining single dish and interferometric observations of the thermal dust continuum emission obtained toward the massive molecular \textit{Hub--Filament System (HFS)} G10.6-0.4, Liu et al. (2012) reported hierarchical fragmentation at $\sim$10 pc scales within this region, which suggests a scenario of OB cluster formation via hierarchical contraction. 
In the meantime, by analyzing the kinematics in the W33A region, Galv{\'a}n-Madrid et al. (2010) proposed that the collision of molecular filaments is conducive to the formation of massive stars. 
Evidence for filament--filament collisions has recently been reported by Jim{\'e}nez-Serra et al. (2010) and Csengeri et al. (2011a) toward the IRDC G035.39-00.33 and DR21 region, respectively, via the detection of widespread SiO and weak and extended CH$_{3}$CN emission.

Despite all this progress, it still remains unclear how the parsec scale molecular filaments connect with the $\lesssim$0.1 pc scale OB star--forming cores.
Is the molecular gas funneled into the ultracompact (UC) H\textsc{ii} regions via molecular filaments? 
This question can be addressed by case studies of morphologies of dense molecular clouds. 

The majority of previous high angular resolution observations in high--mass star forming molecular clumps/cores focuses on the dynamically broadened systems, for the purposes of detecting rotation in massive disks, or global infall motions (e.g. Beltr{\'a}n et al. 2011; Galv{\'a}n-Madrid et al. 2009; Keto et al. 1987; Liu et al. 2010a, 2011; Sandell \& Wright 2010; Zapata et al. 2010; Zhang et al. 1997).
In these cases, the dense molecular gas usually concentrates to a flattened plane due to rotation or the magnetic support.
This geometrical flattening causes the structure to be blended along the line--of--sight, which is adverse for resolving the morphology.

With the still limited angular resolution of the current instruments, we need to look for nearly face--on systems to avoid blending.
Since the face--on orientation of the flattened accretion flows implies that the dominant motion is perpendicular to the line--of--sight, we expect line widths to be much smaller than the 
kinematically broadened line widths of edge--on systems, which are typically several km\,s$^{-1}$.
To select the face--on flattened accretion flows, we have searched for narrow line systems among NRAO\footnote{The National Radio Astronomy Observatory is a facility of the National Science Foundation operated under cooperative agreement by Associated Universities, Inc.} Very Large Array (VLA) archived molecular line data. 
Our target sources were selected from a VLA D--array NH$_{3}$ line dataset of OB cluster forming regions.
Part of our NH$_{3}$ database was obtained for the thesis work of Sollins (2005).
Among the 7 samples in our database ( G10.6-0.4 (or G10.62-0.38), G20.08-0.14N, G28.20-0.05, G28.29-0.36, G33.92+0.11, G35.58-0.03, G43.80-0.13; Galv{\'a}n-Madrid et al. 2009; Keto et al, 1987,1988; Liu et al. 2010a; Sollins 2005), the target G33.92+0.11 shows a line width of $\lesssim$2.5\,km\,s$^{-1}$, which is much narrower than those of rest of the targets, and is the only source where the kinematic broadening does not cause the main and the satellite hyperfine inversion lines of the NH$_{3}$ (J,K)=(1,1) transitions to be blended.
The UC H\textsc{ii} region G33.92+0.11 may represent an OB cluster forming region in a face--on configuration, providing an excellent opportunity to investigate the morphology, possible gravitational instability, and chemical stratification.

G33.92+0.11, at a distance of 7.1$^{+1.2}_{-1.3}$ kpc (Fish et al. 2003) has a bolometric luminosity of 2.5$\times$10$^{5}$ L$_{\odot}$.
Its strong free--free continuum emission at the centimeter bands ($\lambda$=3.6 cm: 1.1 Jy; $\lambda$=1.05 cm: 0.79 Jy), indicates the already existing embedded OB cluster.
Berkeley Illinois Maryland Array (BIMA) observations (10$''$ resolution) of $^{13}$CO/C$^{18}$O 1--0 reveal two associated massive molecular clumps, G33.92+0.11 A and G33.92+0.11 B.
The estimated H$_{2}$ mass  (based on C$^{18}$O emission) of these two molecular clumps are 5100 $M_{\odot}$ and 3200 $M_{\odot}$, respectively.
These values of H$_{2}$ mass are 10 times larger than the virial masses (G33.92+0.11 A: 520 $M_{\odot}$; G33.92+0.11 B: 270 $M_{\odot}$) estimated from the C$^{18}$O line width  (Watt \& Mundy 1999).
This could be due either to the existence of unusually strong support that counters the accretion flow, or to the fact that the majority of motions are perpendicular to the line--of--sight, yielding minimal blueshifted and redshifted velocities (see also Peretto 2006; Schneider et al. 2010; Smith et al. 2012). 
To our knowledge, we do not see any observational evidence for such unusually strong support, and thereby favor the later scenario.

To resolve whether and how the molecular accretion flow continues from the external parsec scale filaments to the embedded UC H\textsc{ii} region, we carried out mosaic observations of the thermal dust continuum and molecular line emission toward G33.92+0.11 using the Submillimeter Array (SMA; Ho, Moran, \& Lo 2004)\footnote{The Submillimeter Array is a joint project between the Smithsonian Astrophysical Observatory and the Academia Sinica Institute of Astronomy and Astrophysics, and is funded by the Smithsonian Institution and the Academia Sinica.}, at (sub--)arcsecond resolution.  
The primary goals of the present paper are to demonstrate the context of the filamentary structures, and to address the internal fragmentation of the molecular filaments  that leads to local intermediate-- or high--mass star formation. 
Details of our observations are summarized in Table \ref{chap_obs}.
Observational results of G33.92+0.11 are presented in Section \ref{chap_result}. 
In Section \ref{chap_discussion}, we discuss the possible interpretations of the G33.92+0.11 data, and the physical implication.
A summary is provided in Section \ref{chap_summary}.


\section{Observations and Data Reduction} 
\label{chap_obs}
\subsection{The SMA Observations}
\label{sub_smadata}
\subsubsection{The 1.3 mm Observations}
We performed observations at the wavelength of 1.3 mm using the SMA in subcompact, compact, extended, and very--extended array configurations. 
Details of the observations are summarized in Table \ref{table_tracks}. 
Tracks on 2011 March 16 and 2011 August 13 were acquired via the regular TAC review processes.
In these two tracks,  we carried out mosaic observations with hexagonal 7--pointings.
The central pointing of these observations is R.A. (J2000)=18$^{\mbox{h}}$52$^{\mbox{m}}$50$^{\mbox{s}}$.272, Decl. (J2000)= 00$^{\circ}$55$'$29$''$.604; pointings are separated by one half of the primary beam (27$''$.5). 
The rest of 1.3 mm observations were carried out using the filler observing time toward one single pointing on the central field.

The observing frequencies of the lower and the upper sidebands were centered at 218.8 GHz and 230.8 GHz, except for the filler observations on 2009 November 02.
The 218.8/230.8 GHz tuning was used to cover the CH$_{3}$CN 12$_{0}$--11$_{0}$ transition at 220.747265 GHz and could not simultaneously observe the H$_{2}$S 2$_{2,0}$--2$_{1,1}$ transition.
Continuum band visibility data at 1.3 mm were constructed by fitting all line--free channels, and were subtracted from the spectral line data.
The combined 1.3 mm continuum data cover a uv sampling range of 4--390 k$\lambda$, and have the maximal sensitivity and uv sampling in the central 1$'$ region.
There is no zero--spacing information available and that maps for spatially extended tracers should be treated with care.
We identify structures with reasonably compact spatial extents and do not address diffuse emission that may suffer from the lack of short spacings (Wilner \& Welch 1994).
The basic calibrations were carried out using the \texttt{MIR IDL} and the \texttt{Miriad} software packages.
Imaging was carried out using the \texttt{CASA} software package. 
We imaged several spectral lines, which are summarized in Table \ref{table_line}.
We note that the spectral transition H$_{2}$S 2$_{2,0}$--2$_{1,1}$ was only observed within the 2009 November 02 filler track. 
Owing to the poor uv coverage of the single short filler observation, we are not able to robustly image this transition. 
Since the H$_{2}$S molecule plays an important role in the sulfur chemistry (Charnley 1997; Hatchell et al. 1998; Buckle \& Fuller 2003; van der Tak et al. 2003; Herpin et al. 2009; Wakelam et al. 2004, 2011), we will still briefly discuss it, based on the amplitude of the uv visibility (see Section \ref{chap_s}).

Parameters of \texttt{CLEAN} (e.g. weighting, uv range, velocity width, etc) were optimized for the S/N ratio and angular/velocity resolution of individual images.
For example, to robustly image the extended emission of CO 2--1 isotopologues, we limited the uv range to 4--180 $k\lambda$, which covers the uv sampling range of the mosaic observations. These choices of the \texttt{CLEAN} parameters will be introduced in the captions of the relevant figures. We note that, if not explicitly indicated, the images presented in the following sections have been corrected for the primary beam response.

\subsection{The NH$_{3}$ Data}
\label{sub_nh3data}
We retrieved the NH$_{3}$ data from the VLA/EVLA online data archive. 
The NH$_{3}$ (J, K) = (1,1) and (2,2) hyperfine transitions were simultaneously observed by the NRAO Very Large Array (VLA) on 2003 February 14 in its D configuration (Project Code: AS0749).
The pointing center of these observations was R.A. (J2000)=18$^{\mbox{h}}$52$^{\mbox{m}}$50$^{\mbox{s}}$.273, Decl. (J2000)= 00$^{\circ}$55$'$29$''$.603.
The observations were performed using the VLA 4 IFs backend setting, with a velocity resolution of 0.6 km\,s$^{-1}$.
The NH$_{3}$ (J, K) = (3,3) hyperfine transitions were observed with the VLA D configuration on 2003 May 13 (Project Code: AS0771).
The observations were performed using the 2 IFs 2AC backend setting, with a velocity resolution of 0.3 km\,s$^{-1}$.
The parameters of the observed NH$_{3}$ transitions are summarized  in Table \ref{table_line}.
The uv sampling ranges of these observations are 2.6--76 $k\lambda$ and 2.1--82 $k\lambda$ for the February 14 track and the May 13 track, respectively.

Calibration and imaging of the data were carried out using the \texttt{AIPS} software package. 
Imaging with the Robust parameter equal to zero yielded the synthesized beams of 3$''$.9$\times$2$''$.8 (P.A. = -43.8$^{\circ}$) and 4$''$.3$\times$2$''$.8  (P.A. = -29.7$^{\circ}$) for the  February 14 track and the May 13 track, respectively.

\section{Results}
\label{chap_result}
This section presents new observational results on G33.92+0.11. 
We arrange this section such that the projected distributions of the selected tracers are introduced in Section \ref{chap_tracer}. 
Structures or regions that have particular excitation or chemical conditions are thereby identified.
The kinematics and spectral profiles of interesting structures or regions will be addressed in Section \ref{chap_kinematics}.

To describe spatial structures, we follow the existing nomenclature in the literature (e.g. Zhang et al. 2009; Liu et al. 2012). In this way, \textit{Massive molecular clumps} refer to structures with sizes of 0.5--1 pc, \textit{massive molecular cores} refer to $\lesssim$0.1 pc--size structures embedded within a clump, and \textit{condensations} refer to distinct molecular substructures within a core. 
\textit{Fragmentation} refers to the dynamical process that produces or enhances multiplicity. 
\textit{Molecular filaments} refer to geometrically elongated molecular structures, and \textit{molecular arms} refer to segments of molecular filaments that are located within the $\lesssim$1 pc radii of molecular clumps and are not internally embedded within molecular clumps.
We note that the structures defined here can be 10--100 times denser than those previously introduced by Bergin \& Tafalla (2007, see their Table 1).

\subsection{Spatial Distribution of Tracers}
\label{chap_tracer}

\subsubsection{Millimeter and Submillimeter Continuum Emission}
\label{chap_ch0}
\begin{figure}
\vspace{-1.5cm}
\begin{tabular}{  p{8.4cm} p{8.4cm} }
\includegraphics[scale=0.43]{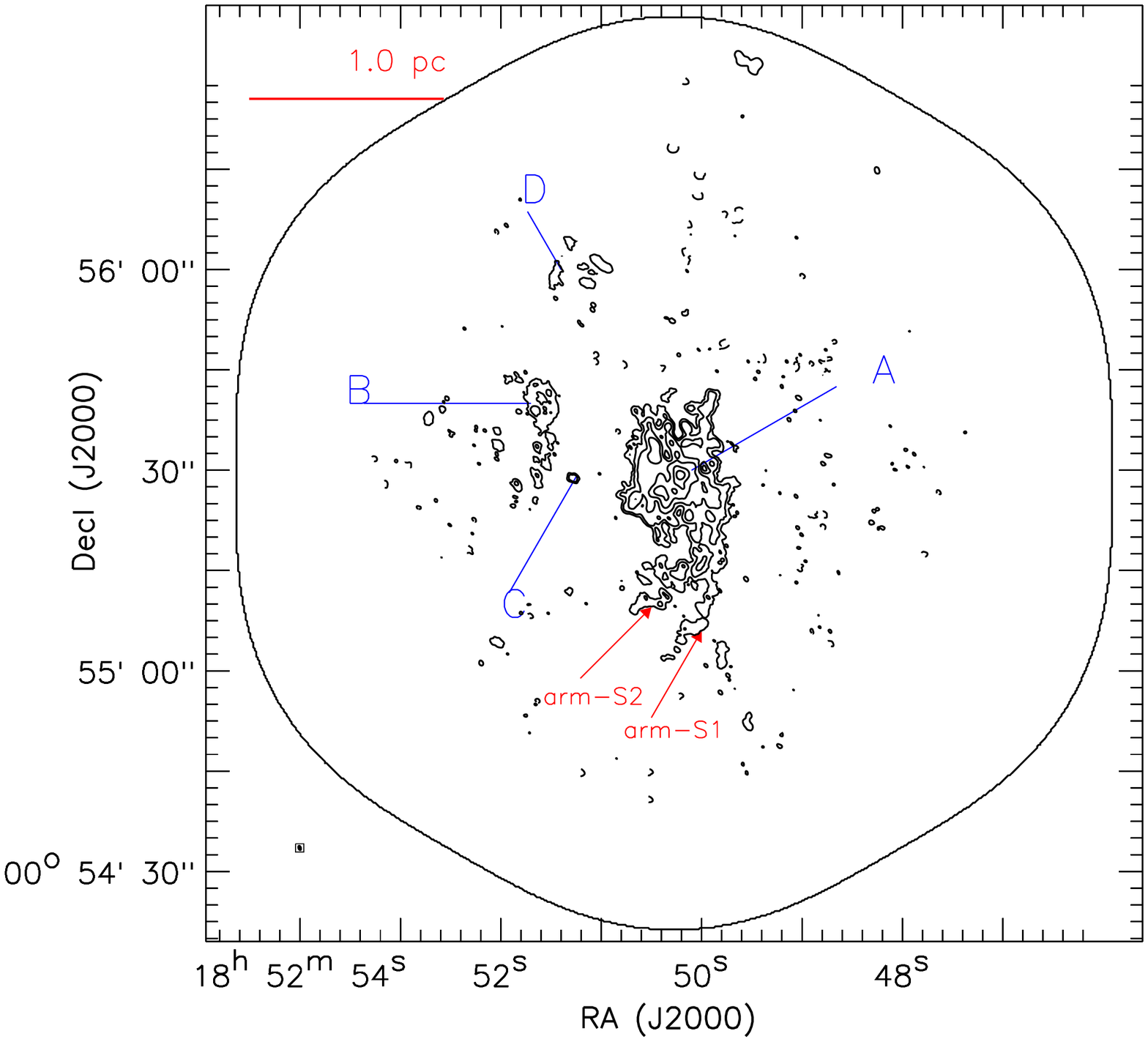} &
\includegraphics[scale=0.43]{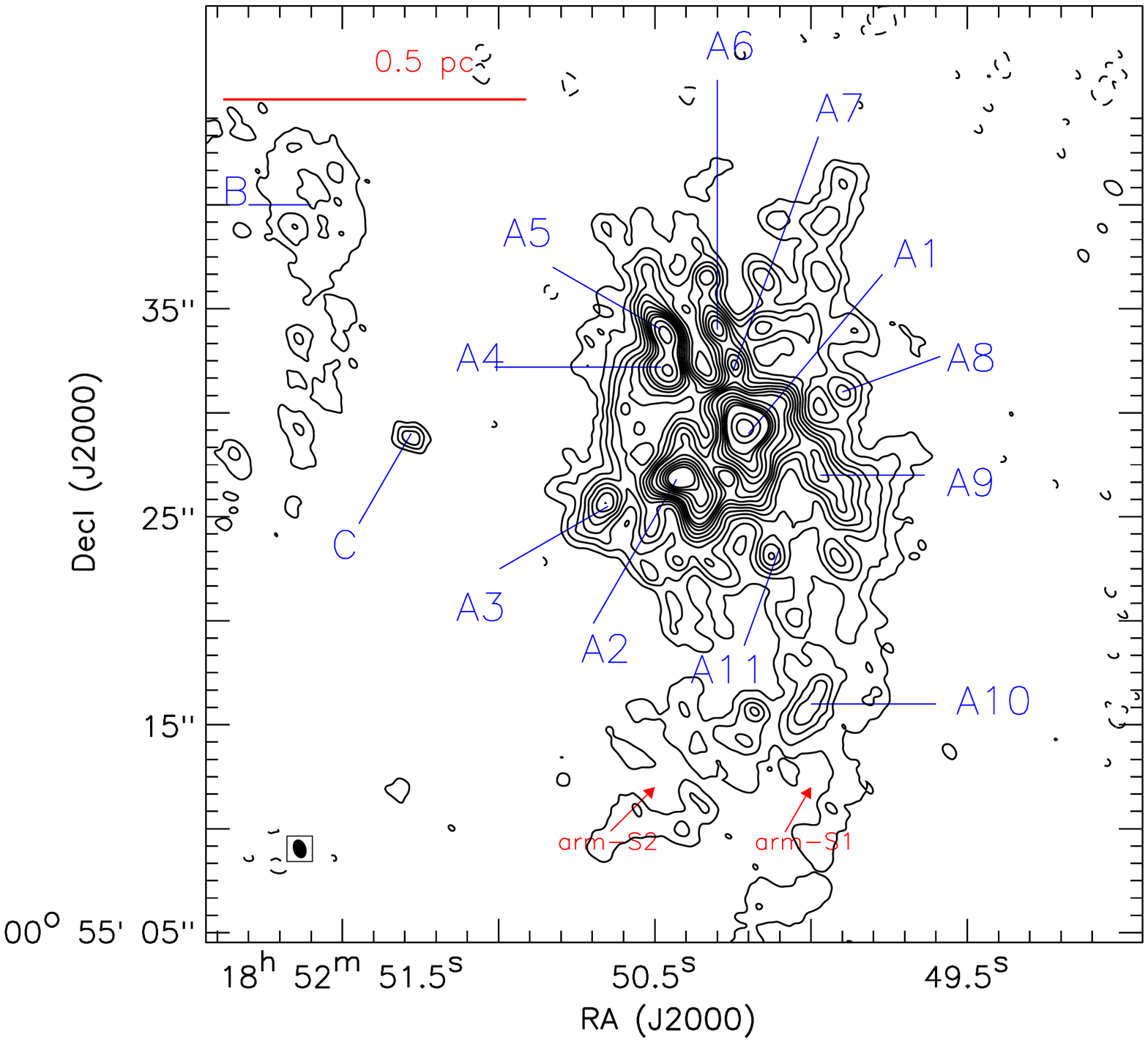} \\
\end{tabular}
\caption{SMA observations of 1.3 mm continuum emissions. 
\textbf{Left: } The SMA observations of the 1.3 mm continuum emission, imaged using natural weighting ($\sigma$$\sim$0.45 mJy\,beam$^{-1}$; $\theta_{maj}$$\times$$\theta_{min}$: 0$''$.92$\times$0$''$.63). Contour levels are: 1.2 mJy\,beam$^{-1}$$\times$[-2, -1, 1, 2, 4, 8, 16, 32]. The peak emission is 57.7 mJy\,beam$^{-1}$ ($\sim$2.3K).  
The mosaic field of view (gain 0.1) is encircled by a black curve. 
\textbf{Right: } The central region of the naturally weighted 1.3 mm continuum image.  Contour levels are: 1.2 mJy\,beam$^{-1}$$\times$[-2, -1, 1, 2, 3, 4, 5, 6, 7, 8, 9, 10, 12, 14, 16, 18, 20, 30]. 
Images in this figure are not corrected for the primary beam response for the sake of demonstrating the significantly detected structure. 
The selection of the labeled cores A($_{1-11}$)--D is based on \texttt{clumpfind} with removal of entities with less background contrast.
Two parsec scale molecular arms are labeled by arm--S1 and arm--S2.
Within the central $\sim$0.6 pc area, we see few elongated structures. 
Geometrically, one elongated structure may appear as a continuation of arm--S1 or S2, toward core A9 and A1. 
Few others may continue from A3$\rightarrow$A2$\rightarrow$A1, from A5$\rightarrow$A4$\rightarrow$A2, and from A6$\rightarrow$A7$\rightarrow$A1.
More systematic structure identification requires single dish observations to improve the sampling in short spacing. 
}
\label{fig_ch0}
\end{figure}

Figure \ref{fig_ch0} shows the SMA observations of the 1.3 mm continuum emission (natural weighting). 
The majority of the continuum emission is due to thermal dust emission, which is one of the most reliable tracers of molecular mass. An exception is the G33.92+0.11 A region, where a significant fraction of free-free continuum emission is found toward the UC H\textsc{ii} region detected by Watt \& Mundy (1999, see also below).

In the left panel of Figure \ref{fig_ch0}, we identify apparent continuum sources, labeled with A--D. 
The continuum sources G33.92+0.11 A and B were resolved by previous BIMA observations of $^{13}$CO/C$^{18}$O 2--1 (Watt \& Mundy 1999).
A compact continuum source G33.92+0.11 C is located in between the continuum sources A and B, and might be confused with the extended structures around sources A and B in previous BIMA observations. 

Two elongated, $\lesssim$1 pc scale structures (i.e. arm--S1 and arm--S2) appear to be connected with the south of G33.92+0.11 A (Figure \ref{fig_ch0}). 
The detected structures are not smooth.
Instead, they show a clumpy and protrusive morphology.  
The blow up of the 1.3 mm continuum images in the central $\sim$1 pc area (right panel in Figure \ref{fig_ch0}) shows that arm--S1 and S2 may continue inward a radius of 0.3 pc.
Each of these arms show embedded local substructures.
We execute the \texttt{clumpfind2D} algorithm (Williams et al. 1994) on the 1.3 mm continuum image made with natural weighting using the parameter: \texttt{levels = 0.0012 * [3,4,5,6,7,8,9,10,12,14,16,18,20,30]}\,Jy\,beam$^{-1}$ .
This yields 28 cores in the G33.92+0.11 A, C region. 
We visually inspect the \texttt{clumpfind2D}--identified cores and reject entities that do not have significant contrast against the local background emission, or have very irregular shapes.
We labeled the remaining local dense cores as A1--11.
Among these \texttt{clumpfind2D}--identified cores, A1--11 contribute to $\sim$89\% of the summed 1.3 mm flux in cores. 
This implies that the average mass of the rest of \texttt{clumpfind2D}--identified 17 cores is only 8\% of the average flux of the 11 most massive cores. 

\begin{figure}
\vspace{-0.5cm}
\includegraphics[scale=0.45]{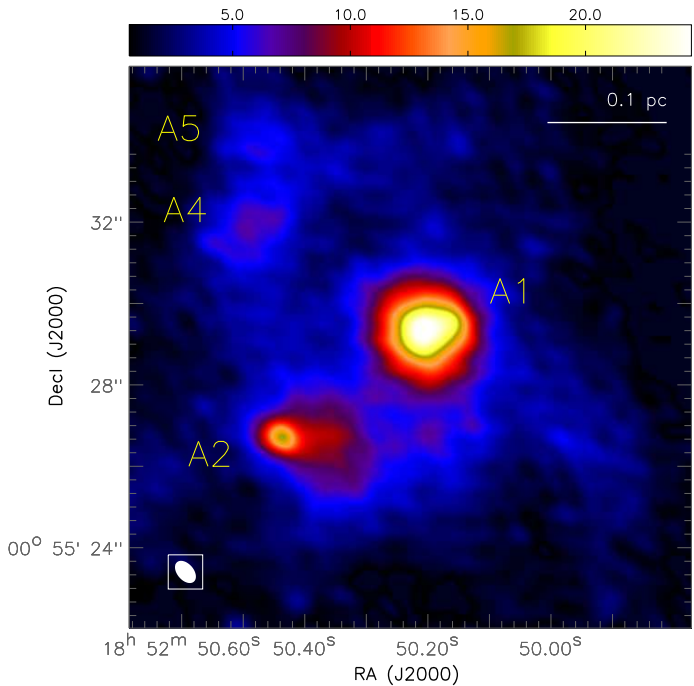}
\caption{ The Briggs Robust 0 weighting 1.3 mm continuum image ($\theta_{maj}$$\times$$\theta_{min}$: 0$''$.63$\times$0$''$.40). The color bar has the unit of mJy\,beam$^{-1}$. The strongest sources A1 and A2 in this figure may be binary, projectively separated by 0.1 pc. 
A2 is connected with elongated molecular structures of $\lesssim$0.05 pc. 
}
\label{fig_345ch0}
\end{figure}

Figure \ref{fig_345ch0} shows the blow up of the Briggs Robust 0 weighting 1.3 mm continuum image. 
From the 1.3\,mm image, we see that core A2 is connected with one or two mini arms toward the west. 
Both cores A1 and A2 are not exactly round in the inner $\lesssim$1$''$ area, but are slightly oblate in the north--south direction.

The sources G33.92+0.11 A and B have 1.3\,mm fluxes of 2.75 Jy and 0.32 Jy, respectively. 
The extrapolation  of the spectral energy distribution reported in Watt \& Mundy (1999) suggests that the free--free continuum emission amounts to 0.6--0.7 Jy (we adopt the value of 0.67 Jy) at 1.3 mm, in the G33.92+0.11 A region. 
According to the spectral energy distribution, the rest of the 1.3 mm flux should be due to thermal dust continuum emission.
More discussions about the emission of the ionized gas are deferred to Section \ref{chap_rrl}. 

Assuming an average temperature of 35 K and a gas-to-dust mass ratio of 100, the 2.1 Jy flux at 1.3 mm toward region G33.92+0.11 A corresponds to a molecular gas mass of 1600, 3000 and 5800 $M_{\odot}$ for $\beta$= 1, 1.5 and 2, respectively (Lis et al. 1998).
Similarly, the 0.32 Jy flux of G33.92+0.11 B corresponds to a gas mass of 180, 350, and 670 $M_{\odot}$ for $\beta$= 1, 1.5 and 2.
The average gas temperature of 35 K is measured based on our NH$_{3}$ observations of multiple (J, K) levels (Section \ref{chap_nh3}).
Our estimates of the molecular mass in G33.92+0.11 A with various assumptions of $\beta$, are consistent with the value reported in Watt \& Mundy (1999) within a factor 4 (see Section \ref{chap_introduction}).
We detected $\sim$5--18 times less molecular mass in G33.92+0.11 B than Watt \& Mundy (1999), which may mainly be due to a large fraction of missing flux implying that G33.92+0.11 B is still relatively lack of structures.

Based on the same assumptions of the gas-to-dust mass ratio and $\beta$, estimates of the molecular mass of cores A1--11 are provided in Table \ref{table_cores}.
We found that in the G33.92+0.11 A region (including arm--S1,S2), about 24.3\% of the molecular mass is already concentrated in the compact cores A1--A11. 
About 11\% of the overall molecular mass is concentrated in the centrally embedded cores A1 and A2. 

The A1 and A2 cores, separated by $\sim$0.1 pc, are massive, and are candidates to form compact OB clusters. 
Massive (binary) cores are also seen in objects in different evolutionary stages. 
For example,  
SMA observations toward G28.34+0.06 P2 ($d\sim$4.8 kpc) also showed a close pair of cores formed by SMA1 (100 $M_{\odot}$) and SMA2 (30 $M_{\odot}$), separated by $\sim$0.1 pc (Zhang et al. 2009). 
Toward the UC H\textsc{ii} region G10.6-0.4, a hot molecular toroid of few tens of solar masses (Liu et al. 2010b; Liu et al. 2011) surrounding an OB cluster of $\sim$175 $M_{\odot}$ was suggested from previous VLA and SMA observations, and may represent a later evolutionary stage. 
Chen et al. (2006) and Csengeri et al. (2011b)  also reported massive protostellar cores embedded in the W3(H$_{2}$O) and the DR21 (OH) region.
The fact that the detected cores tend to show masses higher than the Jeans mass may suggest that this is a common feature in massive molecular clumps.

\subsubsection{The CO 2--1 Isotopologues and N$_2$D$^+$ Emission}
\label{chap_co}
\begin{figure}
\begin{tabular}{  p{7.5cm} p{7.5cm} }
\includegraphics[scale=0.45]{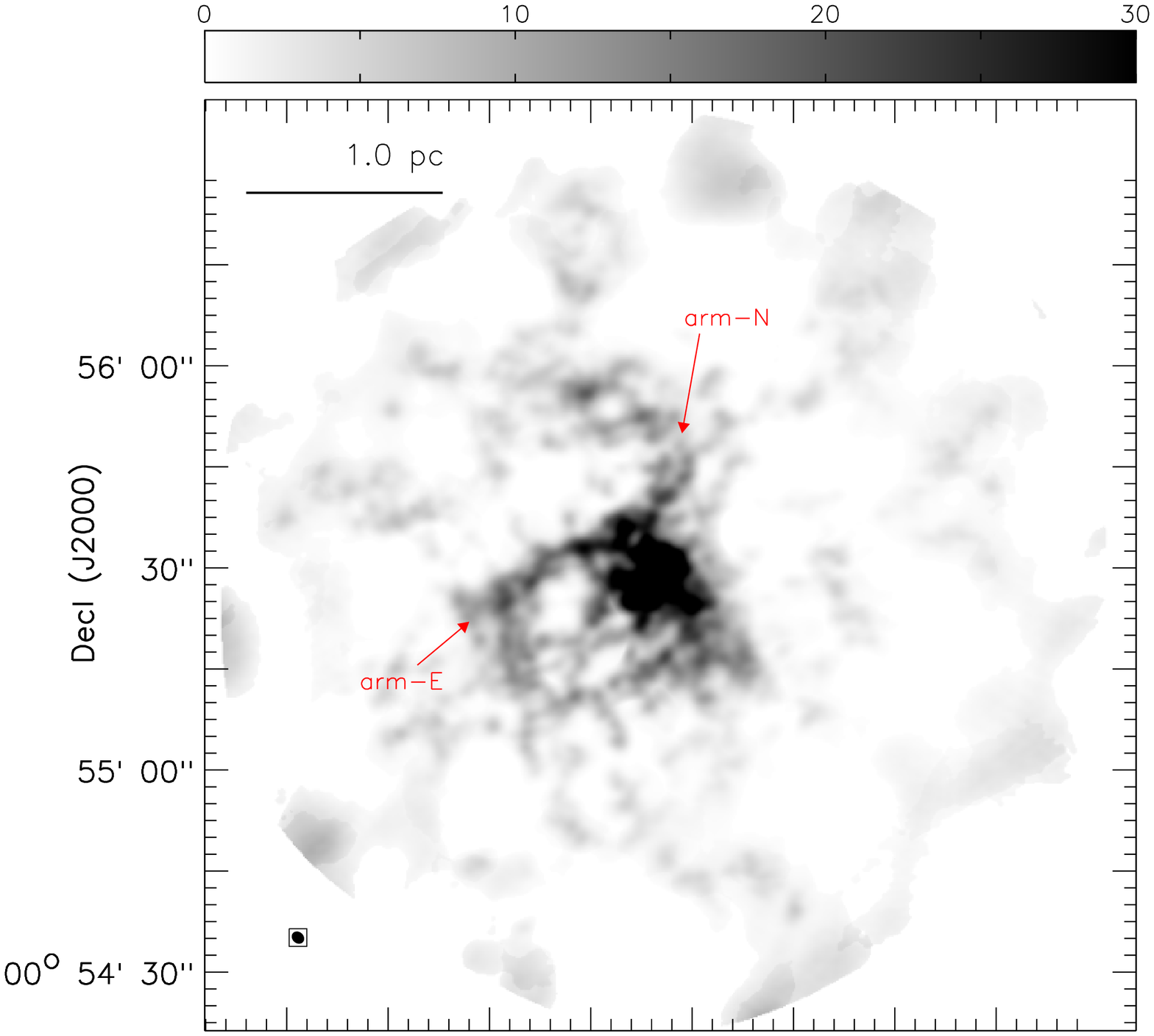} &
\includegraphics[scale=0.45]{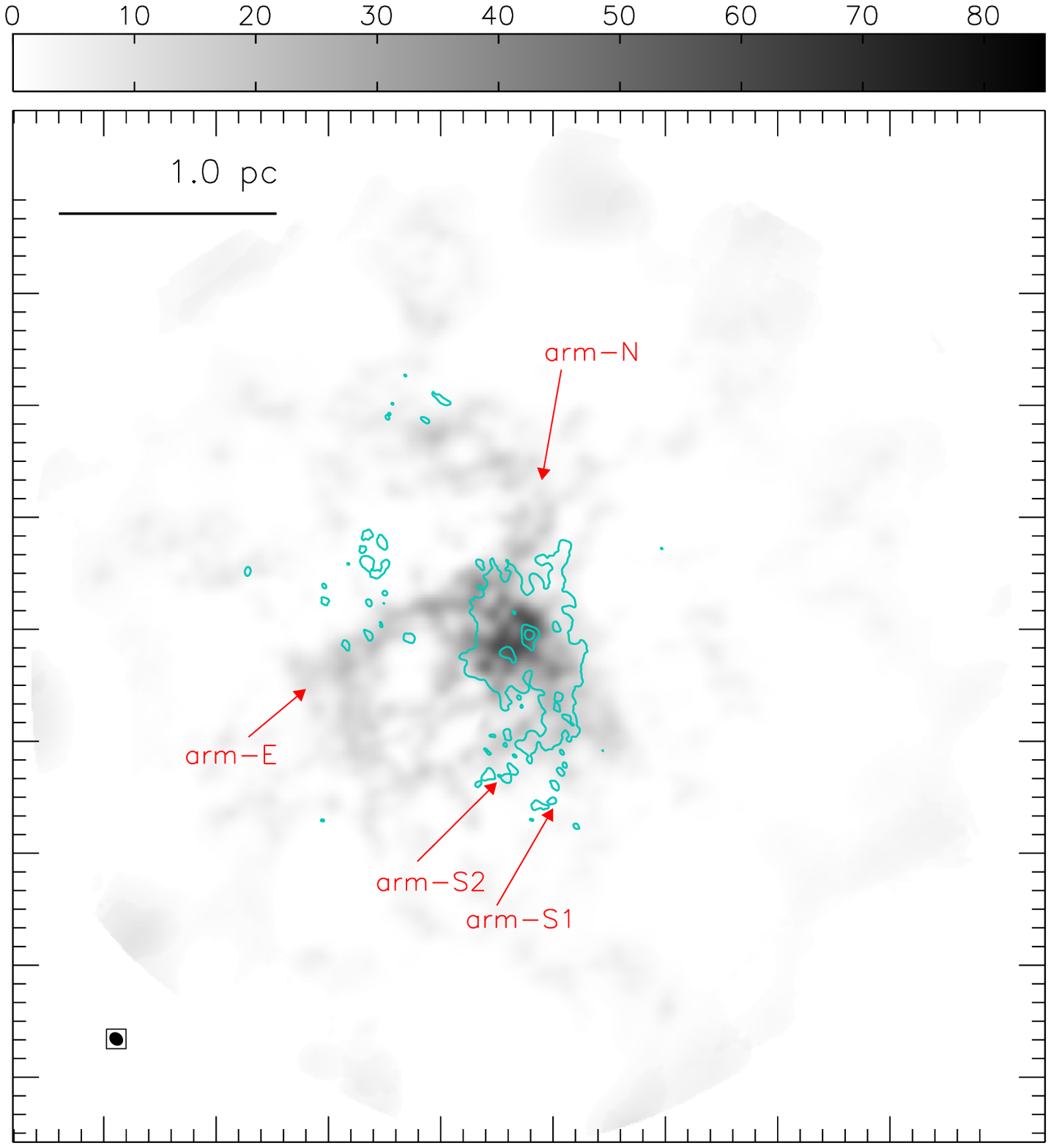} \\
\end{tabular}

\vspace{-1.5cm}

\begin{tabular}{  p{7.5cm} p{7.5cm} }
\includegraphics[scale=0.45]{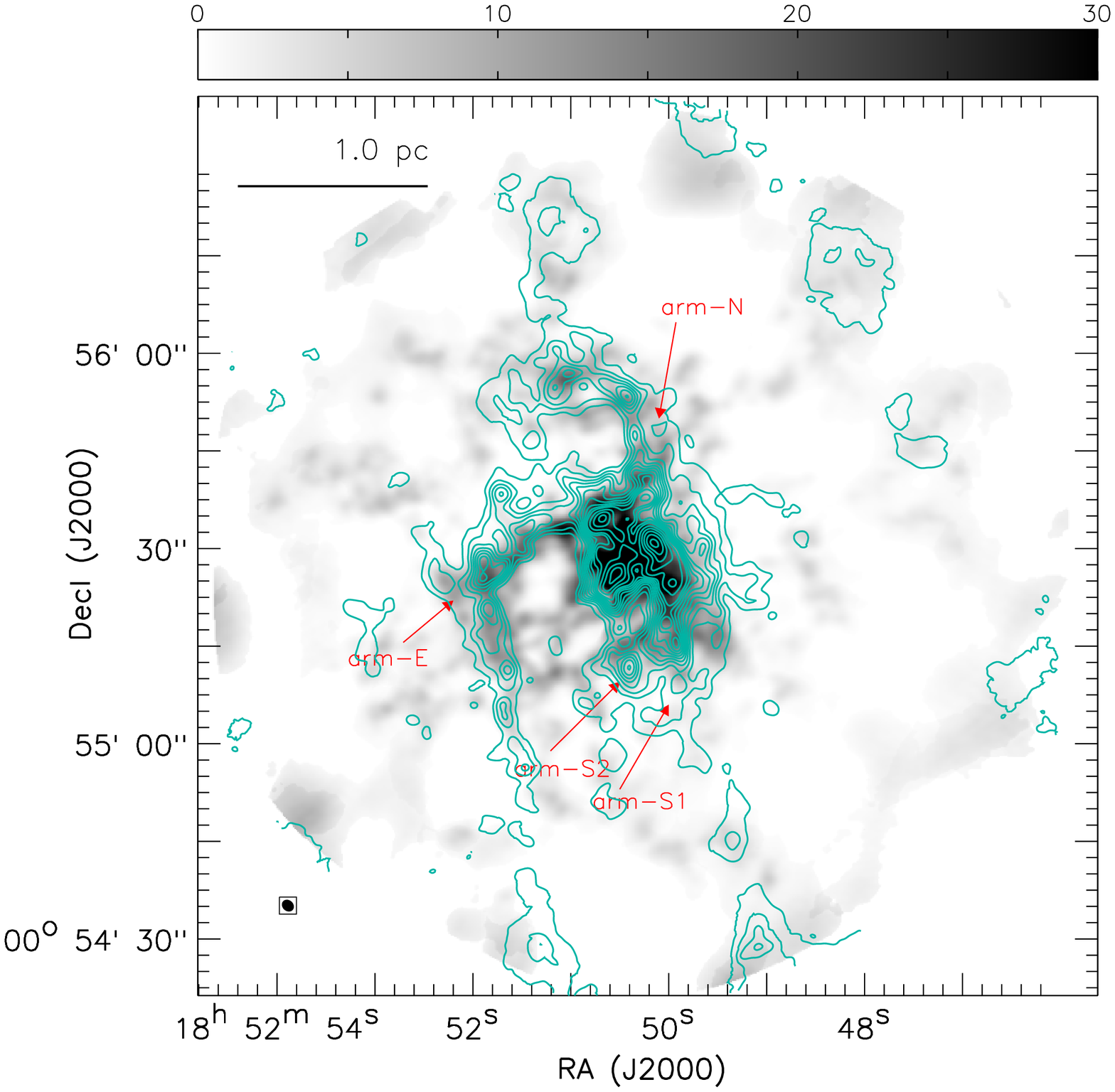}  & 
\includegraphics[scale=0.45]{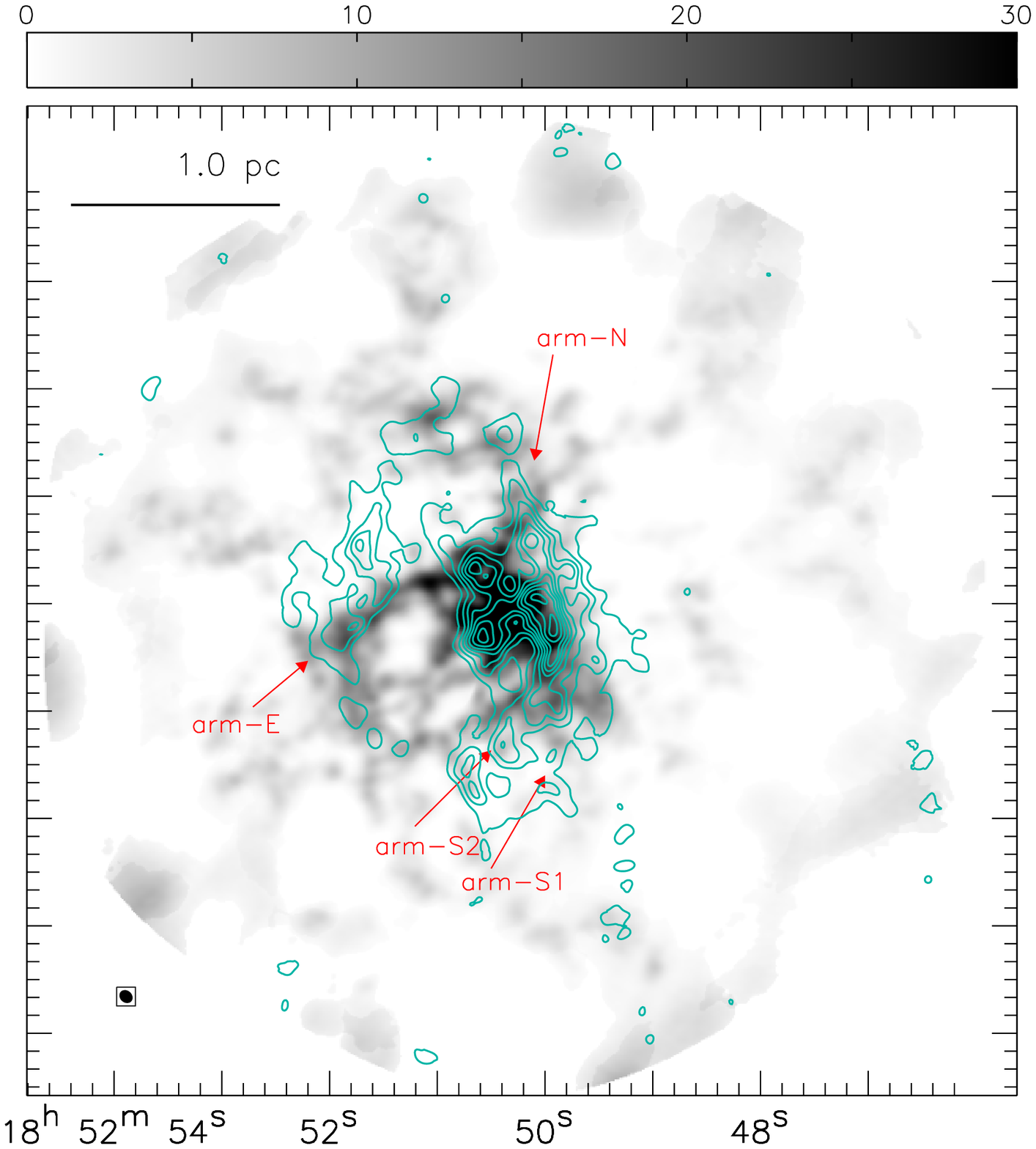}  \\
\end{tabular}
\caption{The velocity integrated intensity maps of CO 2--1 isotopologues. 
\textbf{Top Left: } The $^{12}$CO 2--1 map. (uv range: 4--180 $k\lambda$; $\theta_{maj}$$\times$$\theta_{min}$: 1$''$.96$\times$1$''$.63; velocity range: 100.0 -- 112.4 km\,s$^{-1}$). The rms noise level in each 6 km\,s$^{-1}$ velocity channel is  0.011 Jy\,beam$^{-1}$ (0.08 K) in line free channels. 
\textbf{Top Right: } The $^{12}$CO 2--1 map (grayscale) overlaid with the natural weighted 1.3 mm continuum map (contour). Levels are [1.8, 19.2, 32.4] Jy\,beam$^{-1}$.
\textbf{Bottom Left: } The $^{12}$CO 2--1 map (grayscale) overlaid with the $^{13}$CO 2--1 map (contour; uv range: 4--180 $k\lambda$; $\theta_{maj}$$\times$$\theta_{min}$: 1$''$.96$\times$1$''$.63; velocity range:  104.0--106.4 and 107.6--110.0 km\,s$^{-1}$). The rms noise level in each 1.2 km\,s$^{-1}$ velocity channel is  0.019 Jy\,beam$^{-1}$ (0.15 K) in line free channels. Contour levels are 1.0 Jy\,beam$^{-1}$km\,s$^{-1}$$\times$[1, 2, 3, 4, 5, 6, 7, 8, 9, 10, 12, 14, 16, 18, 20, 22, 24, 26, 28, 30].
\textbf{Bottom Right: } The $^{12}$CO 2--1 map (grayscale) overlaid with the C$^{18}$O 2--1 map (contour; uv range: 4--180 $k\lambda$; $\theta_{maj}$$\times$$\theta_{min}$: 1$''$.96$\times$1$''$.63; velocity range:  104.0--106.4  and 107.6--110.0 km\,s$^{-1}$). The rms noise level in each 1.2 km\,s$^{-1}$ velocity channel is  0.019 Jy\,beam$^{-1}$ (0.15 K) in line free channels. Contour levels are 1.0 Jy\,beam$^{-1}$km\,s$^{-1}$$\times$[1, 2, 3, 4, 5, 6, 7, 8, 9, 10, 12].
Color bars have the unit of Jy\,beam$^{-1}$km\,s$^{-1}$.
When generating these velocity integrated map, pixel values less than 0.03 Jy\,beam$^{-1}$ are masked to suppress the noise, negative sidelobes, and the contribution from the foreground emission. 
By comparing the $^{12}$CO, $^{13}$CO and C$^{18}$O images, we further identify two, parsec scale molecular arm, arm--N and arm--E. 
Arm--E is projectively connected with elongated structures that have roughly north--south alignments. 
}
\label{fig_co}
\end{figure}

\begin{figure}
\includegraphics[scale=0.65]{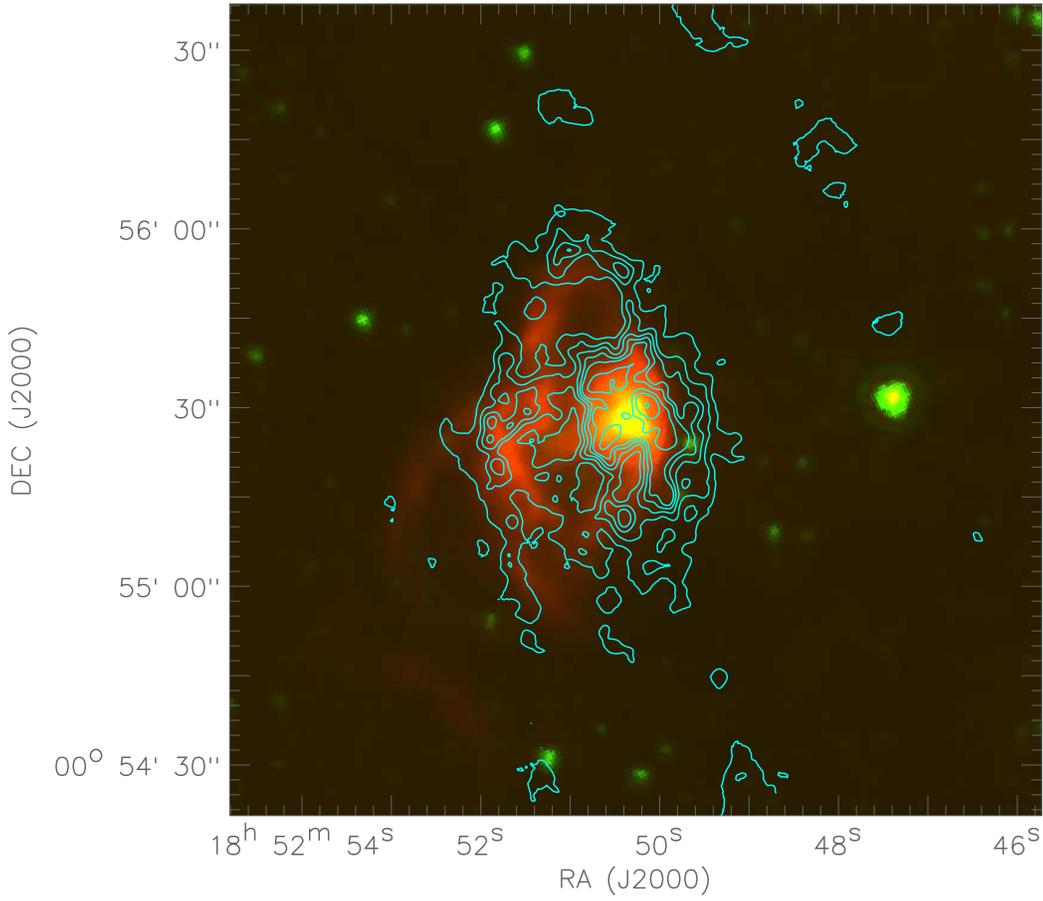}
\caption{The GLIMPSE 4.5 $\mu$m image (green), the GLIMPSE 8 $\mu$m image (red), and the velocity integrated map of $^{13}$CO 2--1 (contours; without being corrected for the primary beam response). Contour levels are 1.0 Jy\,beam$^{-1}$km\,s$^{-1}$$\times$[2, 4, 6, 8, 10, 15, 20, 30, 40, 50, 60].  
In this figure, we see strong 4.5 $\mu$m and 8 $\mu$m emission at the 1.3 mm continuum peak, indicating the centrally embedded OB cluster.
The 8 $\mu$m emission shows extended filamentary features, that are consistent with the PDR regions in arm--N, arm--S2, arm--E, and the north--south elongated structures connected with arm--E. These PDR regions may be illuminated by the centrally embedded OB cluster.}
\label{fig_rgb}
\end{figure}

Figure \ref{fig_co} shows the velocity integrated intensity maps of the $^{12}$CO 2--1, the $^{13}$CO 2--1, and the C$^{18}$O 2--1 transitions. 
Figure \ref{fig_rgb} compares the velocity integrated intensity map of the $^{13}$CO 2--1 transition, with the \textit{Spitzer} Infrared Array Camera (IRAC) observations of the 4.5 $\mu$m and 8 $\mu$m emission. 
The \textit{Spitzer} data are obtained from the archival system of the GLIMPSE survey. 
Both the 8 $\mu$m and 4.5 $\mu$m emission can trace the embedded stellar objects. 
The 8 $\mu$m emission can trace the very warm dust at the surface layer between the H\textsc{ii} region and the neutral gas; the \textit{Spitzer} IRAC 8 $\mu$m band also covers strong Polycyclic Aromatic Hydrocarbons (PAHs) bands that are typically excited in photon--dominated regions (PDRs; see e.g. Berne et al. 2009).
The IRAC 4.5 $\mu$m band covers the shock--excited H$_{2}$ lines and $^{12}$CO (v=1--0) bandhead.
Inspecting the emission detected in these two bands helps diagnose the locations of intermediate mass or high mass stars, and the illumination by UV photons. 

The primary results from these two figures are:
\begin{itemize}
\item[1.] The emission from the CO 2--1 isotopologues consistently trace the extension seen toward the southeast where arm--S1 and arm--S2 are found.  From the velocity integrated emission map of $^{13}$CO 2--1 and C$^{18}$O 2--1, a more extended molecular structure is detected toward the southwest of arm--S1.
\item[2.] In addition to the arm--S1 and arm--S2, we find other two molecular arms, arm--E and arm--N, which are connected with G33.92+0.11 A toward the east and north of this source. 
\item[3.] Perpendicular to arm--E, two other molecular arms, arm--EN and arm--ES, are detected, with arm--EN being projectively connected with G33.92+0.11 B.
\item[4.]  From Figure \ref{fig_rgb}, we see that the molecular arms arm--S2, arm--N, arm--E, and arm--EN, ES are associated with 8 $\mu$m emission probing the PDR.  These molecular arms are likely illuminated by the OB stars embedded in G33.92+0.11 A. We note that all CO molecular arms also appear in other molecular tracers such as SO and NH$_3$ (see Sections 3.1.3 and 3.1.8), indicating that they are coherent structures likely connected to the central hub formed by G33.92+0.11 A and B.
\end{itemize}

Around the systemic velocity $v_{lsr}$ of 107.6 km\,s$^{-1}$, the averaged brightness ratio of C$^{18}$O and the $^{13}$CO 2--1 lines is about 0.62.
Assuming\footnote{The assumption of the X($^{13}$CO)/X(C$^{18}$O) ratio here follows the previously assumptions of X($^{12}$CO) = 10$^{-4}$ and X($^{12}$CO)/X(C$^{18}$O) = 500 used in the literature (Ho, Terebey \& Turner 1994), and the measured $^{12}$C/$^{13}$C$\sim$60 in molecular clouds (Wilson \& Rood 1994 ; Lacy et al. 1994). We make these assumptions to allow a consistent comparison with the case studies of the OB cluster forming region G10.6-0.4 addressed in Ho, Terebey \& Turner  (1994).} that X($^{13}$CO)/X(C$^{18}$O) = 8.3, the $^{13}$CO and C$^{18}$O images have the same fraction of missing flux, and the emission is under local thermaldynamic equilibrium (LTE), the average brightness ratio 0.62 corresponds to optical depths of $\tau_{^{13}co}$$\sim$4.8, and $\tau_{c^{18}o}$$\sim$0.96. 
Since the $^{13}$CO 2--1 transition is optically thick, and C$^{18}$O 2--1 is on average, marginally optically thin/thick, the fact that we see stronger integrated emission near the centrally embedded OB cluster is due to higher local gas temperatures and the presence of localized broad line emission. 

Assuming X($^{12}$CO)/X(C$^{18}$O) = 500, LTE, a gas excitation temperature of 35 K, and $\tau_{c^{18}o}$$\sim$0.96, the derived column density of the $^{12}$CO molecules is N$_{^{12}CO}$$\sim$1.3$\times$10$^{19}$($\frac{\Delta v}{\mbox{1 km\,s$^{-1}$}}$)\,cm$^{-2}$, where $\Delta v$ is the velocity dispersion. 
Assuming X($^{12}$CO)=10$^{-4}$, the estimated H$_{2}$ column density is therefore 
\begin{equation}
\mbox{ N$_{H_{2}}$$\sim$1.3$\times$10$^{23}\times$($\frac{\Delta v}{\mbox{1 km\,s$^{-1}$}}$)\,cm$^{-2}$}.
\end{equation}
The mass in a certain area $A$ is $M_{A}$ = 975 $\times$($\frac{\Delta v}{\mbox{1 km\,s$^{-1}$}}$)($\frac{A}{\mbox{1 pc$^{2}$}}$) $M_{\odot}$.
The error associated with the gas mass and density estimates is mainly caused by uncertainties in the abundance of the CO isotopologues. 
The surface mass density estimated from our $^{13}$CO/C$^{18}$O observations is consistent with the surface mass estimated from the 1.3 mm continuum emission as well as the estimates by Watt \& Mundy (1999).
We do not see evidence of significant CO depletion. 

We can also estimate the volume density n(H$_{2}$) = 4.3$\times$10$^{5}$($\frac{\Delta v}{\mbox{1 km\,s$^{-1}$}}$)($\frac{\ell}{\mbox{0.1 pc}}$)$^{-1}$\,cm$^{-3}$, where $\ell$ is the line--of--sight thickness.
For G33.92+0.11,  $\Delta v$ ranges from 0.5--3 km\,s$^{-1}$ (see also Watt \& Mundy 1999).
The characteristic value of $\ell$ should be comparable to the resolved thickness of the molecular arms, which is $\sim$0.1--0.2 pc (Figure \ref{fig_co}). 
The derived volume density is n(H$_{2}$)=10$^{5}$--10$^{6}$ cm$^{-3}$.
This volume density corresponds to a free--fall timescale of 3.6--12$\times$10$^{4}$ years. 
However, we note that the volume density is probably much higher toward the localized dense structures detected within the arm.
For an average gas kinetic temperature of 35 K, a volume density of n(H$_{2}$)=10$^{5}$--10$^{6}$ cm$^{-3}$, and ignoring other support mechanisms, the Jeans length scale is 0.05--0.16 pc (1$''$.5--4$''$.7).
The estimated Jeans mass is 3--9 $M_{\odot}$, which is smaller than the core masses shown in Table \ref{table_cores}. 
We note that while the Jeans length scale matches scales at which structures are no longer supported by the thermal pressure, the Jeans mass only indicates a lower limit of mass for localized structures at those scales. 

Finally, within our 1.3mm SMA frequency setup, we simultaneously observed the N$_{2}$D$^{+}$ 3--2 transition. 
However, our SMA images yield no significant detection of this molecular line (channel width: 1.2 km\,s$^{-1}$; $\theta_{maj}$$\times$$\theta_{min}$: 2$''$.0$\times$1$''$.9; 3$\sigma$: 72 mJy\,beam$^{-1}$). Since deuterated species such as N$_2$D$^+$ or DCO$^+$ are found to be enhanced in the cold gas of dense pre-stellar and starless cores where CO freezes out onto dust grains (see e.g. Caselli et al. 1999; Crapsi et al. 2005; Busquet et al. 2010), the non-detection of this molecule is consistent with a lack of CO depletion found toward G33.92+0.11.

\subsubsection{The Sulfur--Bearing Molecules}
\label{chap_s}
Our 1.3 mm observations cover line transitions of several important sulfur--bearing molecular species such as H$_{2}$S, SO, SO$_{2}$, $^{13}$CS and OCS.
The abundance ratio of these sulfur--bearing molecules has been used as chemical clocks (see also Li et al. 2011) for the process of high-mass star formation (e.g. Charnley 1997; Viti et al. 2004; Wakelam et al. 2004), although these species are also known to be enhanced in shocked gas in molecular outflows (Bachiller \& Perez Gutierrez 1997; Jimenez-Serra et al. 2005), or found to survive from UV photon illumination toward PDRs (Goicoechea et al. 2006). 

\begin{figure}
\includegraphics[scale=0.4]{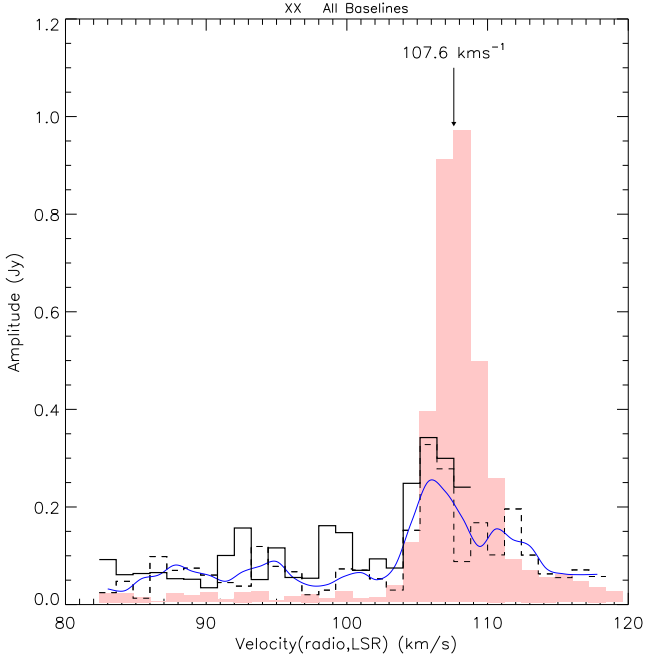}
\caption{The visibility amplitudes of the H$_{2}$S 2$_{2,0}$--2$_{1,1}$ transition (black solid) and the SO 5$_{6}$--4$_{5}$ transition (black dashed). These single polarization data are averaged from all baselines, from the compact array observations (2009 November 02). Due to residing at the band edge, the redshifted emission of the H$_{2}$S 2$_{2,0}$--2$_{1,1}$ transition was not fully covered in the observation. Both transitions show strong emission from 104.0 km\,s$^{-1}$ to 107.6 km\,s$^{-1}$. The SO 5$_{6}$--4$_{5}$ transition show high amplitude wing from 107.6 km\,s$^{-1}$ to 113.6 km\,s$^{-1}$, which can be better seen in the spectrum smoothed with a 3.6 km\,s$^{-1}$ window (blue solid). Filled region shows 0.2 times the visibility amplitudes of SO 5$_{6}$--4$_{5}$, averaged from the central pointing (with all baselines) of the subcompact array observation (2010 March 16).}
\label{fig_h2s}
\end{figure}

As mentioned in Section \ref{sub_smadata}, the H$_{2}$S 2$_{2,0}$--2$_{1,1}$ transition was only observed in the compact configuration (2009 November 02) track, and cannot be robustly imaged due to the poor uv sampling. 
In Figure \ref{fig_h2s}, we compare the scalar averaged visibility amplitudes of the H$_{2}$S 2$_{2,0}$--2$_{1,1}$ transition with the scalar averaged visibility amplitudes of the SO 5$_{6}$--4$_{5}$ transition, taken from the compact array observations. 
We see that for the scales covered by the uv sampling range ($\lesssim$15$''$, i.e. 0.53 pc), the shape and amplitude of these spectra agree well with each other indicating that these two molecules likely trace the same optically thick, compact structures toward G33.92+0.11 A and B.
This is consistent with models of the hot core chemistry where H$_2$S and SO appear to be abundant for time--scales of several times 10$^4$$\,$yrs (e.g. Charnley et al. 1997). 
Whether H$_{2}$S and SO are spatially coincident remains to be verified after supplementing the short spacing observations of the H$_{2}$S 2$_{2,0}$--2$_{1,1}$ transition. 
We note that the spectrum of SO 5$_{6}$--4$_{5}$ further shows a redshifted broad line wing indicating the presence of molecular outflows.
If we compare the compact SO 5$_{6}$--4$_{5}$ visibility amplitude with those taken with the subcompact array (2011 March 16), we find that the compact array observation miss $\sim$90\% of integrated flux, suggesting that the SO emission is very extended across the G33.92+0.11 region.

\begin{figure}
\includegraphics[scale=0.45]{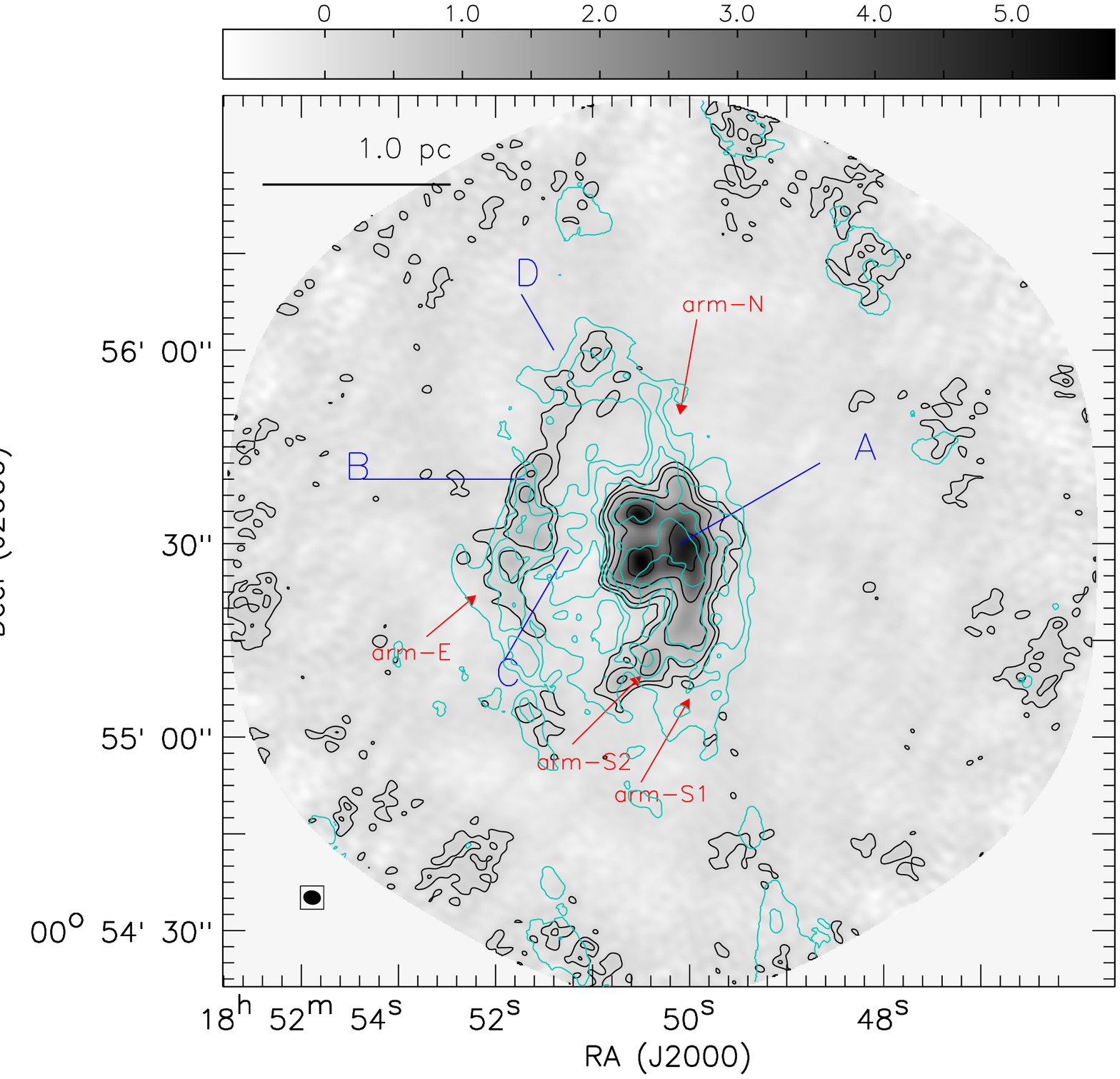}
\hspace{-2cm}
\includegraphics[scale=0.45]{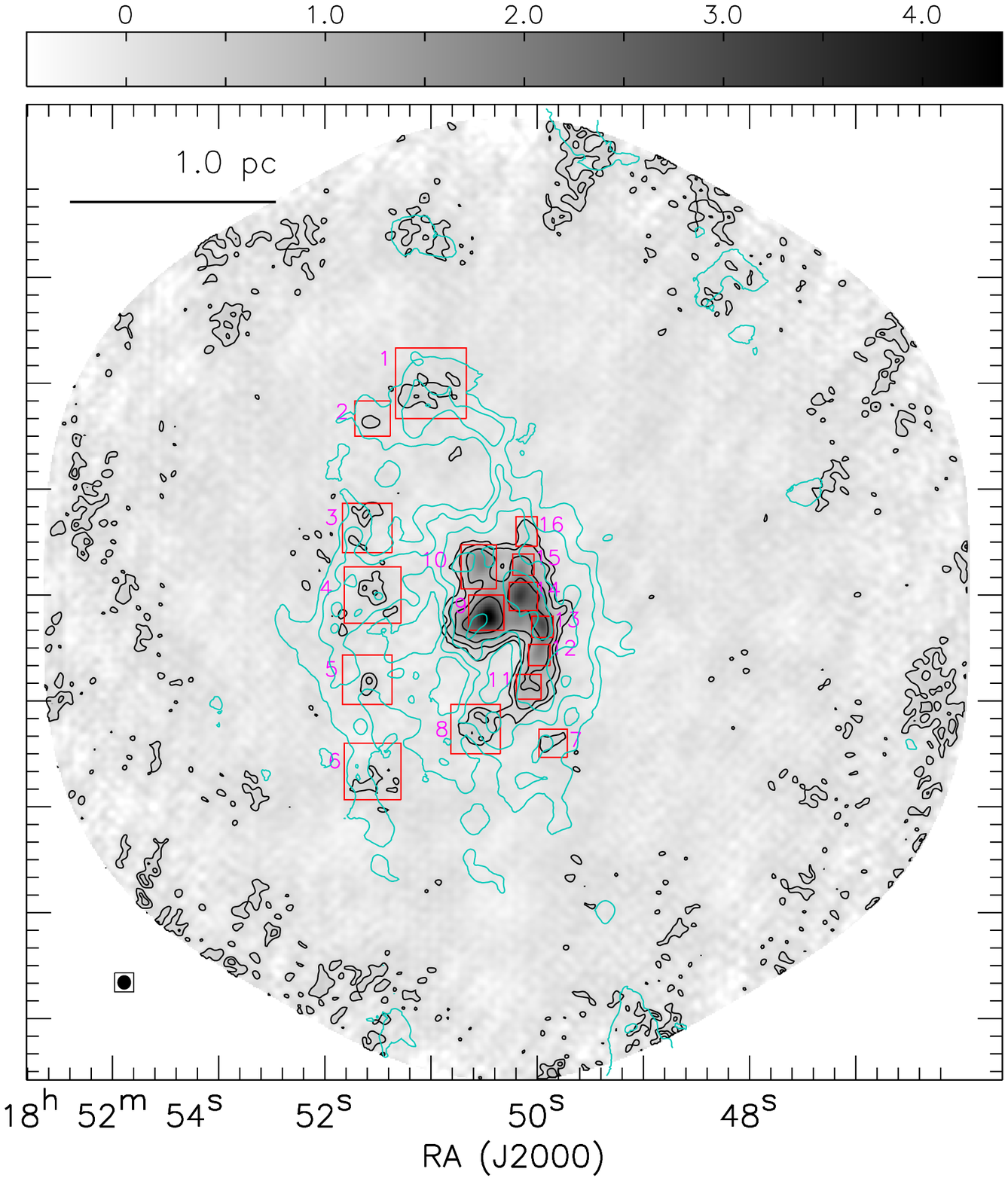}
\caption{
The velocity integrated intensity maps of SO 5$_{6}$--4$_{5}$  and $^{13}$CS 5--4. 
\textbf{Left: } The velocity integrated intensity map of SO 5$_{6}$--4$_{5}$  (black contour; uv range: 4--100 $k\lambda$; $\theta_{maj}$$\times$$\theta_{min}$: 2$''$.7$\times$2$''$.4; velocity range: 102.2--109.4 km\,s$^{-1}$). The rms noise level in each 1.2 km\,s$^{-1}$ velocity channel is  22 mJy\,beam$^{-1}$ in line free channels. Black contour levels are 2.5 Jy\,beam$^{-1}$km\,s$^{-1}$$\times$[1, 2, 4, 8, 16, 32].
\textbf{Right: } The velocity integrated intensity map of $^{13}$CS 5--4 (black contour; uv range: 4--150 $k\lambda$; $\theta_{maj}$$\times$$\theta_{min}$: 2$''$.0$\times$1$''$.9; velocity range:  104.6--113.0 km\,s$^{-1}$). The rms noise level in each 1.2 km\,s$^{-1}$ velocity channel is  24 mJy\,beam$^{-1}$ in line free channels. Black contour levels are 2.5 Jy\,beam$^{-1}$km\,s$^{-1}$$\times$[1, 2, 4, 8].
Color bars have the unit of Jy\,beam$^{-1}$km\,s$^{-1}$.
Cyan contours in both panels show the velocity integrated $^{13}$CO 2--1 map. 
Cyan contour levels are 1.0 Jy\,beam$^{-1}$km\,s$^{-1}$$\times$[2, 4, 8, 16, 32, 64].
These two molecular transitions show significant detections around G33.92+011A, B, D regions, arm--S2, and the elongated PDR region connected with arm--E.
In the right panel, we label the selected regions $^{13}$CS--1$_{,\cdots},$16 and present the spectral profile in Figure \ref{fig_spectra}.
}
\label{fig_so13cs}
\end{figure}

Figure \ref{fig_so13cs} shows the velocity integrated intensity maps of the SO 5$_{6}$--4$_{5}$ transition and the $^{13}$CS 5--4 transition, overlaid with the velocity integrated map of the $^{13}$CO 2--1 transition. 
Strong $^{13}$CS 5--4 emission regions are labeled with $^{13}$CS--1,$_{\cdots}$,16 (Figure \ref{fig_so13cs}). The spectral line profiles measured toward these regions will be discussed in Section \ref{chap_kinematics}. 
From Figure \ref{fig_so13cs}, we find that the  distribution of SO 5$_{6}$--4$_{5}$ and $^{13}$CS 5--4 are correlated with that of the extended 8 $\mu$m emission. 
This behavior is similar to that reported toward the Orion Bar (van der Wiel et al. 2009), where a (chemical) stratification has been reported from the observed 8 $\mu$m, SO and $^{13}$CO/C$^{18}$O emission. 
The $^{13}$CS condensations toward arm--E indeed seem to be found closer in than the SO emission (see e.g. $^{13}$CS--3,4,5), which is consistent with the idea of chemical stratification within a PDR. 
This chemical stratification provides indirect evidence that the molecular filaments are physically associated with the OB star cluster hosted within G33.92+0.11, instead of appearing as a mere projection.

The OCS 19--18 and OCS 18--17 transitions have high ($\gtrsim$100 K) upper--level--energies ($E_{u}$) and show localized distributions. 
Figure \ref{fig_ocs} shows the velocity integrated intensity image of the OCS 19--18 transition. 
Surprisingly, the integrated intensity emission of OCS does not peak at the brightest 1.3\,mm continuum sources A1 and A2, but toward regions A3, A5 and A8. 
In particular, the OCS emission shows a double-peaked structure toward A8. 
The strong OCS emission detected toward these regions in the integrated intensity map, is due to the fact that the OCS line profiles have broad linewidths.
Although the O$^{13}$CS 18--17 transition is also covered in our 1.3 mm observations, we do not detect the O$^{13}$CS 18--17 emission above 3$\sigma$ significance. 
Based on the optically thin assumption and a temperature of $\sim$50\,K (see Section \ref{chap_ch3cn}), the averaged OCS column density in the core A1 and A2 regions is $\sim$1.4$\times$10$^{14}$\,cm$^{-2}$.

From our SMA data, no significant SO$_{2}$ 11$_{5,7}$--12$_{4,8}$ emission is detected (3$\sigma$ brightness$\sim$0.25 K; Table \ref{table_line}). 
Assuming optically thin, LTE conditions, and an average gas excitation temperature of 35K, 
this detection level corresponds to an upper limit of $\sim$4$\times$10$^{14}$$\times$($\frac{\Delta v}{\mbox{1 km\,s$^{-1}$}}$) cm$^{-2}$ for the SO$_{2}$ column density, where $\Delta v$ is the linewidth. 
By comparing this value with the column density N$_{H_{2}}$ derived from the $^{13}$CO/C$^{18}$O ratio (Section \ref{chap_co}, Equation 1), we constrain the averaged fractional abundance of X(SO$_{2}$) to be less than 3.0$\times$10$^{-9}$.
The derived upper limit of the SO$_{2}$ abundance is comparable with the reported SO$_{2}$ abundance toward other young massive star forming regions (van der Tak et al. 2003).

\subsubsection{The SiO Line Emission}
\label{chap_sio}
SiO is known to be an excellent probe of regions affected by shocks (Martin-Pintado et al. 1992; Caselli et al. 1997; Gusdorf et al. 2008; Jim{\'e}nez-Serra et al. 2008).
Indeed, observations of SiO lines toward molecular outflows show that its abundance is enhanced by up to a factor of 10$^{6}$ with respect to the value in quiescent gas (Martin-Pintado et al. 1992; Jim{\'e}nez-Serra et al. 2005).
As a consequence, broad line profiles of SiO emission are expected to arise from these regions. 

Figure \ref{fig_sio} shows the velocity integrated intensity map of the SiO 5--4 transition. 
Strong and localized emission is detected toward regions A5, A6, A8, and few arcseconds north of A8. 
These SiO regions are labeled as SiO--1,$_{\cdots}$,3. 

\begin{figure}
\includegraphics[scale=0.5]{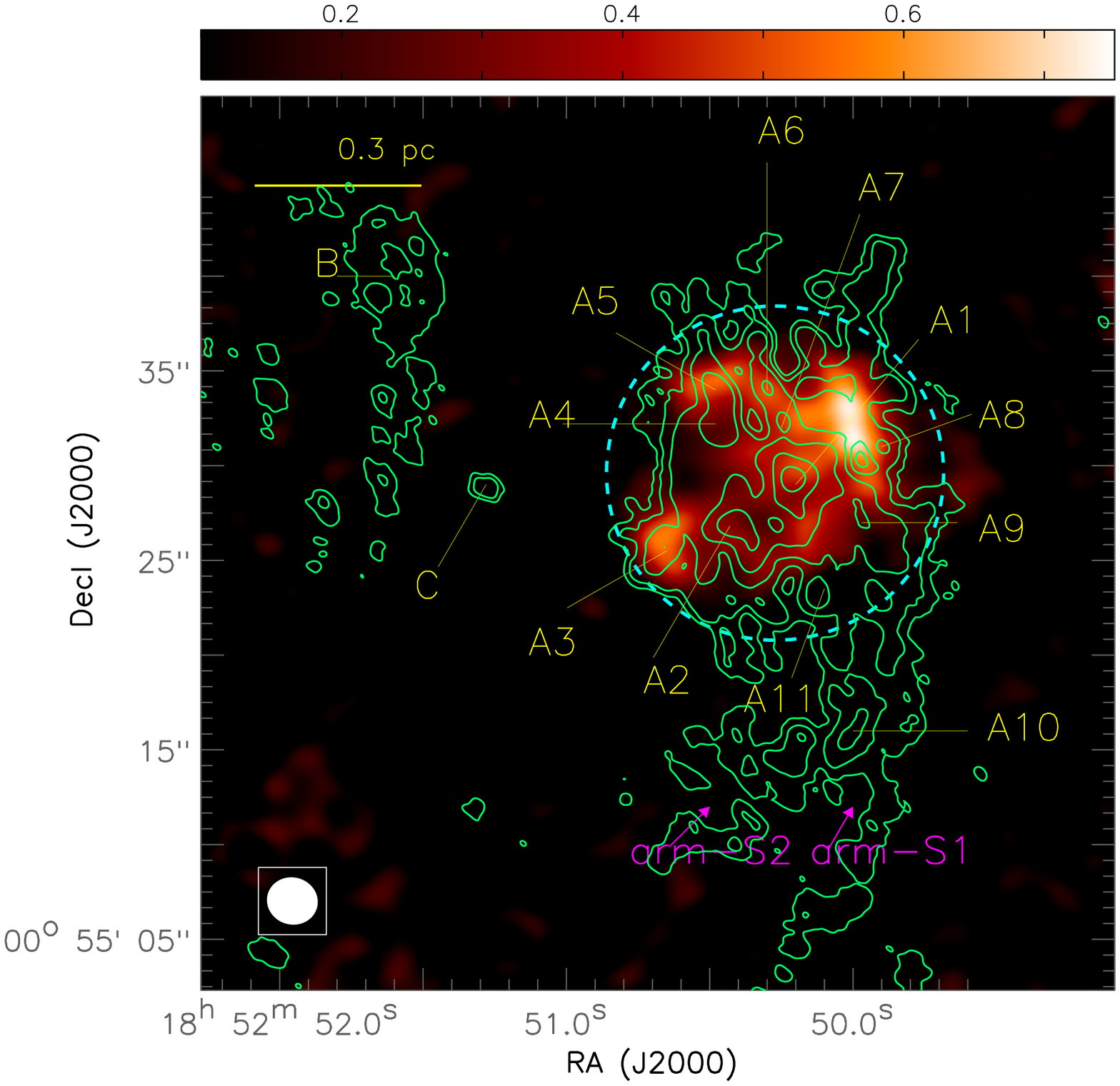}
\caption{
The velocity integrated intensity map of OCS 19--18 (contour; uv range: 4--100 $k\lambda$; $\theta_{maj}$$\times$$\theta_{min}$: 2$''$.7$\times$2$''$.5; velocity range:  102.8--112.4  km\,s$^{-1}$). The rms noise level in each 1.2 km\,s$^{-1}$ velocity channel is  27 mJy\,beam$^{-1}$. 
Color bars have the unit of Jy\,beam$^{-1}$km\,s$^{-1}$.
Green contours show the natural weighting 1.3 mm continuum image. 
Contour levels are 1.2 mJy\,beam$^{-1}$[1, 2, 4, 8, 16, 32, 64].
Dashed circle show the 0.3 pc radius from the pointing center R.A.= 18$^{\mbox{h}}$52$^{\mbox{m}}$50$^{\mbox{s}}$.272, Decl.= $+$00$^{\circ}$55$'$29$''$.604.
}
\label{fig_ocs}
\end{figure}

\begin{figure}
\includegraphics[scale=0.5]{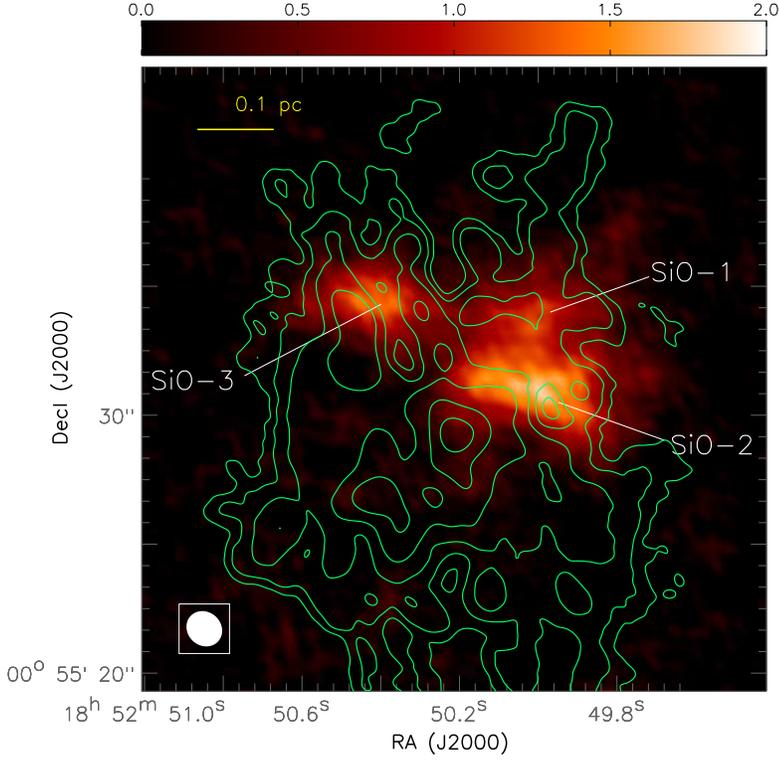}
\caption{
The velocity integrated intensity map of SiO 5--4 (color; uv range: 4--390 $k\lambda$; $\theta_{maj}$$\times$$\theta_{min}$: 1$''$.4$\times$1$''$.3; velocity range:  89.6--122  km\,s$^{-1}$). The rms noise level in each 1.2 km\,s$^{-1}$ velocity channel is 14 mJy\,beam$^{-1}$.
Color bars have the unit of Jy\,beam$^{-1}$km\,s$^{-1}$.
Green contours show the natural weighting 1.3 mm continuum image. 
contour levels are 1.2 mJy\,beam$^{-1}$[1, 2, 4, 8, 16, 32, 64].
Localized SiO emission is seen around A5, A6, A8, and $\sim$3$''$ north of A8.
}
\label{fig_sio}
\end{figure}

By comparing Figure \ref{fig_ocs} with Figure \ref{fig_sio}, we see that the SiO emission regions also show strong emission in high excitation OCS.
This is consistent with the ejection of significant amounts of SiO and OCS from dust grains by the interaction of low-velocity (or C--type) shocks (Caselli et al. 1997; Jimenez-Serra et al. 2005, 2008). 
The measured linewidths of the SiO and OCS lines are $\sim$5-20$\,$km$\,$s$^{-1}$, which supports the idea of low-velocity shocks toward these regions. 

From Figures \ref{fig_ocs} and \ref{fig_sio}, we also find that the SiO and OCS line emission tends to appear at the intersecting points between the large--scale arms and the central molecular clump A of G33.92+0.11. 
As discussed in Section 4.4, we propose that this emission arises from gas that has been recently shocked as a consequence of gas accretion along the molecular filaments/arms onto the central hub. 
The inward temperature gradient detected from the NH$_3$ line emission is also consistent with this scenario (see Section 3.1.8).

We note that we do not detect significant SiO emission in the brightest 1.3 mm continuum sources A1 and A2, suggesting that powerful molecular outflows emanating from these two massive cores may not be present.

\subsubsection{The Hydrogen Recombination Line H30$\alpha$}
\label{chap_rrl}
\begin{figure}

\begin{tabular}{ p{9.4cm} c }
\begin{tabular}{c}
 \includegraphics[scale=0.45]{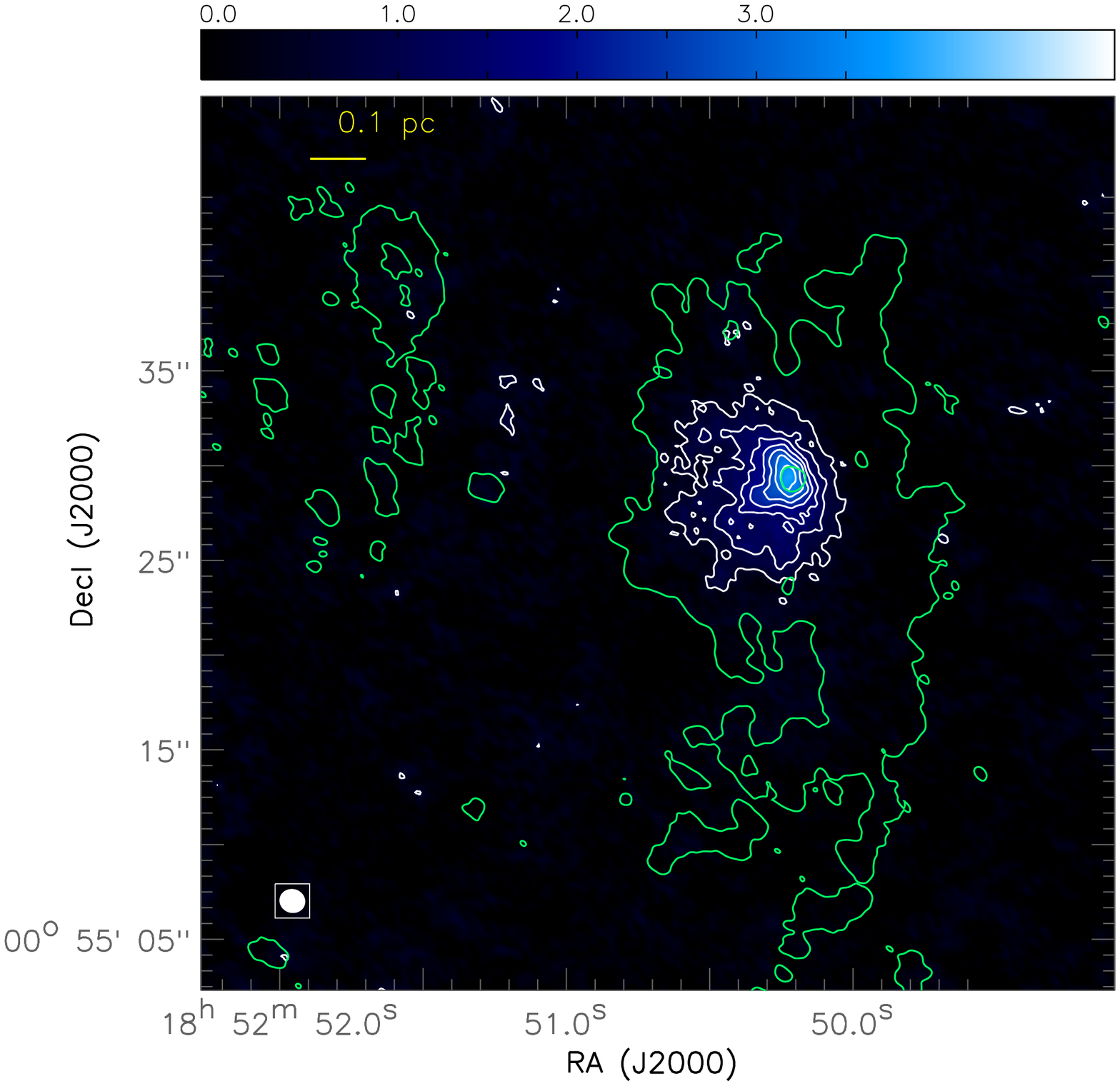} \\
\end{tabular}
&
\begin{tabular}{c}
\includegraphics[scale=0.25]{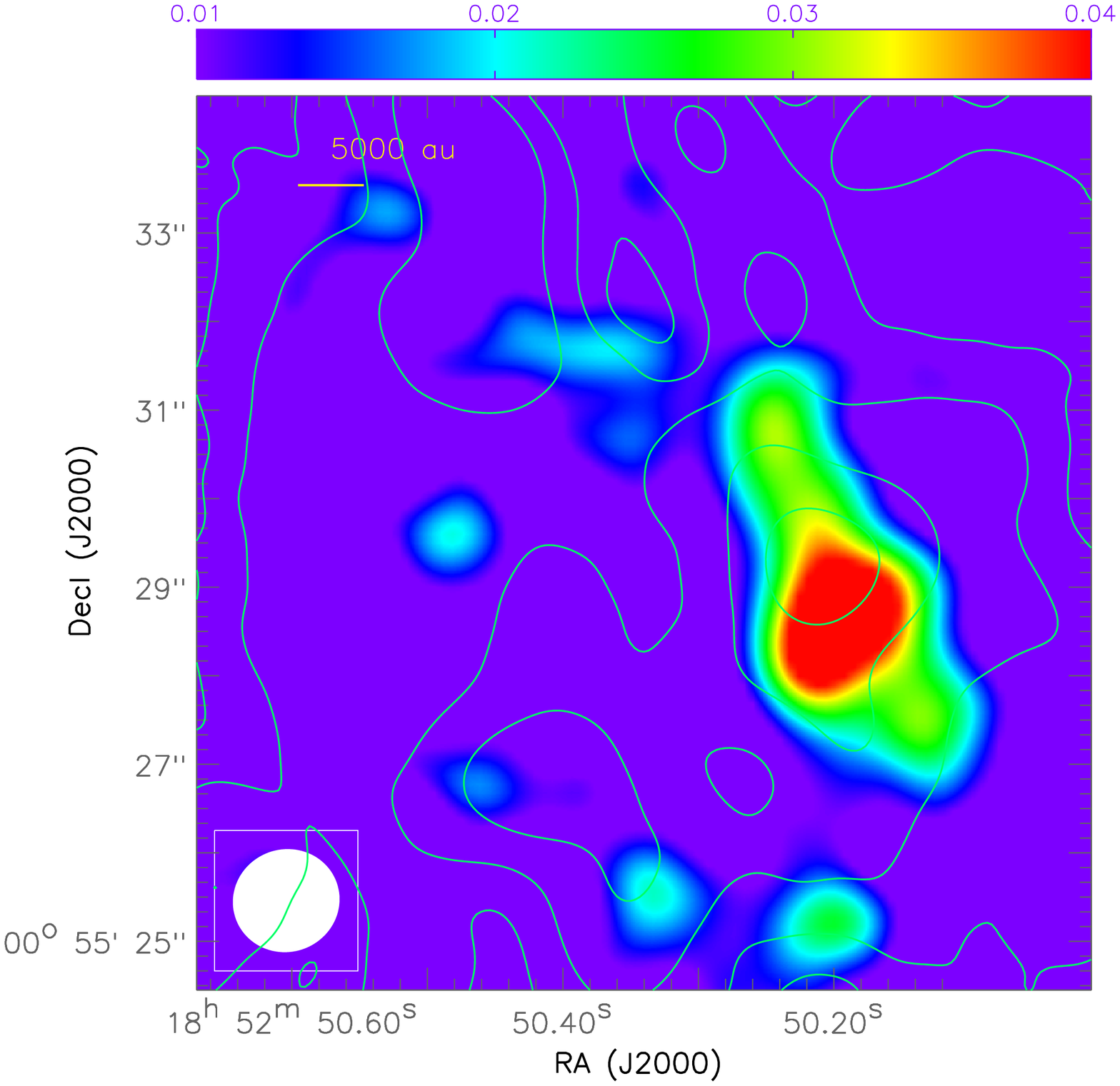} \\
\includegraphics[scale=0.25]{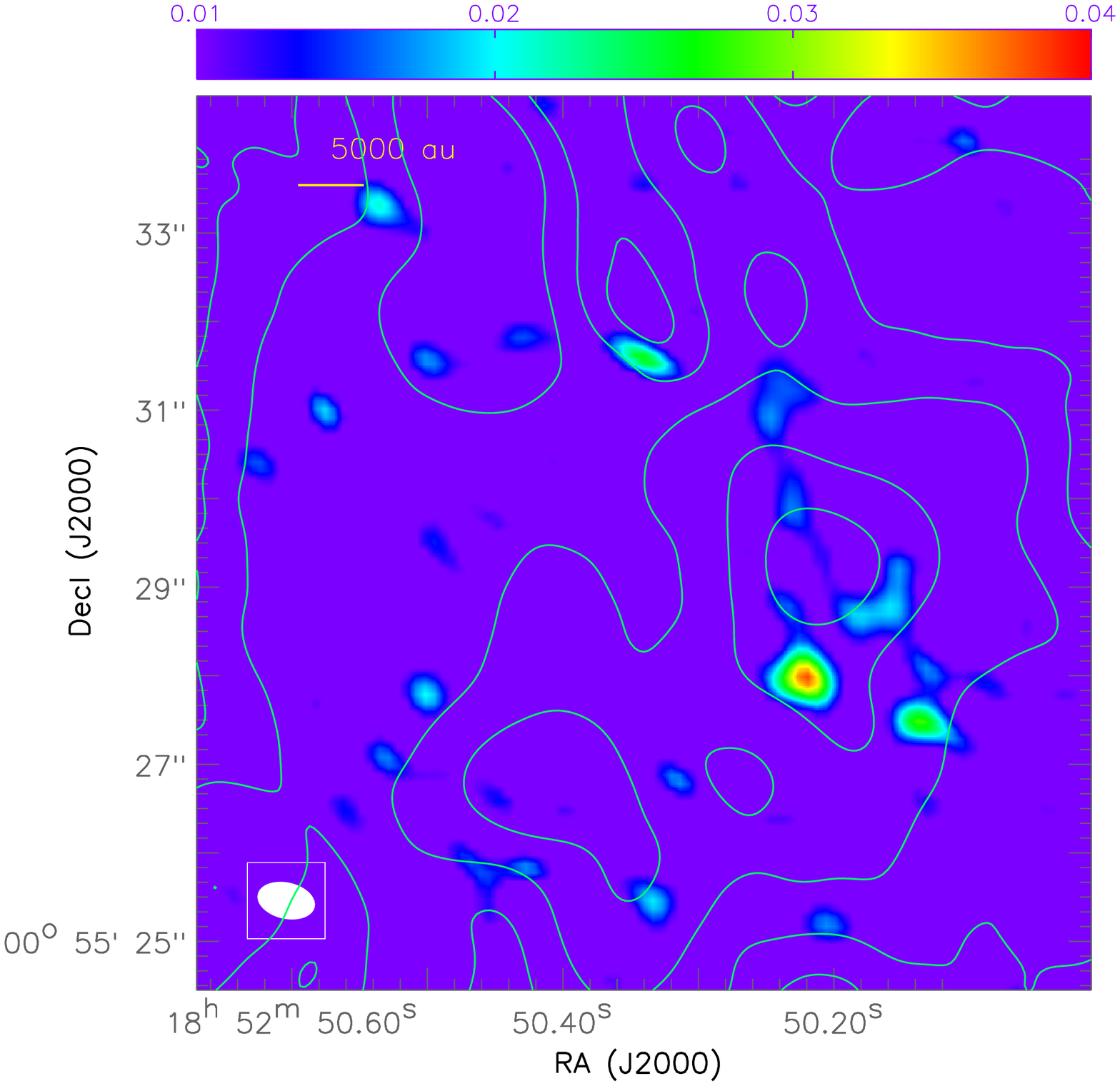} \\
\end{tabular} \\
\end{tabular}
\caption{\textbf{Left: } The velocity integrated intensity map of the hydrogen recombination line H30$\alpha$ (white contour and color; uv range: 4--390 $k\lambda$; $\theta_{maj}$$\times$$\theta_{min}$: 1$''$.4$\times$1$''$.3; velocity range:  78.6--137.1 km\,s$^{-1}$). The rms noise level in each 6 km\,s$^{-1}$ velocity channel is  9.1 mJy\,beam$^{-1}$ in line free channels.  Color bars have the unit of Jy\,beam$^{-1}$km\,s$^{-1}$. Green contours show the natural weighting 1.3 mm continuum map. White contour levels are 0.5 Jy\,beam$^{-1}$km\,s$^{-1}$$\times$[1, 2, 3, 4, 5, 6, 7, 8]. Green contour levels are 1.2 mJy\,beam$^{-1}$$\times$[1, 32].
\textbf{Top Right: } Natural weighting image (color; uv range: 4--250 $k\lambda$; $\theta_{maj}$$\times$$\theta_{min}$: 1$''$.5$\times$1$''$.2; velocity range:  79.5--130  km\,s$^{-1}$). The rms noise level is  3.1 mJy\,beam$^{-1}$. Green contours show the natural weighting 1.3 mm continuum map (1.2 mJy\,beam$^{-1}$$\times$[1, 2, 4, 8, 16, 32, 64]).
\textbf{Bottom Right: } Briggs Robust 0 weighting image (color; uv range: 4--390 $k\lambda$; $\theta_{maj}$$\times$$\theta_{min}$: 0$''$.64$\times$0$''$.4; velocity range:  79.5--130 km\,s$^{-1}$). The rms noise level is  5.2 mJy\,beam$^{-1}$.
Color bars have the unit of Jy\,beam$^{-1}$.
The peak of the H30$\alpha$ emission does not coincide with the peak of the 1.3 mm continuum emission, which suggests that A1 may have unresolved internal structures. 
}
\label{fig_h30a}
\end{figure}

The emission of the hydrogen recombination line H30$\alpha$ traces the distribution and the kinematics of the ionized gas. 
Figure \ref{fig_h30a} shows maps of H30$\alpha$.
We obtain higher angular resolution images of the ionized gas than those previously reported in the literature (e.g. Watt \& Mundy 1999).
From Figure \ref{fig_h30a}, we find a $\lesssim$10$''$ angular scale cometary UC H\textsc{ii} region, which  
peaks toward the A1 core, suggesting the presence of an embedded OB cluster.
We refer to the OB stars that are embedded in either core A1 or  core A2 as the \textit{central OB cluster} hereafter. 
For the rest of localized intermediate-- or high--mass (proto)stars in G33.92+0.11 A we refer to them as the \textit{satellite high--mass stars}.

The brightest region in H30$\alpha$ emission is found toward A (integrated flux density $>$3.5 Jy\,beam$^{-1}$\,km\,s$^{-1}$) and shows a northeast--southwest elongation (see Figure \ref{fig_h30a}, right).
The kinematics of this elongated H\textsc{ii} region will be discussed in Section \ref{chap_kinematics}.
The higher angular resolution map using Briggs Robust 0 weighting shows multiplicity in this bright emission region.
Both the natural weighting map and the Briggs Robust 0 weighting map show that the peak of the H30$\alpha$ emission is offset from the peak of the 1.3 mm continuum emission by $\sim$1$''$ (0.034 pc, or 7000 AU). 
The peak of 1.3 mm continuum emission and the brightest H30$\alpha$ emission region visually form a binary system.
Independent high--mass (proto)stars may be embedded at the peak of the 1.3 mm continuum emission, as the powering source of the elongated UC H\textsc{ii} region. 
This is to be investigated with higher angular resolution observations. 

Based on the H30$\alpha$ data, we provide two simple means of estimating the flux of the 1.3 mm free--free continuum emission. 
Assuming optically thin free--free continuum emission, electron temperature T$_{e}$$\sim$10,000 K, line FWHM $\delta v$ of 30\,km\,s$^{-1}$ (from measurement of G33.92+0.11), at a frequency of 230 GHz ($\lambda\sim$1.3 mm), the peak radio recombination line to free--free continuum flux ratio estimated by $\frac{T_{L}}{T_{C}}$$\sim$7.0$\times$10$^{3}$$(\frac{\delta v}{\mbox{\footnotesize{km\,s$^{-1}$}}})^{-1}$$(\frac{\nu}{\mbox{\footnotesize{GHz}}})^{1.1}$$(\frac{T_{e}}{\mbox{\footnotesize{K}}})^{-1.15}$$[1+\frac{N(\mbox{\footnotesize{He}}^{+})}{N(\mbox{\footnotesize{H}}^{+})}]^{-1}$ is about 2.3.   
By summing the flux of the H30$\alpha$ line around the systemic velocity (107.6 km\,s$^{-1}$) channel, and by dividing this flux by a factor of 2.3, gives a free--free continuum emission contribution of $\sim$35 mJy for the A1 region, and of 10 mJy for the A2 region. 
Based on the same assumptions, the 3$\sigma$ noise level of the H30$\alpha$ images in each 6 km\,s$^{-1}$ velocity bins (27.3 mJy\,beam$^{-1}$; $\theta_{maj}$$\times$$\theta_{min}$: 1$''$.4$\times$1$''$.3) corresponds to the 1.3\,mm free--free continuum flux density of 11\,mJy\,beam$^{-1}$.
This flux density corresponds to fluxes of 3.5\,mJy and 1.5\,mJy within the smaller beams of the natural weighted and Briggs Robust 0 weighted 1.3 mm continuum maps (Figure \ref{fig_ch0}).
With these estimates, we conclude that the 1.3 mm continuum emission is primarily dominated by thermal dust emission.  
We note that T$_{e}$ and $\delta v$ are varying spatially, and therefore, a detailed modeling of these quantities requires observations of hydrogen recombination lines and continuum emission in multiple frequencies (e.g. Keto, Zhang, \& Kurtz 2008).

From the total flux at 1.3 mm and the brightness distribution of the H30$\alpha$ line, we can provide a second estimate of the flux density of the free--free emission at 1.3 mm. 
Assuming the same distribution for the free--free emission and the H30$\alpha$ line emission, the flux of the free--free continuum emission can be estimated by extrapolating the fluxes of the centimeter continuum emission reported in Watt \& Mundy (1999).  
The upper limits to the 1.3 mm free-free continuum emission toward cores A1 and A2, are 180 and 45 mJy, respectively.
We note that to derive the molecular mass of these cores, we have subtracted the upper limits of the 1.3 mm free--free continuum emission in order to obtain a lower limit for the gas mass estimates (see Table \ref{table_cores}).

\subsubsection{High Velocity Molecular Gas}
\label{chap_hv}
We identify the high velocity molecular gas (HVG) based on the emission from $^{12}$CO and from shock tracers such as SO, OCS and SiO. 
In particular, SO, whose critical density is higher than that of $^{12}$CO, permits us to  
follow the outflow closer to the systemic velocity $v_{lsr}$, around which the $^{12}$CO emission is too extended to be robustly imaged in the SMA observations.
In addition, SO allows us to follow the HVG in the blueshifted velocity range, for which the CO 2--1 isotopologues are confused with the emission from the foreground molecular clouds (see also Fish et al. 2003). 
Observations of multiple molecular species in the Orion--KL region showed that SO or SO$_{2}$ are particularly good tracers of low velocity outflows (Wright et al. 1996). 

\begin{figure}
\hspace{-1cm}
\includegraphics[scale=0.45]{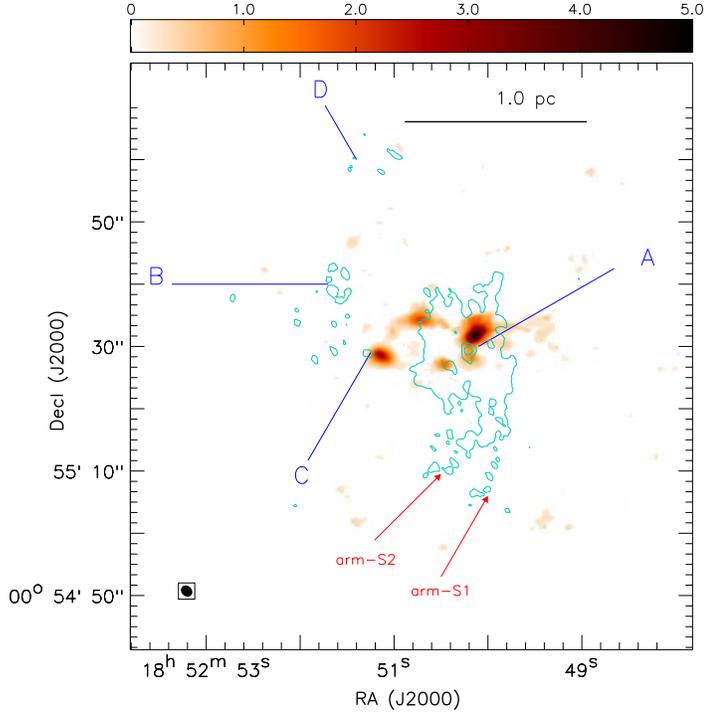}
\caption{
The redshifted high velocity $^{12}$CO 2--1 emission (color; uv range: 4--180 $k\lambda$; $\theta_{maj}$$\times$$\theta_{min}$: 1$''$.96$\times$1$''$.63; velocity range: 124.5 -- 142.5 km\,s$^{-1}$). The rms noise level in each 6 km\,s$^{-1}$ velocity channel is  0.011 Jy\,beam$^{-1}$ (0.08 K) in line free channels.
Color bar has the unit of Jy\,beam$^{-1}$km\,s$^{-1}$.
Cyan contours show the natural weighting 1.3 mm continuum image. 
Contours are [1.8, 19.2, 32.4] Jy\,beam$^{-1}$. 
}
\label{fig_red}
\end{figure}

\begin{figure}
\includegraphics[scale=0.9]{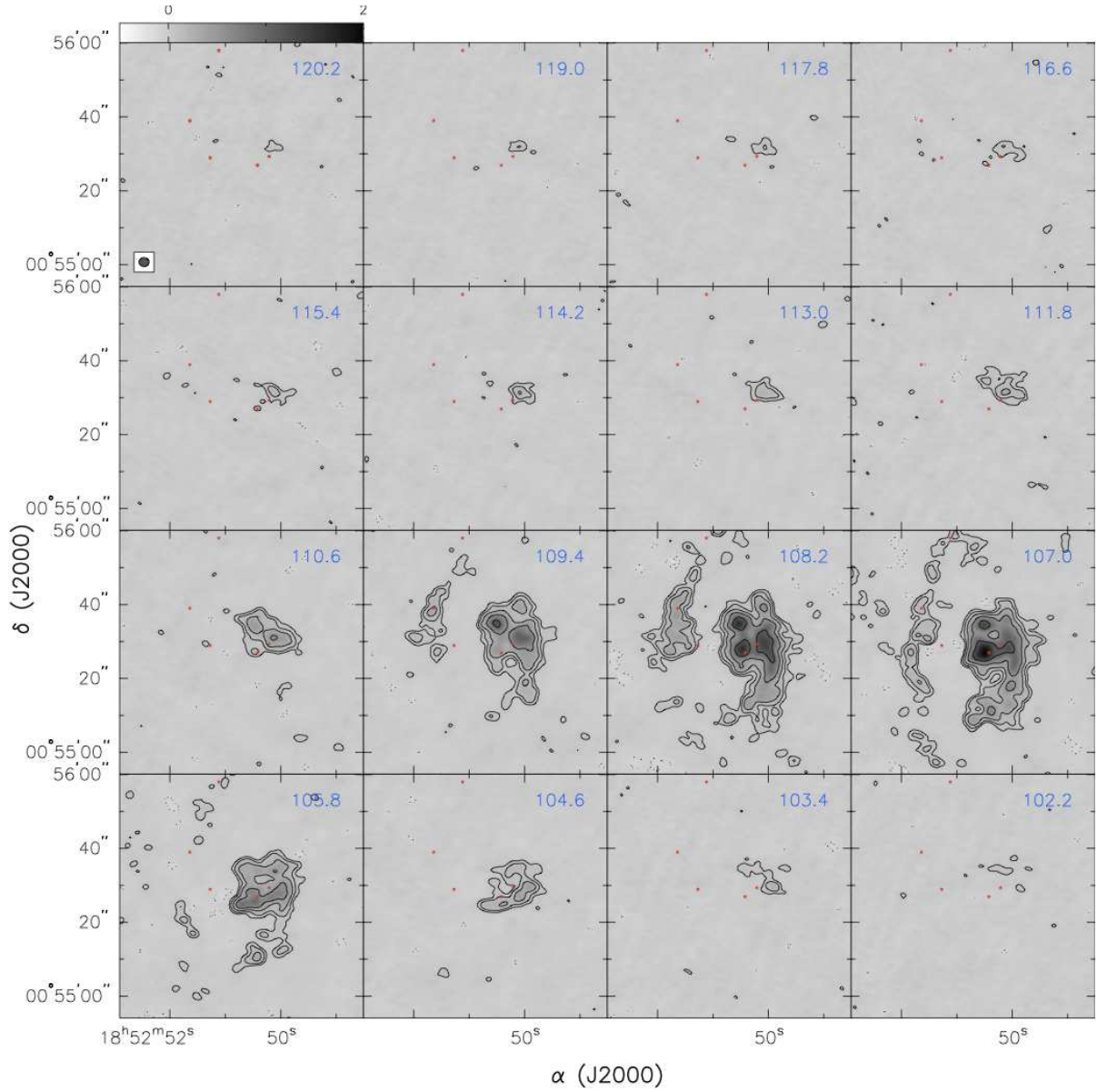}
\caption{Channel maps of the SO 5$_{6}$--4$_{5}$ line. Positive contours (solid) are 66 mJy\,beam$^{-1}$(3$\sigma$)$\times$[1, 2, 4, 8, 16, 32, 64, 128]. Negative contours (dotted) are -66 mJy\,beam$^{-1}$$\times$[1, 2, 3]. The synthesized beam is $\theta_{maj}$$\times$$\theta_{min}$: 2$''$.7$\times$2$''$.4. Locations of G33.92+0.11 A1, A2, B, C and D are marked with red stars. }
\label{fig_sochan}
\end{figure}

\begin{figure}
\includegraphics[scale=0.5]{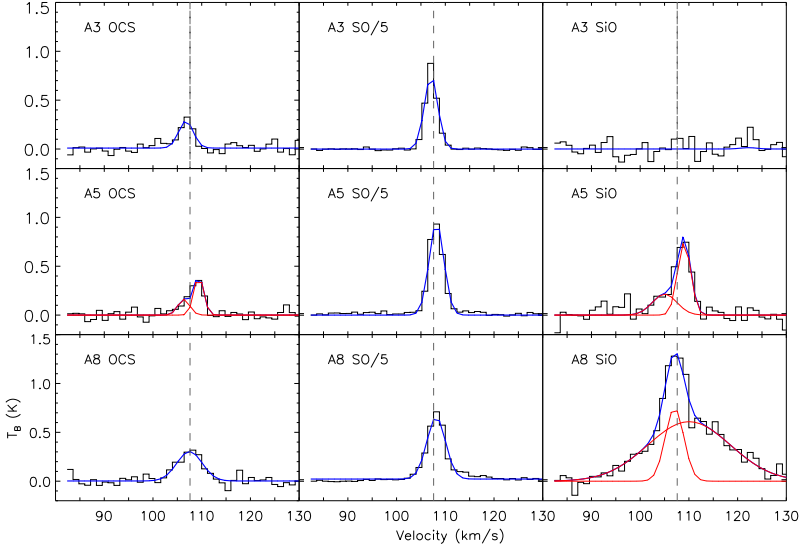}
\caption{The OCS 19--18, SO 5$_{6}$--4$_{5}$, and SiO 5--4 spectra at the core A3, A5, and A8 regions where we see peaks of OCS 19--18 emission. The SO 5$_{6}$--4$_{5}$ spectra are scaled by a factor of 0.2. We fit these spectra with one or two Gaussian components, which are summarized in Table \ref{table_highvel}. The dashed line labels the systemic velocity $v_{lsr}$ of 107.6 km\,s$^{-1}$. We note the excess of redshifted emission in the SO 5$_{6}$--4$_{5}$ spectra in the  core A8 (maybe also A5) region, in the velocity range of 113--118\,km\,s$^{-1}$.}
\label{fig_highvelspec}
\end{figure}

The velocity integrated intensity map of the highly redshifted $^{12}$CO 2--1 emission, and the channel maps of the SO 5$_{6}$--4$_{5}$ transition are shown in Figures \ref{fig_red} and \ref{fig_sochan}.
Figure \ref{fig_highvelspec} compares the SO spectra with those of OCS and SiO, at the location of core A3, A5 and A8, where the OCS emission peaks. 
Discussions about the spectra are deferred to Section \ref{sub_diversity}.

The most significant compact component of high velocity $^{12}$CO 2--1 emission is seen toward the northwest of source A1 and toward the west of source G33.92+0.11 C.
We refer to these two HVG regions as A1--HVG and C--HVG, respectively. 
A1--HVG and C--HVG likely host high velocity molecular outflows, with A1--HVG showing multiplicity in the SiO images  (i.e. SiO--1,2; Figure \ref{fig_sio}). In addition, A1--HVG is spatially resolved in the SO 5$_{6}$--4$_{5}$ channel maps (Figure \ref{fig_sochan}), and shows terminal velocities of $v_{lsr}+$12.6 km\,s$^{-1}$ and $v_{lsr}-$5.4 km\,s$^{-1}$ for the redshifted and the blueshifted components, respectively.
The irregular shape of the A1--HVG in the SO 5$_{6}$--4$_{5}$ channel maps is potentially caused by more than one powering source, and structures impacted by shocks. In $^{12}$CO 2--1, A1--HVG is detected up to velocities of 139.5 km\,s$^{-1}$ ($v_{lsr}+$31.9 km\,s$^{-1}$).

Other localized molecular outflows could be present toward regions A2, A5, and A10, as detected from the $^{12}$CO 2--1 and SO 5$_{6}$--4$_{5}$ channel maps. 
We refer to these outflow sources as A2--HVG, A5--HVG, and A10--HVG, respectively. 
A5--HVG is seen in the velocity channels of 103.4--117.8 km\,s$^{-1}$ in SO (Figure \ref{fig_sochan}). 
For A2--HVG, the $^{12}$CO 2--1 emission is also found up to velocities of 139.5 km\,s$^{-1}$ ($v_{lsr}+$31.9 km\,s$^{-1}$), and for A5--HVG, up to velocities of 133.5 km\,s$^{-1}$ ($v_{lsr}+$25.9 km\,s$^{-1}$).
The blueshifted $^{12}$CO and SO emission around the core A2 may partially be influenced by the expansion of the ionized gas. 
Systematic and more robust identification of more localized high velocity gas will also require better understanding of the structures of the parent molecular cloud.

Assuming optically thin $^{12}$CO 2--1 emission, local thermodynamical equilibrium (LTE), a gas excitation temperature of 50 K,  and a $^{12}$CO abundance ratio X($^{12}$CO)=10$^{-4}$, we can estimate the mass, momentum, and energy of the high velocity gas (see Table \ref{table_outflow}).
The assumed gas excitation temperature is comparable to the highest kinetic temperature we detected from the NH$_{3}$ observations and the CH$_{3}$CN observations (Section \ref{chap_nh3}, \ref{chap_ch3cn}), and is similar to those derived toward molecular outflows (see e.g. L1157-mm; Bachiller et al. 1993).
In the optically thick regions, the peak emission in the $^{12}$CO 2--1 channel maps is about 57 K (i.e. 8 Jy\,beam$^{-1}$).
This peak $^{12}$CO 2--1 emission is consistent with the assumed gas excitation temperature and LTE.  
The estimated total energy is $\sim$4$\times$10$^{45}$ erg (Table \ref{table_outflow}).
These high velocity gas components in G33.92+0.11 are not as energetic as the high--mass bipolar molecular outflow in G240.31+0.07 (Qiu et al. 2009).

\subsubsection{The CH$_{3}$CN J=12--11 Transitions}
\label{chap_ch3cn}
The CH$_{3}$CN molecule is observed to be abundant in hot molecular cores, and is believed to be formed either on the surface of dust grains or via gas-phase photo-chemistry after the evaporation of large amounts of NH$_3$ from the grains (MacKay 1999). 

The previous BIMA observations of the CH$_{3}$CN J=6--5 transitions, detected significant emission in K = 0, 1 components toward G33.92+0.11 (Watt \& Mundy 1999). 
Although our 1.3 mm observations cover all CH$_{3}$CN J=12--11 transitions, we do not obtain significant detections for the lines with K ladders higher than 3.

\begin{figure}
\begin{tabular}{  p{8.3cm} p{8.3cm}  }
\includegraphics[scale=0.47]{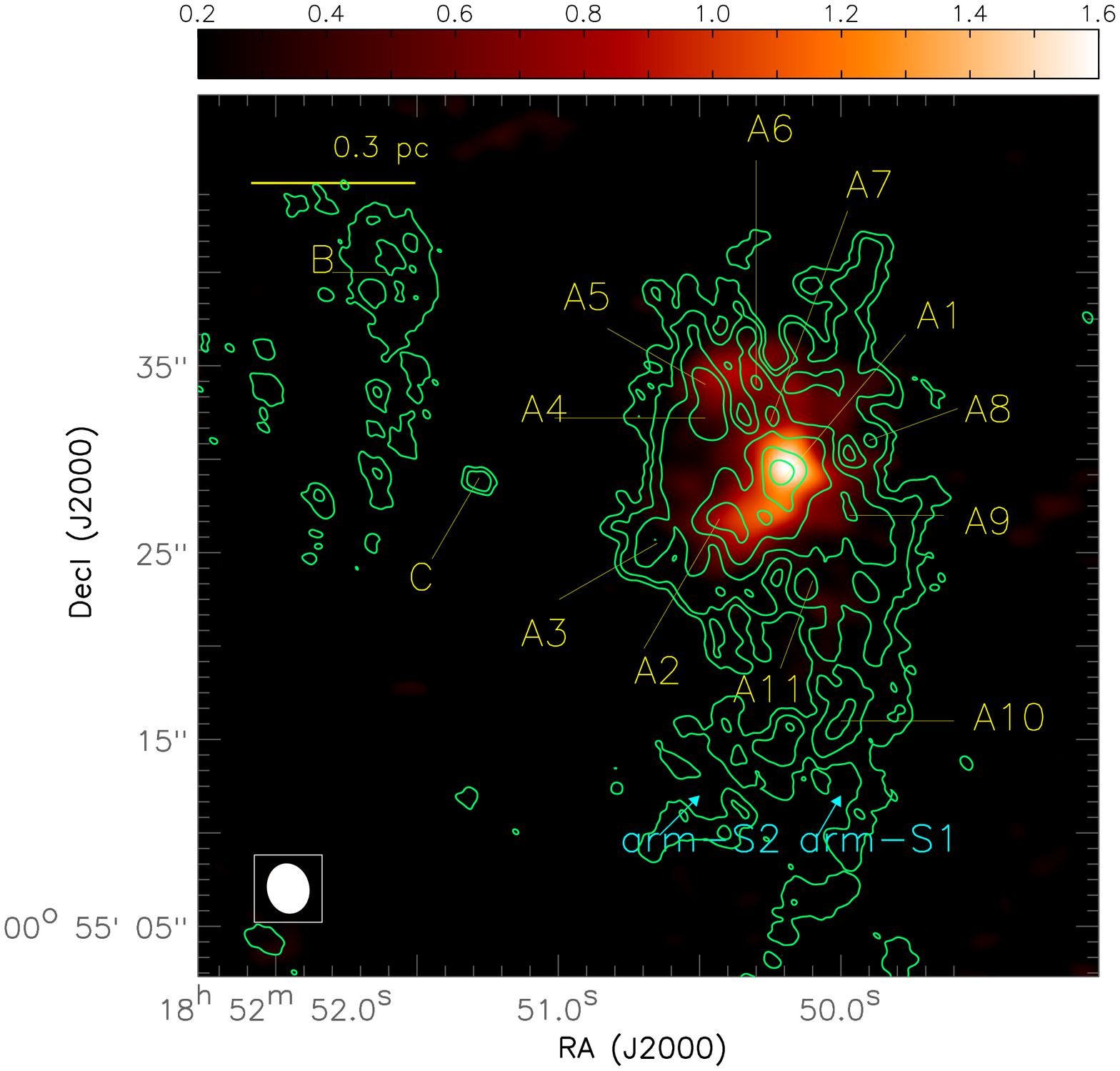}   &
\includegraphics[scale=0.47]{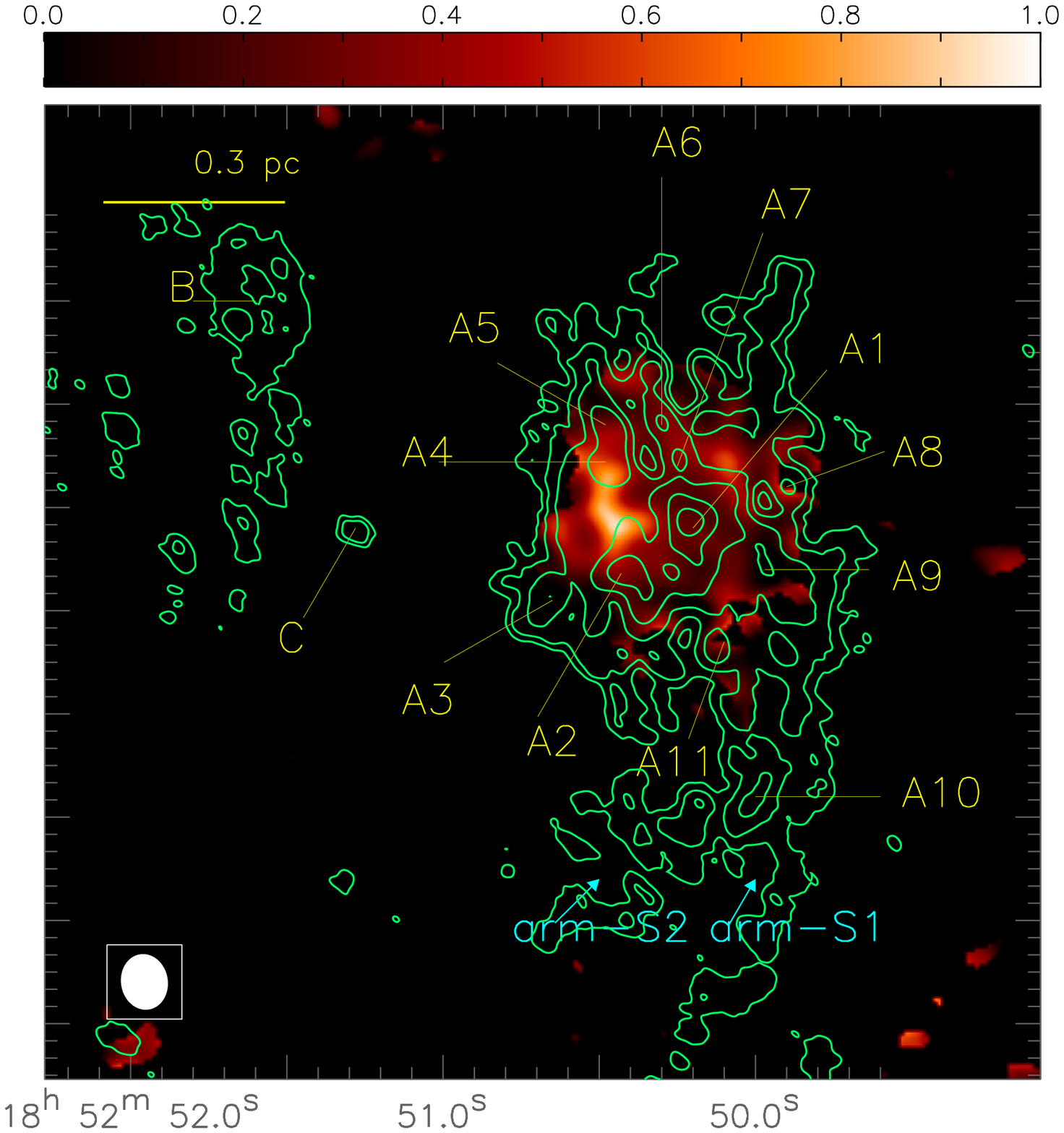}      
\end{tabular}
\caption{ The velocity integrated intensity maps of CH$_{3}$CN J=12-11 transitions. The CH$_{3}$CN J=12$_{0}$--11$_{0}$ and CH$_{3}$CN J=12$_{1}$--11$_{1}$ transitions trace comparable excitation conditions, and are integrated to one map. 
\textbf{Left: } The velocity integrated intensity map of CH$_{3}$CN J=12$_{0}$--11$_{0}$ and CH$_{3}$CN J=12$_{1}$--11$_{1}$ (color; uv range: 4--100 $k\lambda$; $\theta_{maj}$$\times$$\theta_{min}$: 2$''$.7$\times$2$''$.2; velocity range:  105.2--117.2  km\,s$^{-1}$). The rms noise level in each 1.2 km\,s$^{-1}$ velocity channel is 23 mJy\,beam$^{-1}$.
Color bar has the unit of Jy\,beam$^{-1}$km\,s$^{-1}$.
\textbf{Right: } The integrated intensity ratio map of the K=0 and 1 components to the K=3 component of the CH$_{3}$CN J=12--11 transition.
Pixels in the CH$_{3}$CN J=12$_{0}$--11$_{0}$ $+$ CH$_{3}$CN J=12$_{1}$--11$_{1}$ map are masked if the values are smaller than 0.24 Jy\,beam$^{-1}$km\,s$^{-1}$.
}
\label{fig_ch3cn}
\end{figure}

Figure \ref{fig_ch3cn} shows the velocity integrated intensity maps of the K = 0, 1 (combined) components of the CH$_{3}$CN J=12--11 transition.
In contrast to the distribution of the emission of sulfur--bearing molecules, the integrated intensity maps of the brightest K = 0, 1 components peak toward the 1.3 mm continuum sources A1 and A2.
This is consistent with the results of the H30$\alpha$ emission (Section \ref{chap_rrl}), indicating that at least core A1 hosts a cluster of OB stars which emit UV photons.
Diffused emission of the K = 0, 1 components is seen toward the northwest of the A1 core, and the A5, A6, A9 regions, which could be associated with shocked gas (Csengeri et al. 2011a).
Weak detections of CH$_{3}$CN K = 2, 3 are seen near A1 and A2. 
The ratio of the integrated flux of the K = 3 component with the combined K = 0, 1 components (Figure \ref{fig_ch3cn}) show the non--uniform heating between the A2 and the A5 regions, and around the A8 region.  
We note that the region between A2 and A5 is the conjunction of the central massive cores A1,2 with one molecular filament (see also Figure \ref{fig_345ch0}). 
As discussed in Section 4, low-velocity accretion shocks could be interacting with the central hub toward this region, heating the gas and enhancing the abundance of molecules such as SiO, OCS, and CH$_3$CN.  

Although the higher excitation K = 2, 3 components are detected at cores A1 and A2, the relatively strong K = 0, 1 components of the CH$_{3}$CN J=12--11 emission at the same places does not imply a much higher excitation temperature than the temperature of ambient molecular gas. 
In the region which shows significant detection of the K = 3 component (primarily just the A1 and A2 regions), the fitting of the CH$_{3}$CN rotational diagram based on the optically thin assumption yields an averaged temperature of 49.3 K.
The optically thin assumption may not be valid in cores A1 and A2, and the fitted averaged temperature should be viewed as an upper limit. 
The moderate excitation temperature found toward A1 and A2 is consistent with the weak OCS 19--18 and 18--17 emission, and the absence of SO$_{2}$ 11$_{5,7}$--12$_{4,8}$ line emission.
Based on the derived temperature, in the optically thin limit, the averaged CH$_{3}$CN column density in the core A1 and A2 region is $\sim$14$\times$10$^{14}$\,cm$^{-2}$.

\subsubsection{NH$_{3}$ Observations}
\label{chap_nh3}
The NH$_{3}$ molecule is one of the most reliable probes of distribution kinematics, and of kinetic temperature of the molecular gas in star--forming cores or clumps. 
It has relatively small molecular weight, and does not deplete onto the grains even when the density is as high as $\gtrsim$10$^{6}$\,cm$^{-3}$ (Bergin \& Langer 1997).
The abundance of the NH$_{3}$ molecule is more stable than that of other heavier molecules, which can be severely enhanced in outflows or other shock heated regions. 
The line ratio of the main hyperfine transition and the much fainter satellite hyperfine transition of the same (J,K) level, can be used to constrain the opacity; the opacity of multiple (J,K) levels can further be used to derive the molecular gas excitation temperature (e.g. Ho \& Townes 1983; Longmore et al. 2007; Rosolowsky et al. 2008; Osorio et al. 2009).
 
\begin{figure}
\begin{tabular}{  p{7.5cm} p{7.5cm} }
\includegraphics[scale=0.45]{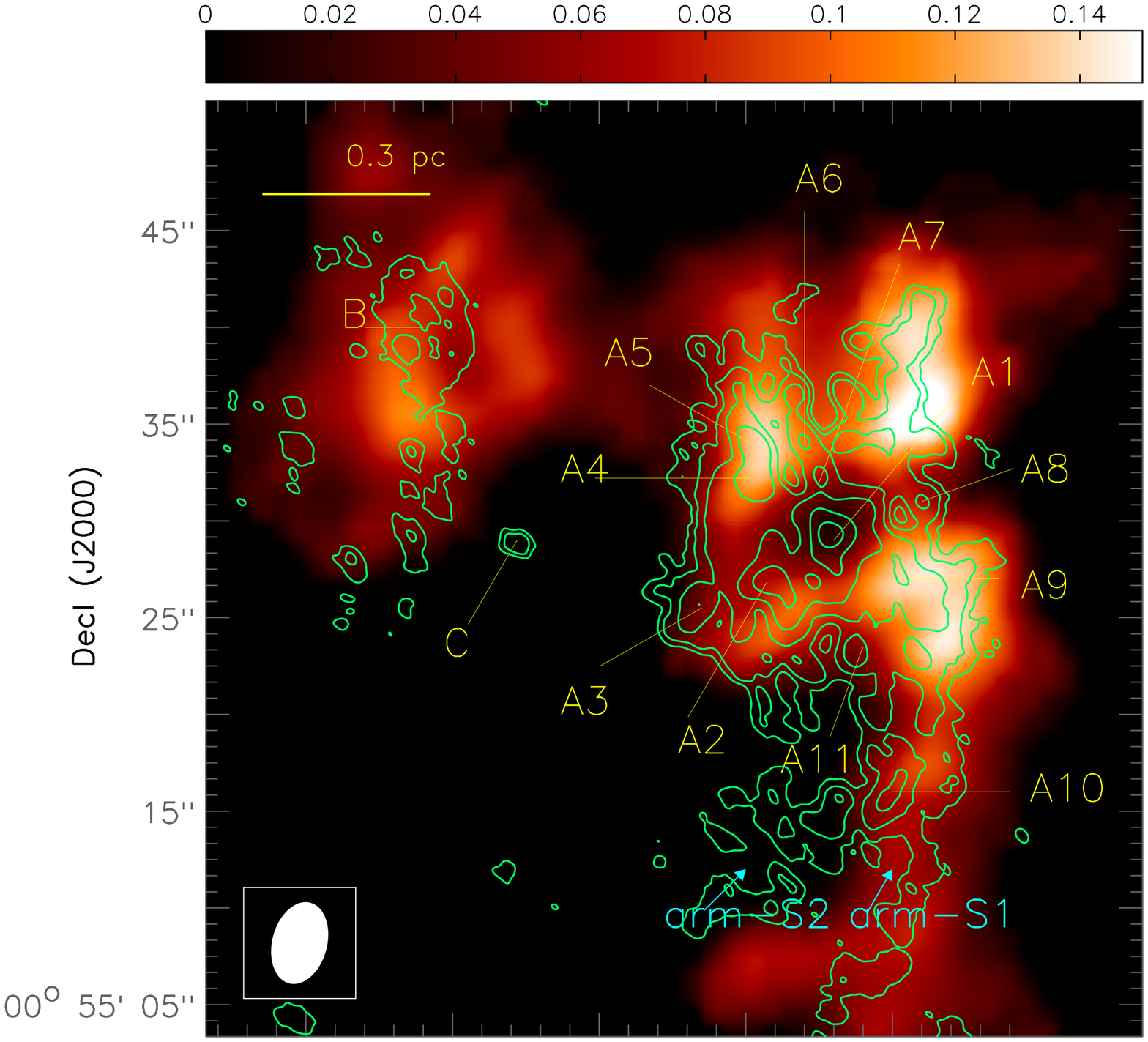} &
\includegraphics[scale=0.45]{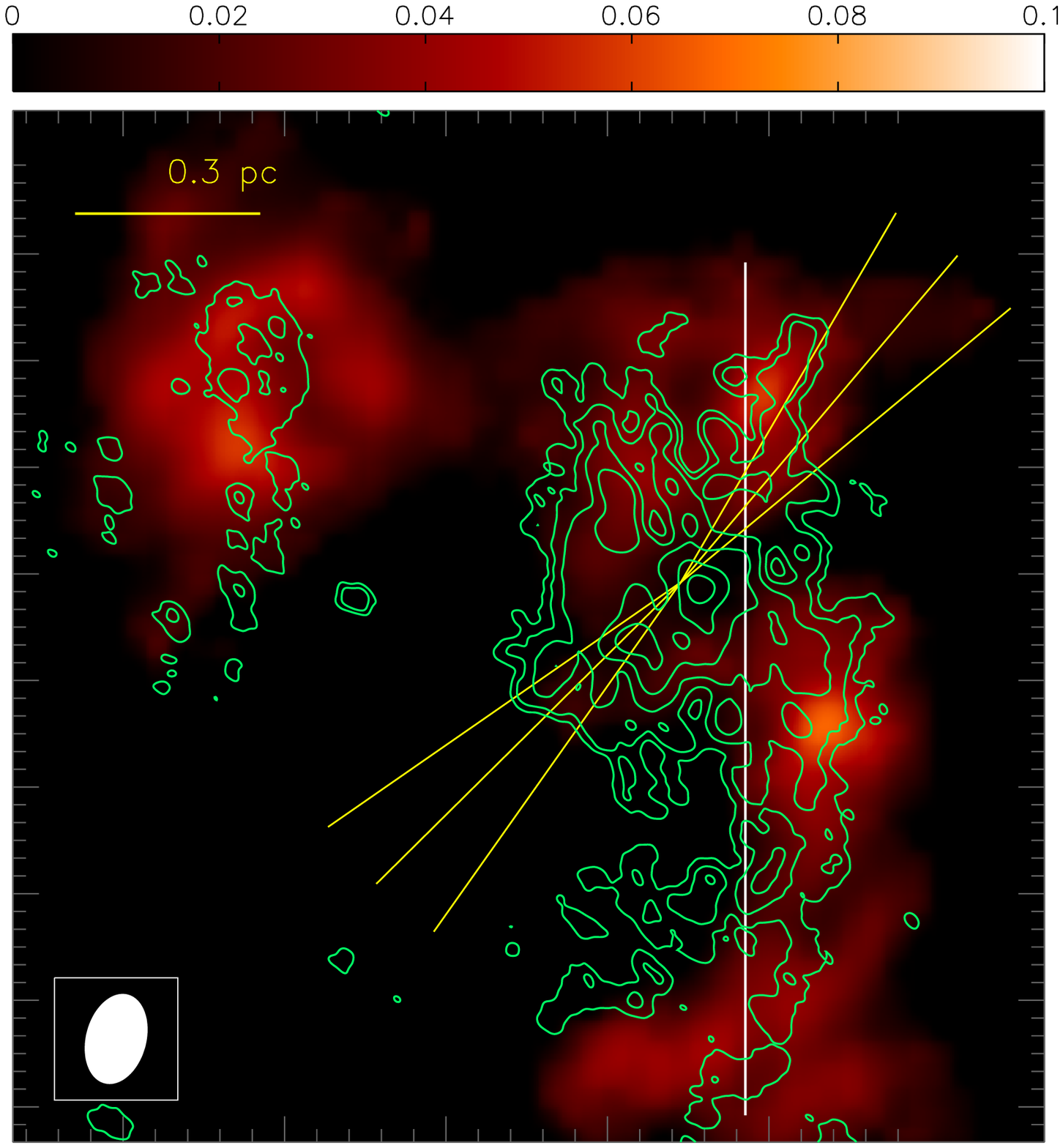} \\
\end{tabular}

\vspace{-1.2cm}

\begin{tabular}{  p{7.5cm} p{7.5cm} }
\includegraphics[scale=0.45]{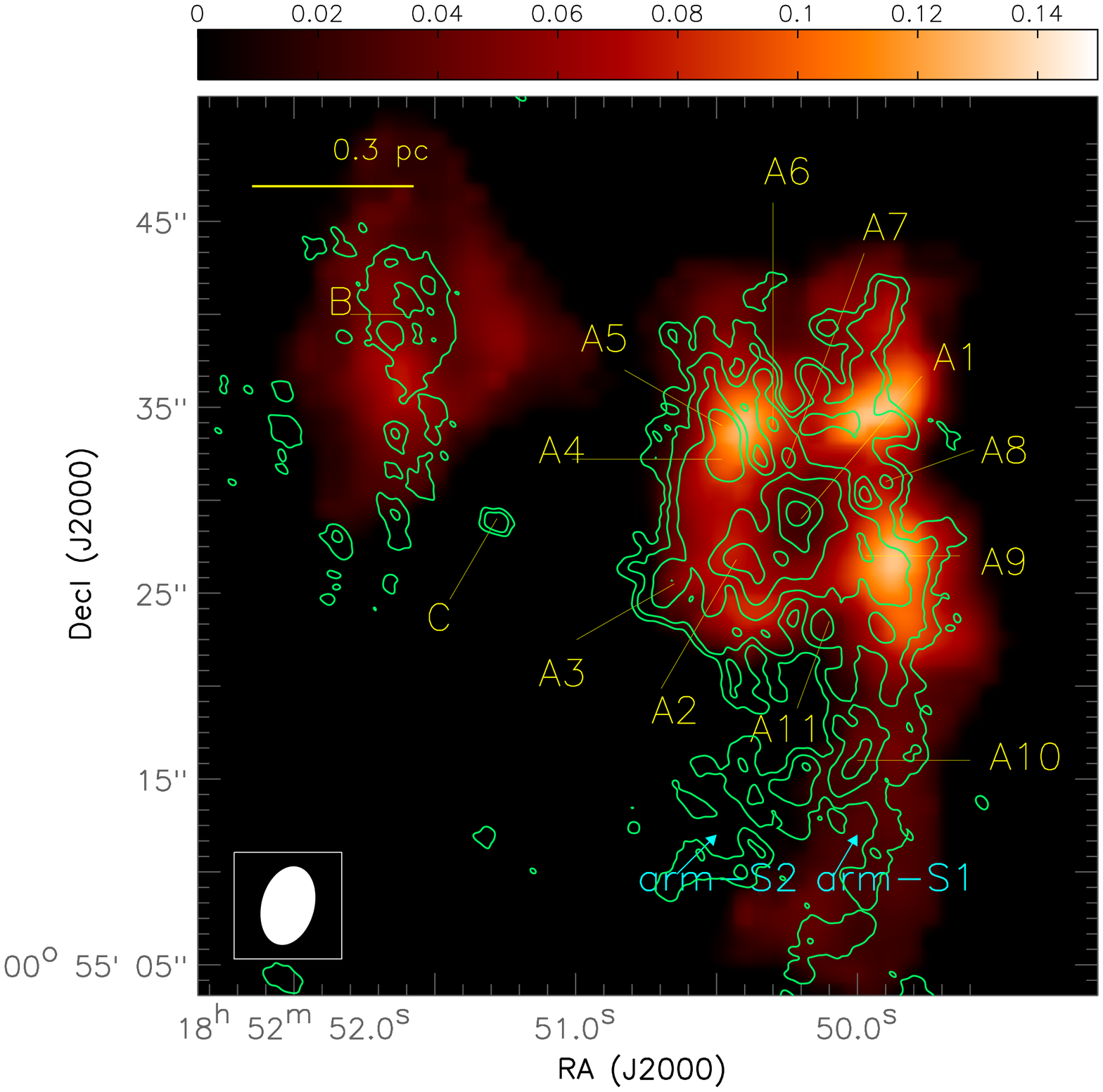} & 
\includegraphics[scale=0.45]{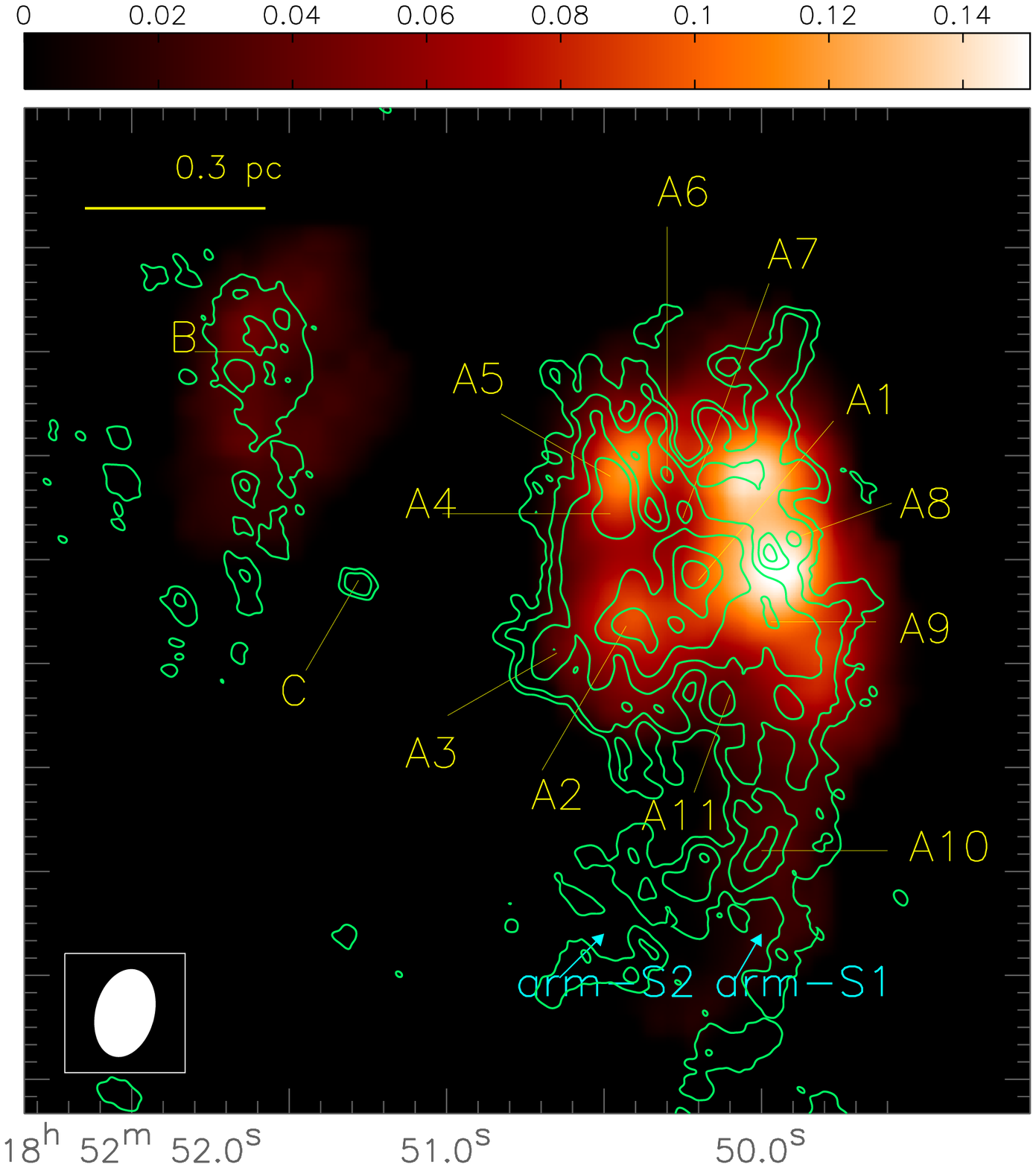} \\
\end{tabular}
\caption{SMA observations of 1.3 mm continuum emissions overlaid with the VLA observations of NH$_{3}$ velocity integrated intensity. \textbf{Top Left: } NH$_{3}$ (J,K)=(1,1) main hyperfine inversion line velocity integrated intensity (color) and SMA observations of 1.3 mm continuum emission. 
\textbf{Top Right: } NH$_{3}$ (J,K)=(1,1) satellite hyperfine inversion line velocity integrated intensity (color) and SMA observations of 1.3 mm continuum emission. 
\textbf{Bottom Left: } NH$_{3}$ (J,K)=(2,2) main hyperfine inversion line velocity integrated intensity (color) and SMA observations of 1.3 mm continuum emission. 
\textbf{Bottom Right: } NH$_{3}$ (J,K)=(3,3) main hyperfine inversion line velocity integrated intensity (color) and SMA observations of 1.3 mm continuum emission. The unit of color bars is Jy\,beam$^{-1}$\,km\,s$^{-1}$. Contour levels are 1.2 mJy\,beam$^{-1}$$\times$[1, 2, 4, 8, 16, 32]. The synthesized beams of the NH$_{3}$ images are shown in the bottom left.
The lines in the top right panel show the PV cuts used in Figure \ref{fig_nh3pv_offx3} and \ref{fig_nh3pv_offnon}.
}
\label{fig_nh3}
\end{figure}

Our VLA observations yield significant detections of the main and satellite hyperfine inversion transitions of NH$_{3}$ (J,K)=(1,1), and the main hyperfine inversion transition of NH$_{3}$ (J,K)=(2,2) and (3,3).
Figure \ref{fig_nh3} shows the velocity integrated emission maps of the detected lines. 
The geometry traced by the NH$_{3}$ transitions generally agrees with that seen in the 1.3 mm continuum image. 
Due to the poor angular resolutions of the NH$_{3}$ observations, the expected absorption line in the center of G33.92+0.11 A is smeared over a $\sim$4$''$ beam area, that leads to a low integrated flux in the center. 
From Figure \ref{fig_nh3}, we find that the NH$_3$ emission tends to peak closer into the hub when going from the (J,K)=(1,1) transition to the (J,K)=(2,2) and (J,K)=(3,3) lines (see regions NW and SW in this Figure). As shown below, this translates into a temperature gradient along the detected arms (i.e. arm--N and arm--S1) as one moves inward to cores A1 and A2.

In particular, toward the arm--NW region in between Decl.=19$^{\circ}$55$'$35$''$ and Decl.=19$^{\circ}$55$'$40$''$, the (J,K)=(1,1) satellite hyperfine line shows much lower integrated flux than the main hyperfine line. This could be related to the fact that the main hyperfine line shows a broad line profile, similar to that detected in SiO or OCS, explaining the contrast between the integrated flux of the NH$_{3}$ main and satellite hyperfine lines. 
In the same region, we also see significant emission of the NH$_{3}$ (J,K)=(2,2) and (3,3) main hyperfine lines, implying the localized heating. 
In the A8 region, we see significantly stronger (J,K)=(3,3) line emission, also implying the high (3,3) to (1,1) and (3,3) to (2,2) line ratio.  
This is consistent with the results from observations of SiO, SO, and OCS that suggest localized heating sources, outflows, and shocks around A8  (e.g. SiO--2). 
We also note the enhanced NH$_{3}$ rotational temperature north of A2, which is consistent with the results of the CH$_{3}$CN line ratio (Figure \ref{fig_ch3cn}, bottom left).

\begin{figure}
\includegraphics[scale=0.7]{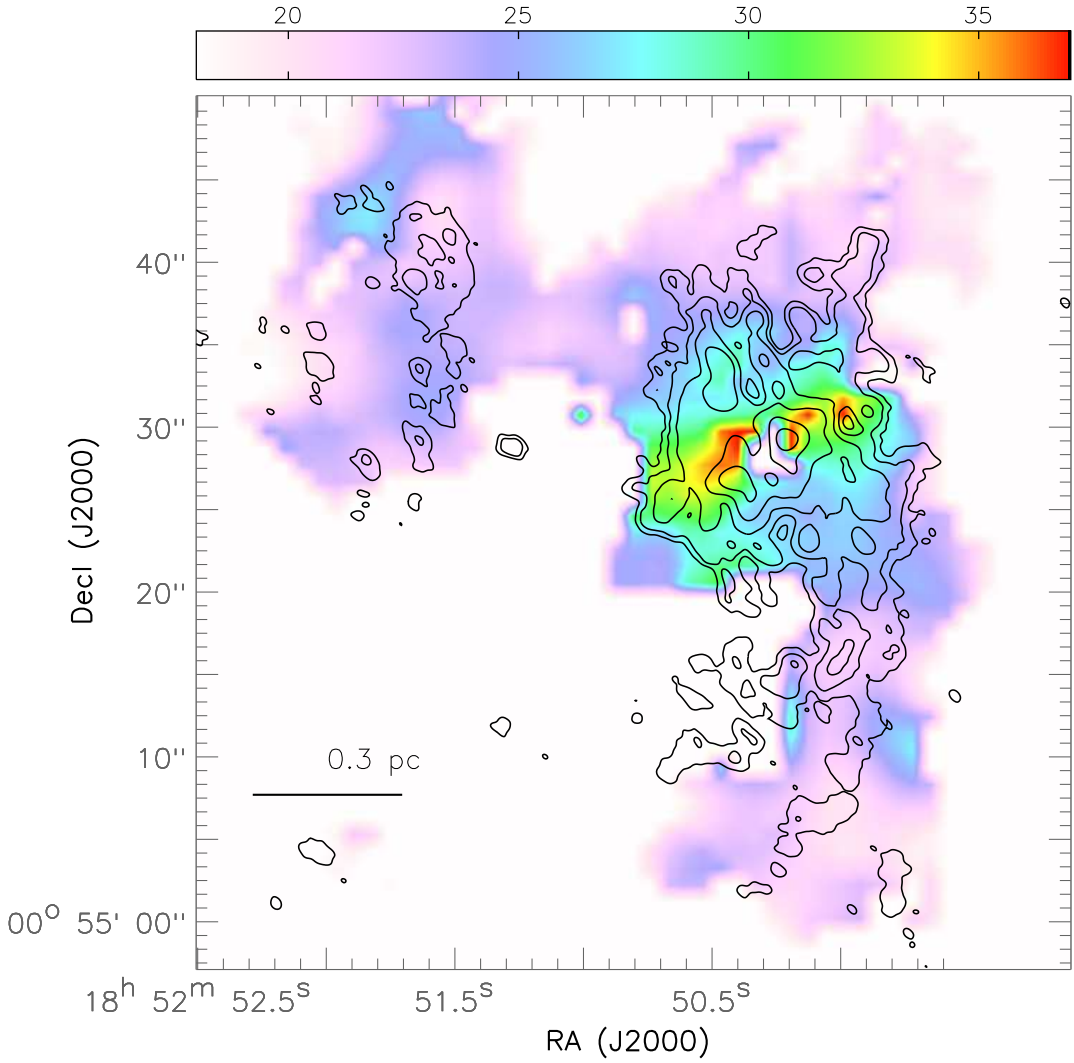}
\caption{The rotational temperature T$_{rot}$ derived from the NH$_{3}$ (J,K)=(1,1) and (2,2) hyperfine transitions. In the range of presented T$_{rot}$. The gas kinetic temperature T$_{k}$ is comparably higher than T$_{rot}$ (by $\lesssim$10 K). Contours show the natural weighted 1.3 mm continuum image. Contour levels are 1.2 mJy\,beam$^{-1}$$\times$[1, 2, 4, 8, 16, 32]. The central 0.6 pc region shows higher temperature than the extended area. The temperature is not smoothly enhanced toward the peak 1.3 mm continuum emission. Localized higher temperature is seen at A8, and can be marginally seen toward the west of A1. This nonuniform temperature enhancement can be interpreted by localized stellar and outflow/shock heating.}
\label{fig_temperature}
\end{figure}

We derived the NH$_{3}$ rotational temperature using the  NH$_{3}$ (J,K)=(1,1) and (2,2) line data, based on the formalism described in Ho \& Townes (1983).  
Figure \ref{fig_temperature} shows the distribution of the derived NH$_{3}$ rotational temperature.
For locations that show distinct velocity/temperature components, we present the higher temperature component. 
The estimate of the NH$_{3}$ rotational temperature toward the regions with smeared line absorption (as in e.g. A1) is not reliable.
Nevertheless, this part of information is already complemented by the rotational diagram analysis of the CH$_{3}$CN data (Section \ref{chap_ch3cn}; Figure \ref{fig_ch3cn}). 
From Figure 16, we find that the G33.92+0.11 A and B regions have an overall NH$_{3}$ rotational temperature of 20--35 K, with the molecular gas being warmer in the central 0.5 pc area around G33.92+0.11 A. In addition, Figure 16 reveals a temperature gradient along the detected arms when moving inward to cores A1 and A2. Indeed, while the temperature at the outer regions is $\sim$23 K, it increases to $\sim$30 K at the interaction regions between the arms and the central hub, and it finally rises to temperatures of $\sim$35 K closer to cores A1 and A2. This progressive increase of the gas temperature, together with the detection of shock tracers such as SiO or OCS, suggests an scenario where the molecular gas funnels along the arms into the central hub and gets shocked at the surface of the clump hosting the cluster of OB stars.

We note that the (J,K)=(3,3) transition is detected within a more limited region as compared to the detections of  (J,K)=(1,1) and (2,2) transitions. 
The NH$_{3}$ rotational temperature based on the NH$_{3}$ (J,K)=(1,1) and (3,3) lines yields a similar distribution in the central 0.6\,pc area in G33.92+0.11 However, with additional uncertainties caused by the ortho to para ratio (Ho \& Townes 1983). 
The image of the NH$_{3}$ (J,K)=(3,3) main hyperfine inversion line (Figure \ref{fig_nh3}) still serves as a good reference to identify high temperature regions. 
 
In Figure \ref{fig_temperature}, we also see the elongated heated region in the southeast and northeast of G33.92+0.11 A, which follows the morphology of the champagne flow (Churchwell 2002) seen in the H30$\alpha$ image (Figure \ref{fig_h30a}).
From the results introduced in this section, we argue that not only the centrally embedded OB stars play a role in the heating of the molecular clumps via UV photon illumination, but also the interaction of accretion shocks generated by the funneling of gas along the arms into the central hub. 
Our case study of G33.92+0.11 region complements the higher mass end of discussions about stellar heating, e.g. in Offner et al. (2009) and Longmore et al. (2011a).

\subsection{Kinematics and Spectral profiles}
\label{chap_kinematics}
We introduce the resolved kinematics of the molecular gas in Section \ref{sub_moleculark}.
The motion of the ionized gas in the UC H\textsc{ii} regions are marginally resolved in the observations of the H30$\alpha$ emission, and are discussed in Section \ref{sub_ionizedk}.

\subsubsection{Molecular Gas} 
\label{sub_moleculark}
\begin{figure}
\includegraphics[scale=0.5]{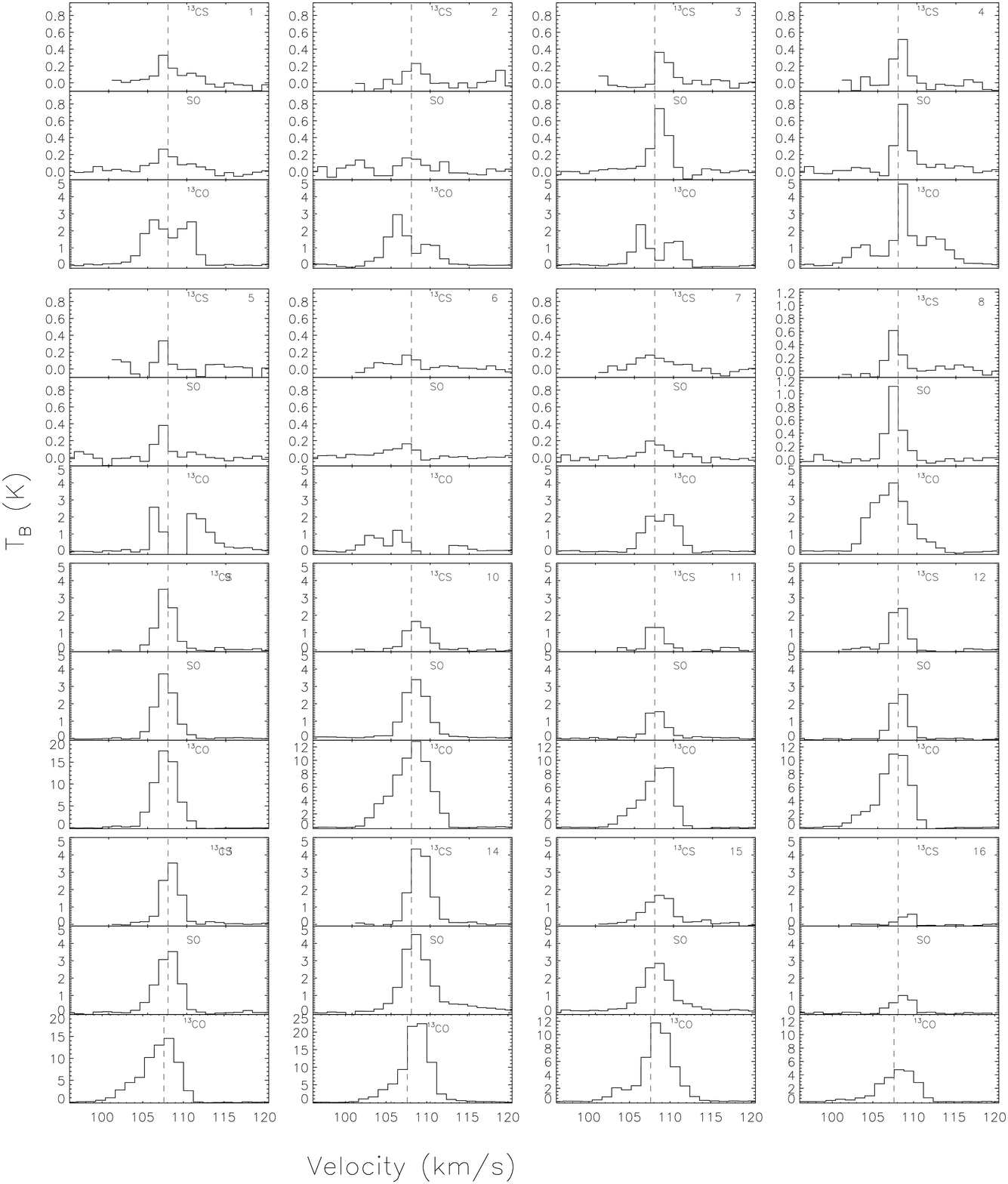}
\caption{The averaged $^{13}$CS 5--4, SO 5$_{6}$--4$_{5}$, and $^{13}$CO 2--1 spectra in the selected $^{13}$CS--1$_{,\cdots,}$16 regions. The dashed lines label the systemic velocity $v$=107.6 km\,s$^{-1}$. 
We note that the double peak profiles of $^{13}$CO 2--1 in the $^{13}$CS--1$_{,\cdots,}$7 regions may be caused by both the missing flux in the interferometric observations and the foreground (self--)absorption, which degrade the intensity around the cloud velocity.}
\label{fig_spectra}
\end{figure}

In our observations, the emission of the CO 2--1 isotopologues, the SO 5$_{6}$--4$_{5}$ transition, and the $^{13}$CS 5--4 transition trace the kinematics out to a scale of $\gtrsim$1\,pc radius. 
As discussed in Section \ref{chap_s}, the detected chemical stratification suggests that the large-scale molecular filaments are physically associated with the centrally embedded OB cluster in G33.92+0.11.
Figure \ref{fig_spectra} shows the spectra of $^{13}$CO 2--1, SO 5$_{6}$--4$_{5}$, and $^{13}$CS 5--4 from selected regions (see Figure \ref{fig_so13cs}).

From Figure \ref{fig_spectra}, we see a general trend that the dense molecular gas in the north (or northeast) is redshifted, while the dense molecular gas in the south  (or southwest) is blueshifted. 
This is also observed in the channel maps of SO 5$_{6}$--4$_{5}$ (Figure \ref{fig_sochan}).
At the $\sim$1 pc scale radius, the recognized systemic motion has a velocity scale of $|v|\lesssim$0.6 km\,s$^{-1}$.
Following the regions $^{13}$CS--8$\rightarrow$9$\rightarrow$10 and $^{13}$CS--7$\rightarrow$11$\rightarrow$12$\rightarrow$13$\rightarrow$14$\rightarrow$15$\rightarrow$16, without corrections for the inclination, the velocity scale of the systemic motion is within $\sim$1.2 km\,s$^{-1}$.

\begin{figure}
\rotatebox{-90}{
\includegraphics[scale=0.7]{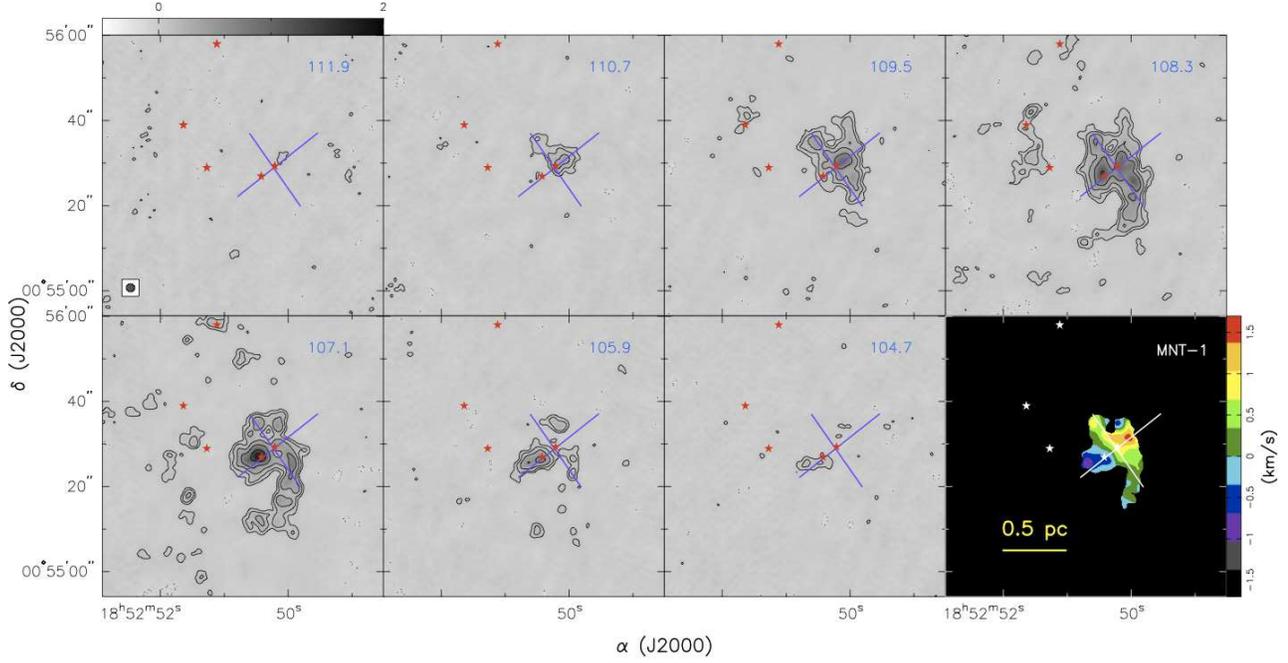}
}
\caption{Channel maps of the $^{13}$CS 5--4 line. Positive contours (solid) are 72 mJy\,beam$^{-1}$(3$\sigma$)$\times$[1, 2, 4, 8, 16, 32, 64, 128]. Negative contours (dotted) are -72 mJy\,beam$^{-1}$$\times$[1, 2, 3]. The synthesized beam is $\theta_{maj}$$\times$$\theta_{min}$: 2$''$.0$\times$1$''$.9. Locations of G33.92+0.11 A1, A2, B, C and D are marked with stars. The bottom right panel shows the intensity--weighted (only selected the $>$0.2\,Jy\,beam$^{-1}$ pixels) average velocity (i.e. moment 1) map of the $^{13}$CS 5-4 line. We display the velocity relative to the $v_{lsr}=107.6$\,km\,s$^{-1}$ in the average velocity map.}
\label{fig_13cschan}
\end{figure}

\begin{figure}
\rotatebox{-90}{
\includegraphics[scale=0.7]{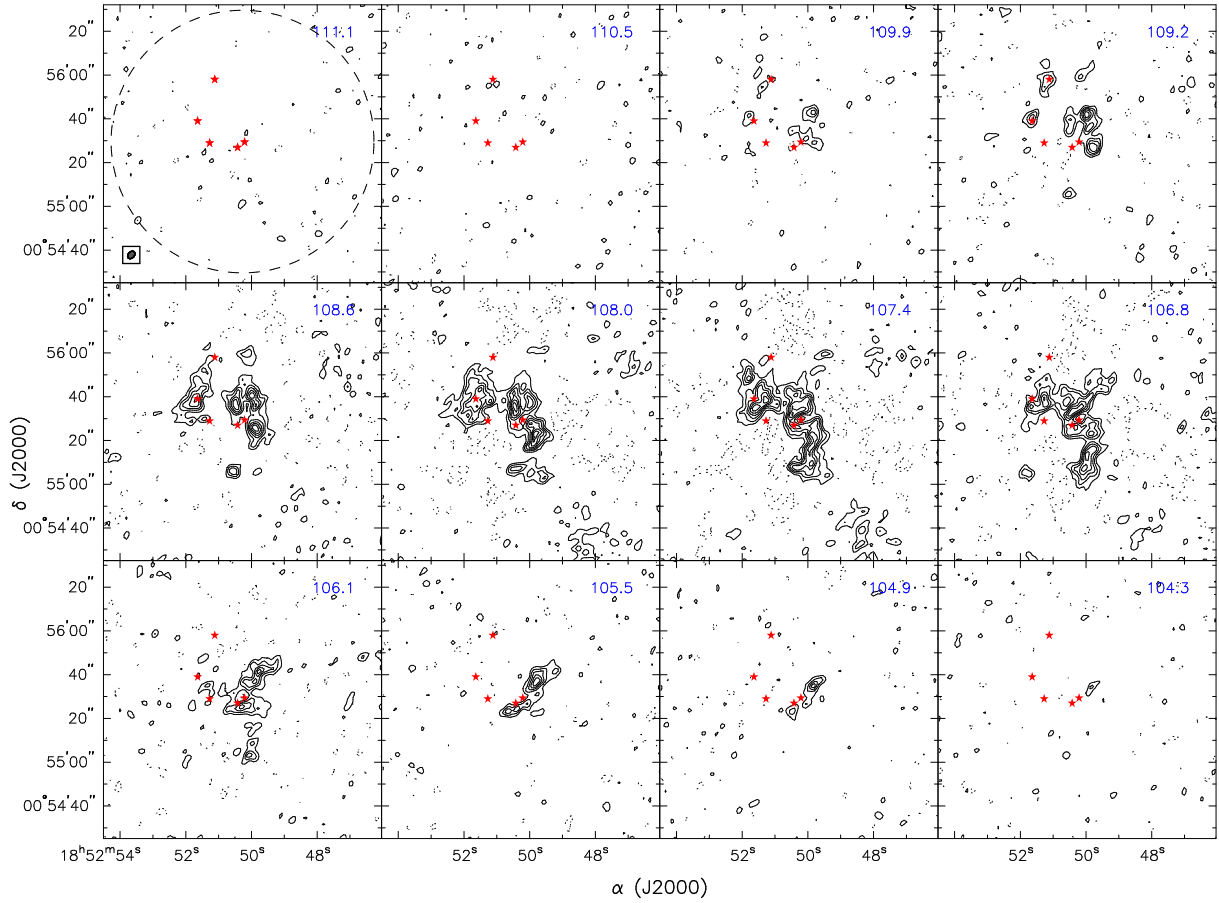}
}
\caption{Channel maps of the NH$_{3}$ (1,1) main hyperfine inversion transition. This image is not corrected for the primary beam response for the sake of showing the significantly detected extended structures. Locations of G33.92+0.11 A1, A2, B, C and D are marked with red stars. The primary beam and the synthesized beam of this observation are shown in the first panel. The observation track the rest frequency of the main hyperfine transition. One of the inner satellite hyperfine transitions is covered in the channels of 97.5--101.8 km\,s$^{-1}$. Positive (solid) and negative (dotted) contours start from $\pm$9 mJy\,beam$^{-1}$ (2$\sigma$), with separations of $\pm$9 mJy\,beam$^{-1}$.}
\label{fig_11chan}
\end{figure}

\begin{figure}
\rotatebox{-90}{
\includegraphics[scale=0.7]{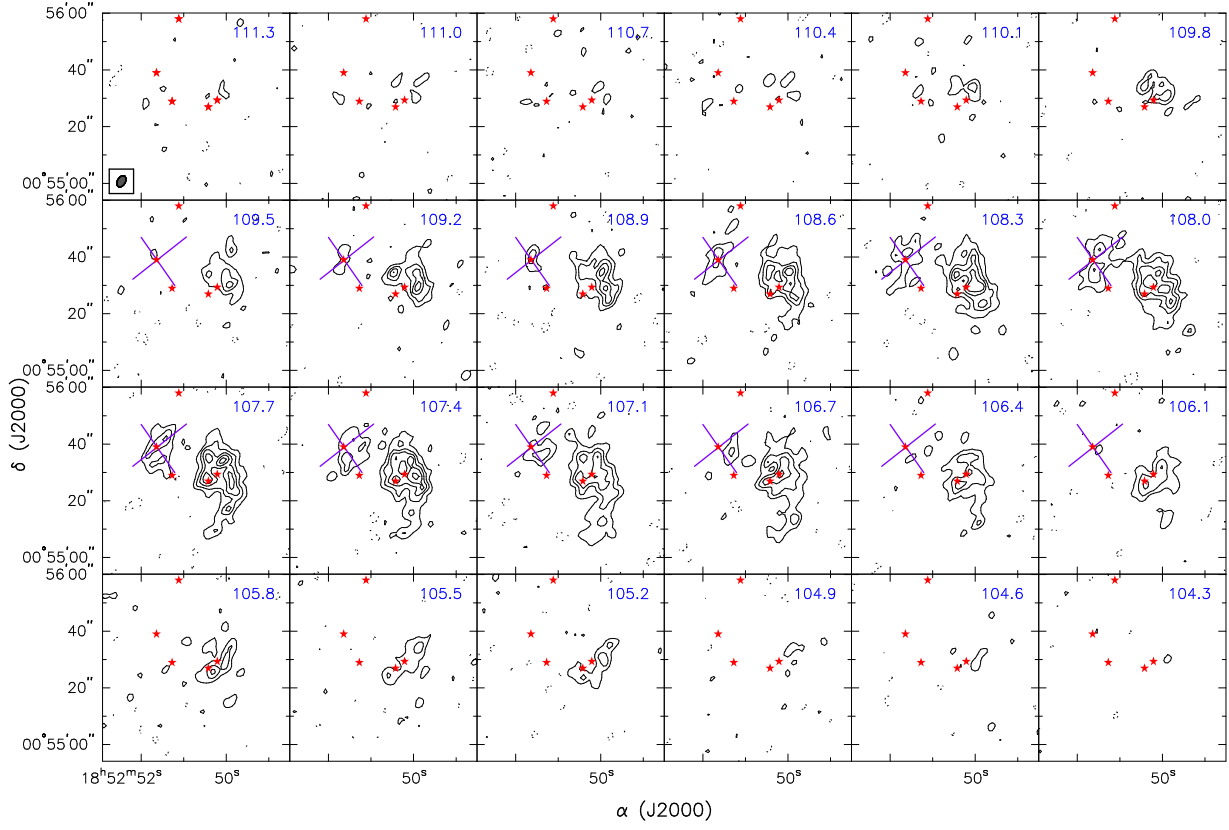}
}
\caption{Channel maps of the NH$_{3}$ (3,3) main hyperfine inversion transition. Locations of G33.92+0.11 A1, A2, B, C and D are marked with red stars. The synthesized beam of this observation are shown in the first panel. The plotted area is much smaller than the primary beam of this observation (2$'$). Positive (solid) and negative (dotted) contours start from $\pm$9 mJy\,beam$^{-1}$ (2$\sigma$), with separations of $\pm$9 mJy\,beam$^{-1}$. Purple lines provide the references for the relative motions within G33.92+0.11 B.}
\label{fig_33chan}
\end{figure}

Within the central $\sim$1 pc region around G33.92+0.11 A and B, we resolve more complicated structures and kinematics.
Figures \ref{fig_13cschan}, \ref{fig_11chan}, and \ref{fig_33chan} show the channel maps of the $^{13}$CS 5--4 transition and the  NH$_{3}$ (J, K) = (1,1) and (3,3) hyperfine transitions.
From the channel maps we see:
\begin{itemize} 
\item[1.] From the velocity channel of 106.1 km\,s$^{-1}$ to the velocity channel of 108.0 km\,s$^{-1}$, an extended ($\gtrsim$20$''$) arm--S component appears to converge from the south of G33.92+0.11 A toward the north (Figures \ref{fig_11chan}, \ref{fig_33chan}). 
\item[2.] We resolve an overall northwest--southwest velocity gradient in G33.92+0.11 A in the channel maps of the $^{13}$CS 5--4 line. As indicated in Figure \ref{fig_13cschan}, the motion within G33.92+0.11 A resembles an inclined disk--like structure. The highest blueshifted and redshifted line emission regions are compact, and are located very close to the 1.3\,mm continuum peaks, marked by the stars. Around the cloud velocity (e.g. the channels $v=107.1$ km\,s$^{-1}$ and $v=108.3$ km\,s$^{-1}$), the most significant line emission region shows an elongated southwest--northeast distribution. The distribution of molecular gas appears to be non--uniform. 
A similar rotation profile is also seen in the G33.92+0.11 B region in the NH$_{3}$ channel maps, although the direction of the velocity gradient is nearly perpendicular to that in G33.92+0.11 A (Figure \ref{fig_33chan}).
At the cloud velocity, the G33.92+0.11 B region might be confused with the elongated molecular arm--EN,ES (Section \ref{chap_co}), which contribute to some excess of emission with north--south orientation. Over this G33.92+0.11 B region, we see a low gas temperature (Section \ref{chap_nh3}), without H\textsc{ii} regions (Section \ref{chap_rrl}) or massive molecular outflows (Section \ref{chap_hv}). The blueshifted and redshifted gas are more likely to be accreting toward the center, rather than moving outward which requires a not yet detected powering mechanism.
\end{itemize}

\begin{figure}
\rotatebox{-90}{
\includegraphics[scale=0.69]{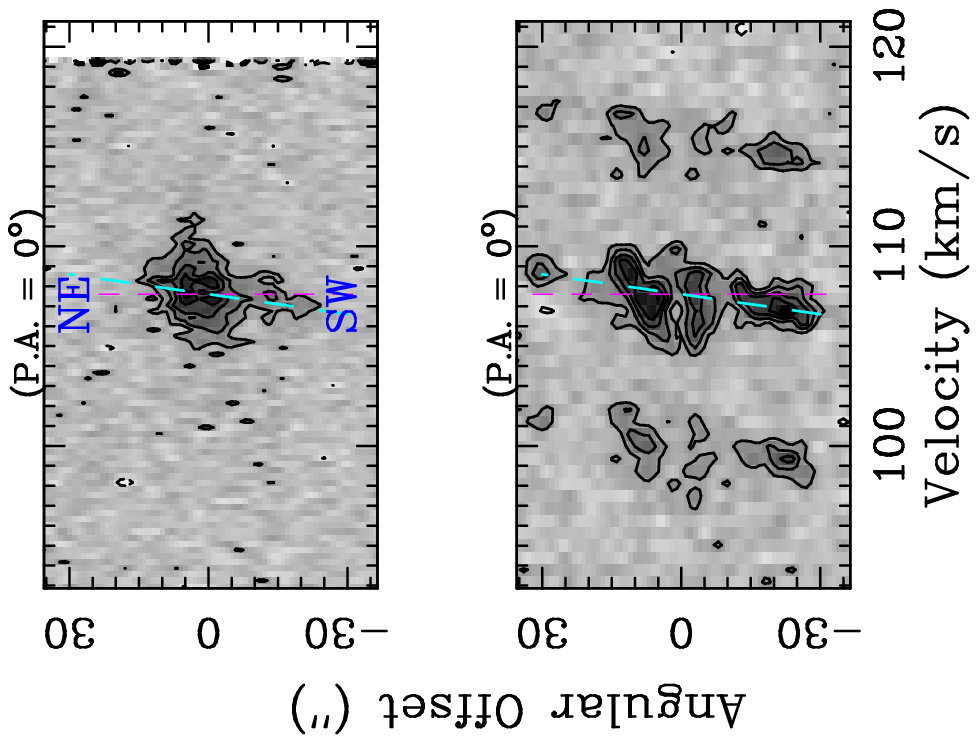}
}
\caption{The Position--Velocity diagram of the NH$_{3}$ (J,K)=(1,1) and (3,3) lines. 
The zero offset represents the center of the PV cuts: R.A.=18$^{\mbox{h}}$52$^{\mbox{m}}$50$^{\mbox{s}}$.073, Decl.= 00$^{\circ}$55$'$29$''$.603, which is -3$''$ offset from the  pointing center of the NH$_{3}$ observations (Figure \ref{fig_nh3}). 
The position angles of the respective PV cuts are labeled in individual panels. 
Positive (solid) and negative (dotted) contours start from $\pm$9 mJy\,beam$^{-1}$ (2$\sigma$), with separations of $\pm$9 mJy\,beam$^{-1}$. The purple dashed line labels the systemic velocity $v_{lsr}$ of 107.6 km\,s$^{-1}$.
The cyan dashed line indicates the resolved velocity gradient of dense gas.}
\label{fig_nh3pv_offx3}
\end{figure}

\begin{figure}
\rotatebox{-90}{
\includegraphics[scale=0.75]{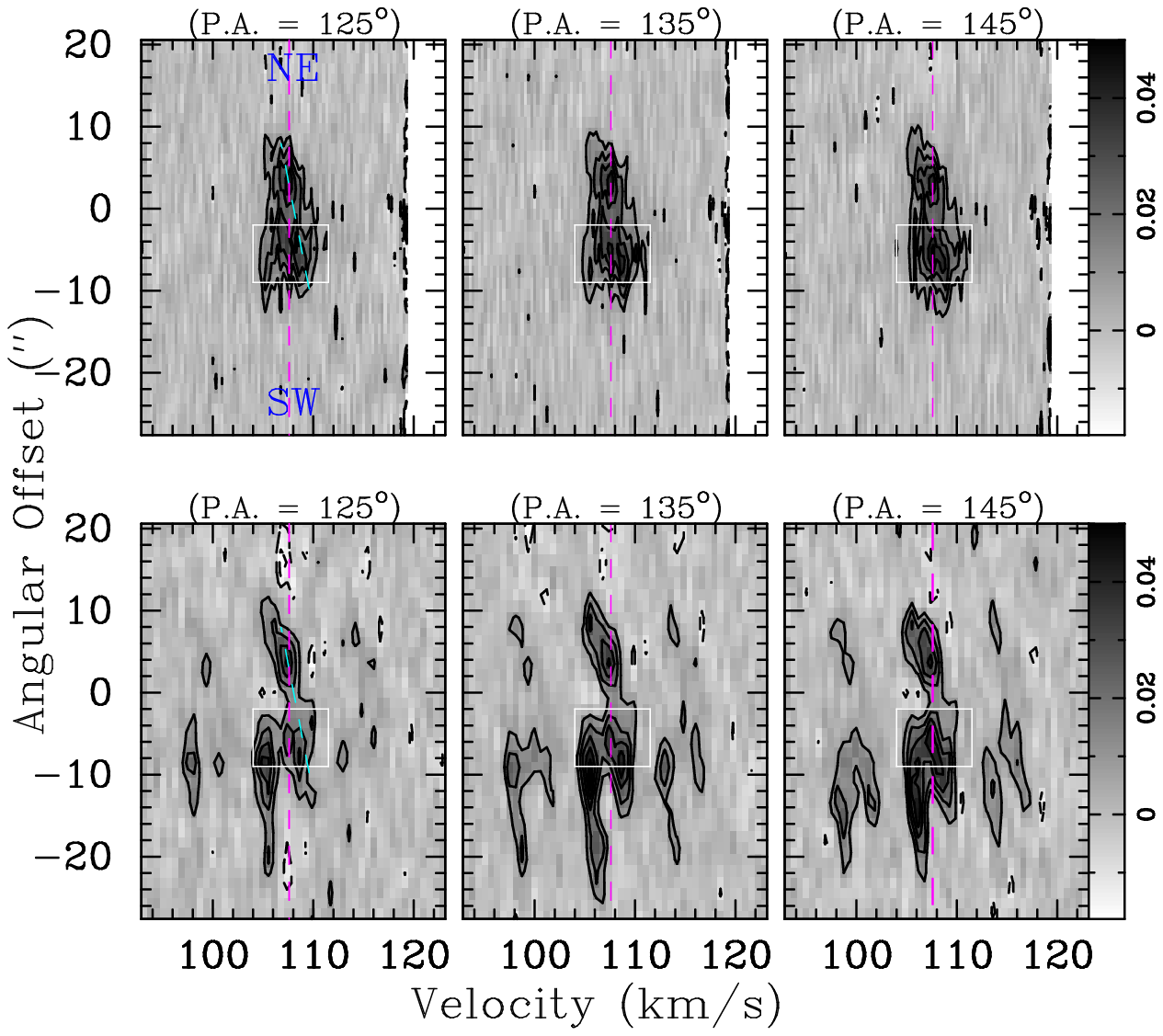}
}
\caption{The Position--Velocity diagram of the NH$_{3}$ (J,K)=(1,1) and (3,3) lines. 
The zero offset represents the center of the PV cuts: R.A.=18$^{\mbox{h}}$52$^{\mbox{m}}$50$^{\mbox{s}}$.273, Decl.= 00$^{\circ}$55$'$29$''$.603, which is also the pointing center of the NH$_{3}$ observations (Figure \ref{fig_nh3}). 
The position angles of the respective PV cuts are labeled in individual panels. 
Positive (solid) and negative (dotted) contours start from $\pm$9 mJy\,beam$^{-1}$ (2$\sigma$), with separations of $\pm$9 mJy\,beam$^{-1}$. The purple dashed line labels the systemic velocity $v_{lsr}$ of 107.6 km\,s$^{-1}$.
The cyan dashed line indicates the resolved velocity gradient of dense gas.
The white box in each panel indicates the spatial and the velocity ranges that are potentially contaminated by the A1--HVG.}
\label{fig_nh3pv_offnon}
\end{figure}

\begin{figure}
\includegraphics[scale=0.5]{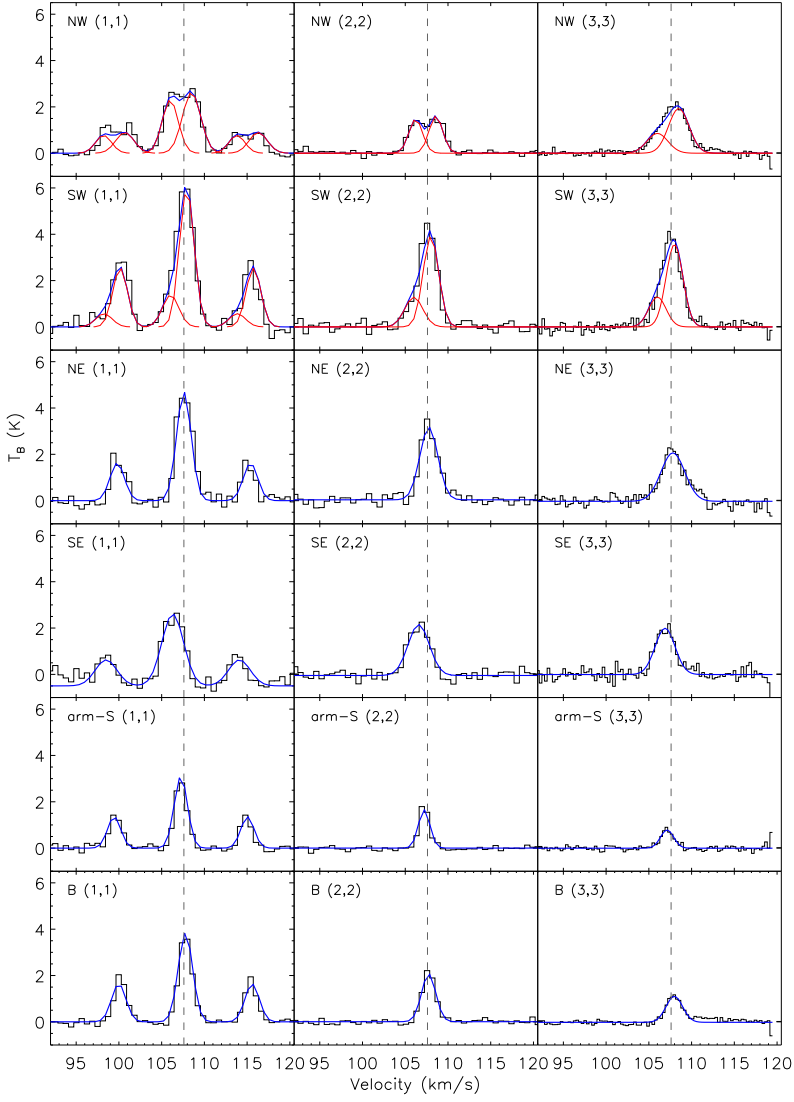}
\vspace{-2cm}
\caption{The averaged NH$_{3}$ spectra in respective 6 regions (black). The regions are labeled in the bottom left panel of Figure \ref{fig_nh3}.  
For each spectrum, we use 1 or 2 Gaussian components to quantitatively characterize the line profile.
The overall fitted spectra are plotted in blue color. Individual Gaussian components in the NW and SW regions are shown in red color. The Gaussian components to fit these spectra are summarized in Table \ref{table_nh3gau}.}
\label{fig_nh3spectra}
\end{figure}

Consecutive PV cuts of the NH$_{3}$ (1,1) and (3,3) data cubes are generated, centered on the UC H\textsc{ii} region (i.e. R.A.=18$^{\mbox{h}}$52$^{\mbox{m}}$50$^{\mbox{s}}$.273, Decl.= 00$^{\circ}$55$'$29$''$.603) and are centered on 3$''$ west of the UC H\textsc{ii} region, to examine the velocity gradient from G33.92+0.11 A to the extended arm--S1 and S2, and the velocity components within the central 0.6 pc area in G33.92+0.11 A.
The PV diagrams of the NH$_{3}$ (J, K) = (1,1) and (3,3) lines are shown in  Figures \ref{fig_nh3pv_offx3} and \ref{fig_nh3pv_offnon}.
Fitting of the spectra profiles of the NH$_{3}$ emission in five selected regions (see Figure \ref{fig_nh3}, bottom left panel) is also provided in Figure \ref{fig_nh3spectra} and Table \ref{table_nh3gau}, to quantitatively characterize motion and excitation conditions. 
In Figure \ref{fig_nh3pv_offx3}, we verify a moderate velocity gradient of 0.96 km\,s$^{-1}$\,pc$^{-1}$ from arm--S toward the north of G33.92+0.11 A (i.e. P.A. = 0$^{\circ}$).

The most blueshifted motion is seen in the SE and the SW regions, with a velocity of $\sim$-1.5 km\,s$^{-1}$ relative to the cloud velocity of 107.6\,km\,s$^{-1}$.
The arm--S region has a lower blueshifted velocity of -0.5 km\,s$^{-1}$ relative to the cloud velocity of 107.6\,km\,s$^{-1}$.
The redshifted motion relative to 107.6\,km\,s$^{-1}$ is seen in the northern region (i.e. NW, NE, and G33.92+0.11 B), with velocities no higher than +1\,km\,s$^{-1}$.
The fitted velocity of G33.92+0.11 B is systematically redshifted from the cloud velocity of 107.6 km\,s$^{-1}$, which is consistent with the overall velocity gradient at $>$1 pc scale.  

The fitted NH$_{3}$ FWHM in the selected five regions is 1.6--3.5 km\,s$^{-1}$ with a medium velocity of $\sim$2.5 km\,s$^{-1}$.
Our results are consistent with the previously reported C$^{18}$O 1--0 line widths of 2.5--3.2 km\,s$^{-1}$ (Watt \& Mundy 1999).
At the $\sim$0.3 pc size NW and SE region (Figure \ref{fig_nh3}), the resolved velocity gradient potentially broadens the spectra by 4.6\,(km\,s$^{-1}$\,pc$^{-1}$)$\times$0.3\,(pc)=1.38\,km\,s$^{-1}$.

\subsubsection{Ionized Gas}
\label{sub_ionizedk}
\begin{figure}
\includegraphics[scale=0.7]{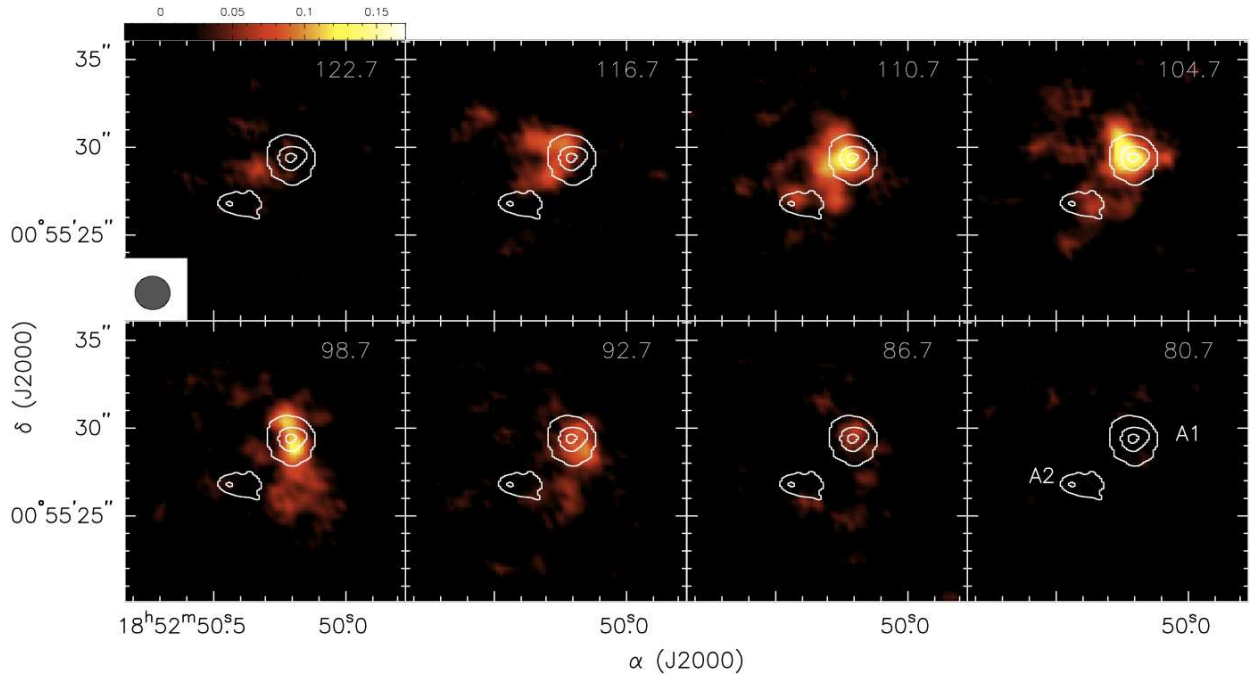}
\caption{Channel maps of the H30$\alpha$ line (color), overlaid with the Briggs Robust 0 weighting 1.3 mm continuum image (contour), around the massive binary cores A1 and A2. Color bar has the unit of Jy\,beam$^{-1}$. Contours are 4 mJy\,beam$^{-1}$$\times$[2, 4, 6]. }
\label{fig_h30achan}
\end{figure}

Figure \ref{fig_h30achan} shows the channel maps of the H30$\alpha$ line (natural weighting). 
In velocity range of 92.7--116.7 km\,s$^{-1}$, the brightest component of the H30$\alpha$ emission moves from the southwest of core A1 for blueshifted velocities to the northeast of core A1 for redshifted velocities. 
The overall distribution of these bright component is an elongated UC H\textsc{ii} region around the most massive core A1 (see also Section \ref{chap_rrl}; Figure \ref{fig_h30a}, right).
The detected velocity range of -10.9 to +9.1 km\,s$^{-1}$ (relative to 107.6\,km\,s$^{-1}$) of the ionized gas can be explained by the thermal expansion of the ionized gas, toward the lower density bipolar directions (see discussions in Liu, Zhang \& Ho 2011).
It can also be explained by an ionized jet. 
In the high angular resolution image of the H30$\alpha$ line (Figure \ref{fig_h30a}, bottom right), the bright emission in this elongated UC H\textsc{ii} region is concentrated to a narrow distribution, which implies that the ionized gas may be collimated. 
The distribution of the ionized gas remains to be constrained by more sensitive observations of hydrogen recombination lines. 
We note that the OB stars embedded in the brightest H30$\alpha$ emission region (Figure \ref{fig_h30a}, bottom right) potentially ionize all adjacent molecular jets. 
We also note that other cases of (not yet spatially resolved) high velocity ($\gtrsim$60 km\,s$^{-1}$) ionized jets in the $L$$\gtrsim$10$^{6}$ L$_{\odot}$ OB cluster forming regions have been reported torward G10.6-0.4 by Keto \& Wood (2006) or NGC 7538 IRS 1 (Sandell et al. 2009).


\section{Discussion}
\label{chap_discussion}
In this section, we discuss physical implications of our observational results. 
In Section \ref{sub_g33pv}, we argue a small inclination angle of the target source G33.92+0.11 based on the detected kinematics. 
In Section \ref{sub_fragmentation}, we discuss the evolutionary context of OB cluster forming molecular clumps, based on the resolved geometry/morphology and kinematics in G33.92+0.11.
In Section \ref{sub_angular momentum}, we address how the asymmetrical matter distribution in G33.92+0.11 alleviates the angular momentum problem in the accretion process.
The resolved chemical diversities in G33.92+0.11 are discussed in Section \ref{sub_diversity}.

\subsection{G33.92+0.11 as a Nearly Face--On Spinning--Up Massive Molecular Clump?}
\label{sub_g33pv}
With a $\sim$1.6--5.8$\cdot$10$^{3}$\,$M_{\odot}$ embedded molecular gas (see Section \ref{chap_ch0}), the expected rotational velocity ($\sim\sqrt{GM/r}$) at a 0.3 pc radius around G33.92+0.11 A should be 4.6--8.8\,km\,s$^{-1}$, while the expected free infall velocity ($\sim\sqrt{2GM/r}$) is 6.5--12\,km\,s$^{-1}$. 
Further in, i.e. at a radius of 0.05\,pc, the expected rotational or infall velocities around G33.92+0.11 A1 and A2 cores are $\gtrsim\pm$2.8 km\,s$^{-1}$.
These velocities are large compared to the previously reported C$^{18}$O 1--0 line width of 2.5--3.2 km\,s$^{-1}$ (Watt \& Mundy 1999),  and our NH$_{3}$ observations constrain the systematic motion of the dense gas to be within a velocity range of [$v_{lsr}$$-$1.6 km\,s$^{-1}$, $v_{lsr}$$+$1.0 km\,s$^{-1}$] toward G33.92+0.11 A (Section \ref{sub_moleculark}).

The detected small velocities in G33.92+0.11 A can be explained by planar motions with an inclination angle much smaller than $\sim$30$^{\circ}$. 
Assuming that all OB cluster--forming massive molecular clumps are flattened either due to rotation or due to the magnetic support, and considering that their rotation axes do not have preferential orientations, the probability that the inclination angle (i.e. the angle between the rotational axis and the line of sight) of a target is smaller than 30$^{\circ}$ is about 13\%.
Among the 7 targets in our database of VLA NH$_{3}$ observations (Section \ref{chap_introduction}), the probability to find at least one such highly inclined object can be estimated by $P$ = $1-(1-13\%)^{7}$ = $62.3\%$.
As addressed in Section \ref{sub_moleculark}, the resolved motion in G33.92+0.11 A, B resembles an inclined disk--like structure as well.

If G33.92+0.11 A, B are face--on massive molecular clumps, the global systematic rotation/infall will minimally contribute to the measured NH$_{3}$ line widths (e.g. by $\ll$1.4\,km\,s$^{-1}$).
In the NE and SW regions which are less affected by these dynamical broadening (see also Section \ref{sub_moleculark}), as well as the G33.92+0.11 B, the detected NH$_{3}$ (1,1) line widths are FHWM$\sim$2\,km\,s$^{-1}$.
At linear scales of $\sim$0.1\,pc, the detected $\sim$50\,$M_{\odot}$ localized dense molecular cores  (Table \ref{table_cores}) have virial velocities of 1.4\,km\,s$^{-1}$, although they are expected to contribute to only $\sim$30\% of the measured flux in the case of optically thin emission (e.g. the NH$_{3}$ (1,1) satellite hyperfine lines). 
We suggest that the detected 2\,km\,s$^{-1}$ NH$_{3}$ (1,1) line width is mainly caused by turbulence in the bulk of the molecular gas. The NH$_{3}$ (3,3) line preferentially traces warmer and broader heated components, either associated with the central warm cores A1 and A2, or shocks.
High angular resolution observations of NH$_{3}$ can help constrain the broadening by the kinematics of the known localized cores as well as the protostellar outflows, and provide a more accurate measure of turbulent velocity. 

We note that the previous observations on a similar type of object G10.6-0.4 have found high rotational/infall velocities of 3--4 km\,s$^{-1}$, and a turnover radius r$_{T}\sim$0.3\,pc within which specific angular momentum experiences a significant loss (Liu et al. 2010a; see Figure \ref{fig_ch0} and Figure \ref{fig_ocs} for a sense of structures at the 0.3\,pc radius in G33.92+0.11).
Still, the smaller redshifted and blueshifted velocities detected in G33.92+0.11 are not conclusive evidence that G33.92+0.11 is a face--on system. 

Our new observations on G33.92+0.11 do not rule out the possibility that any support mechanism (e.g. turbulent, magnetic or thermal support) retards the molecular accretion flow. 
However, strong turbulence cannot be the dominant mechanism to support the molecular gas because turbulence itself also leads to the same level of spectral line broadening, which is discount by the observed narrow NH$_{3}$ lines. 
The evidence of strong magnetic support so far is absent.
Thermal support is most unlikely since it also leads to the spectral line broadening, and is contradictory to the modest NH$_{3}$ and CH$_{3}$CN excitation temperature of 30--50\,K measured from this region (Section \ref{chap_ch3cn}, \ref{chap_nh3}). 
Additionally, strong thermal support is adverse to fragmentation, which is contradictory to the detection of rich localized structures in the 1.3 mm continuum images (Figure \ref{fig_ch0}).  

\subsection{The Evolutionary Context of the Hub--Filament System}
\label{sub_fragmentation}
The resolved geometry in G33.92+0.11 resembles a $\sim$1 pc scale, molecular \textit{Hub--Filament System} (c.f. \textit{HFS}: Myers et al. 2009, 2011).
This system is embedded with a close pair of high--mass cores (Section \ref{chap_ch0}) in the central 0.1 pc region.
The extended filaments in this system show numerous localized cores, and the \textit{in situ} intermediate-- or high--mass star formation which exerts (proto)stellar feedback. 
Over the observed area, the abundance of the late type molecule SO$_{2}$ is very low, implying that both the centrally embedded high--mass cores, and the intermediate-- or high--mass star forming cores located in the $\lesssim$1 pc scale molecular filaments  (\textit{satellite cores} hereafter), are at an early evolutionary stage (see Section \ref{sub_diversity}). 
Recalling the observational facts that:
\begin{itemize}
\item[1.] We do not detect evidence of energetic massive molecular jets emanating from the central massive cores.  
\item[2.] From the observations of H30$\alpha$ emission, it seems that the UV photon illumination or the expansion of the ionized gas mostly affects the molecular gas southeast and northeast of the central massive cores. 
\item[3.] The high angular resolution observations of the thermal dust continuum emission show organized filamentary structures, rather than irregular fragments. Fine structures with small size scale can exist, even near the central OB cluster. For example, we see mini arms east of the core A2. This may indicate that either the feedback from the central OB cluster is not yet strong enough to significantly affect or sweep up the structure; or that the majority of molecular gas is (self--)shielded. 
\item[4.] Kinematically, the detected linewidths of the molecular gas toward G33.92+0.11 are small. The kinematics of the dense molecular gas appears to be still organized, without showing expansion signatures which typically have velocities of several km\,s$^{-1}$. These seem to be in contradiction to the expected strong influence by the central OB cluster.  
\end{itemize}
We think that not all those satellite cores are secondary clumpy structures, whose formation is related to the influence of the strong feedback exerted from the central OB cluster (reference to Section \ref{chap_rrl} for the definition of the central OB cluster). 
Instead, they may form around the same time.
We hypothesize that filamentary molecular structures already exist in a much early evolutionary stage (i.e. the time before the formation of the centralized massive cores and the satellite cores), and contribute to the accretion of molecular gas into the central hub (see below).
The matter concentration in the initial Hub--Filament configuration leads to a shorter local free--fall timescale than the timescale of the global contraction, which allows the simultaneous formation of the satellite cores and the centralized massive cores.
While dense molecular cores evolve in a much quicker dynamical timescale and form (proto)stellar objects (see also the discussion in Section \ref{sub_diversity}), the geometry of extended molecular filaments remains similar to their initial geometrical configuration.

Given/assuming the comparable ages, the fact that the molecular mass of each of the central massive cores is several times higher than the molecular mass of the satellite cores (Table \ref{table_cores}) requires the centralized molecular cores to accrete more efficiently. 
This requirement can be realized because of two unique properties of the central region:  (1) the deeper gravitational potential well in the center yields higher infall velocities, and (2) with the HFS configuration, molecular gas is channelled to the central region more efficiently through the detected molecular filaments. 
The fossil record of this continuous accretion onto the central molecular clump would be the detection of broad line emission from typical shock tracers such as SiO, OCS or CH$_{3}$CN (Bachiller \& Perez Gutierrez 1997; Jimenez-Serra et al. 2005; 2010; Csengeri et al. 2011a).
Even with comparable accretion velocities, the more adequate molecular gas reservoir for the central region will allow higher accretion rate than the satellite cores that are embedded in isolated, single molecular filaments. 
  
The molecular filaments may persist until later evolutionary stages. 
As an example, toward G10.6-0.4 a centralized 175 $M_{\odot}$ OB cluster (Sollins \& Ho 2005), few expanding UC H\textsc{ii} regions (Liu, Zhang \& Ho. 2011), and many satellite high--mass stars (Liu et al. 2010b) are already present. However, its central $\sim$2 pc scale massive molecular clump still keeps the geometrical configuration that looks similar to the Hub--Filament configuration in G33.92+0.11 (Liu et al. 2011b). 
Since filamentary structures seen toward G10.6-0.4 have small cross sections, gas infall can overcome the strong feedback from the central OB cluster, helping the gas accretion onto it. 
This scenario is supported by advanced numerical MHD simulations (Smith et al. 2009; Wang et al. 2010), which show that in the \textit{clump-fed} configuration, the luminous O--type stars are fed by the extended ($\gg$0.1 pc) dense filaments connected to the parsec scale, central massive clump.
The relevant recent analytical framework can be found in Hartmann et al. (2011), Myers et al. (2009), (2011), Pon et al. (2011), Toal{\'a} et al. (2011), and therein.

\subsection{The Angular Momentum Issue}
\label{sub_angular momentum}
In addition to how the accretion flow overcomes the stellar feedback, another fundamental issue for the OB star formation is how to propagate angular momentum out, or how to redistribute the angular momentum. 
The angular momentum problem becomes important even before stellar feedback comes into play.
The angular momentum is a conserved quantity that persistently affects the dynamical evolution, and may limit the mass of the massive core as well as the subsequently formed OB stars.  
Earlier analytical work suggested that strong (e.g. a few mG), organized magnetic field may play some role of dissipate the angular momentum (e.g. Keto et al. 1987).

With high angular resolution observations of the polarized thermal dust emission and observations of multiple molecular lines on OB cluster forming region G31.41+0.31, Girart et al. (2009) reported the decrease of the specific angular momentum toward the center of the system, and the competitively strong organized magnetic field that helps propagate out the angular momentum.  
Contrary to the case of G31.41+0.31, deeper observations of polarized thermal dust emission on OB cluster forming region G10.6-0.4 yields a null detection, implying that either the magnetic field is weak, the orientation of the magnetic field lines are random, or inefficient grain alignment (Liu et al. 2011b). 
The molecular filaments around G10.6-0.4 have radiative orientations, which contradictory to the flattened structure expected for a strongly magnetized system with an organized B field.
Observations of multiple spectral lines consistently indicate a significant decrease of the specific angular momentum toward the center of G10.6-0.4 (Liu et al. 2010a).

Other than the scenario of dissipating angular momentum via magnetic field braking, Liu et al. (2010a) suggested fragmentation can help resolve this angular momentum issue.   
Theoretical studies further suggested that in strongly self--gravitational systems, molecular arms help propagate out the angular momentum (Lodato \& Rice 2005).
In turbulent systems that have local distribution of velocity and specific angular momentum, the accretion can also take place in a diffusive manner, for which only the gas with lower specific angular momentum trickles in.

The possibilities of redistributing angular momentum via fragmentation, or propagating out the angular momentum via molecular arms, are now being explored with our higher angular observations. 
If the comparably massive core A1 and A2 are orbiting with respect to each other, the gas accreting into the individual cores  A1 and A2 only needs to lose excess specific angular momentum relative to that in the orbits of A1 and A2.
Our results of the H30$\alpha$ line observations and the 1.3 mm continuum emission suggest that the core A1 itself may show multiplicity (Section \ref{chap_rrl}; Figure \ref{fig_h30a}, right). 
The subsequent fragmentation in core A1 may alleviate the angular momentum issue at scales $\ll$0.1 pc.
On the other hand, we suspect that the aforementioned diffusive accretion process is not efficient enough at this 0.1--0.2 pc scale.

The molecular arm in the north of core A2 (Figure \ref{fig_345ch0}) as well as the more extended ($\sim$1 pc) arm--S1 and arm--S2 may also help propagate out the angular momentum.
At 1 pc scale, the size and curvature of the cavity wall have the evolutionary timescales of 10$^{4}$--10$^{5}$ years (Liu, Zhang, \& Ho 2011), which may be also sufficiently long for the SO$_{2}$ abundance to be enriched and become abundant in the gas phase (see Section$\,$\ref{sub_diversity}).
However, we cannot rule out the possibility that some of the apparent parsec scale molecular arms are in fact curved cavity walls of the H\textsc{ii} region (e.g. arm--N), not the trailing arms that may form in a rotating system.

\subsection{The Chemical Diversity and Its Origin}
\label{sub_diversity}
Our SMA images of different molecular tracers toward G33.92+0.11 suggest the presence of different chemical regimes across this high-mass star forming region. 
Within the central $\sim$0.1$\,$pc of G33.92+0.11, cores A1 and A2 present bright CH$_3$CN J=12-11 K=0,1 line emission, suggestive of hot-core chemistry (bright CH$_3$CN lines are typically found toward high-mass hot cores such as W3(H$_{2}$O), NGC7538 IRS1 or G24.78; see Chen et al. 2006; Bisschop et al. 2007; Galvan-Madrid et al. 2010). 
Other hot--core molecules (e.g. H$_2$S and SO) are also found toward the inner $<$15$''$ of G33.92+0.11 (see Section \ref{chap_s}), supporting the idea that cores A1 and A2 are hot--core--like objects where the molecular gas has been chemically enriched by the evaporation of the grain mantles by the heating from the central OB stars. 
The lack of detection of SO$_2$ (a late-type product of the sulfur chemistry), however, indicates that the time-scales involved in the formation of cores A1 and A2 are only some 10$^4$ yr (Charnley et al. 1997; Viti et al. 2004; Wakelam et al. 2004). 
This implies that cores A1 and A2 are likely young. 
This hypothesis is consistent with the moderate CH$_3$CN temperatures measured toward cores A1 and A2 (of $\sim$50$\,$K), and suggests that these cores could be at an intermediate evolutionary stage for which the central OB star cluster has not had enough time to heat up the surrounding molecular gas to the gas temperatures of $\sim$100-300 K typically found toward more evolved hot cores (see e.g. van der Tak et al. 2000). 
We note that the more evolved OB cluster forming clump G10.6-0.4 (a system with similar physical properties to those of G33.92+0.11, but seen at a different viewing angle), shows strong SO$_{2}$ emission lines (Liu et al. 2010a), supporting the idea that G33.92+0.11 is at an early stage in its evolution. Finally, toward cores A1 and A2, we do not find any line emission from SiO or OCS that could be associated with outflowing activity from the central OB star cluster (Section 3.1.4). 

At distances of $\sim$0.3-0.4 pc from the central cores A1 and A2, the SMA images reveal localized emission from typical shock tracers such as SiO and OCS (Martin-Pintado et al. 1992; Jimenez-Serra et al. 2005), showing broad line profiles (linewidths from $\sim$5-20$\,$km$\,$s$^{-1}$). As shown in Section 3.1.4, the regions where SiO and OCS have been enhanced tend to be located near the conjunctions of the massive molecular clump G33.92+0.11 and the external molecular arms/filaments. 
NH$_{3}$ is also significantly enhanced around the SiO emission peaks SiO--1,2 and toward the region where arm--S1,S2 continue into G33.92+0.11 A, indicating that a significant amount of molecular gas has been injected into the gas phase from dust grains. Since the NH$_3$ inversion lines reveal a temperature gradient as one moves inward the central clump in G33.92+0.11, we propose that the SiO, OCS and NH$_3$ enhancement found at distances of 0.3-0.4 pc from the central OB cluster, is due to gas accretion shocks at the regions where the external molecular arms/filaments are entering the massive molecular clump G33.92+0.11 A. 
We note that weaker CH$_3$CN emission is also detected toward these regions, suggesting that this emission could be associated with low-velocity shocks as found toward DR21(OH) by Csengeri et al. (2011a) and NGC 7538 S (Sandell \& Wright 2010).

Alternatively, gas shock compression could increase the gas density and pressure toward these localized regions, accelerating their collapse and subsequently leading to the formation of new protostars. 
The molecular outflows driven by these new objects would interact with the surrounding medium within the central G33.92+0.11 clump, sputtering dust grains and injecting large amounts of molecular material into the gas phase (Caselli et al. 1997; Jimenez-Serra et al. 2008; Gusdorf et al. 2008).

At even larger spatial scales from the central G33.92+0.11 A core (i.e. at $\sim$0.7 pc), our SMA data show that the large-scale filaments/arms present emission from sulfur-bearing species such as $^{13}$CS and SO. 
In particular, from Figures 4 and 6 there seems to be a systematic spatial shift between the 8$\,$$\mu$m emission (associated with PAH emission and found closer in to the central OB star cluster) and the $^{13}$CS, $^{13}$CO and SO lines. 
This chemical stratification has also been observed toward typical PDR regions such as the Orion bar, the Horse-Head Nebula or MonR2 (see respectively van der Wiel et al. 2009; Goicoechea et al. 2006; Bern\'{e} et al. 2009). 
Therefore, it is possible that the strong UV field arising from the central cluster of OB stars is escaping the dense gas of the central molecular clump, and illuminates the closer walls of the molecular filaments/arms leading to the formation of a typical PDR. 
Indeed, the H\textsc{ii} region detected toward G33.92+0.11 shows a cometary morphology, with the less-density gas (and therefore easier to escape from) toward the east (see Figure 9).  
The scenario where G33.92+0.11 shows multiple chemical regimes, will be tested in the near future with a deep and broad bandwidth molecular line survey toward the hot cores in G33.92+0.11 A, toward the shocked regions at the confluence of the molecular arms with the central clump, and toward the PDR found in arm-E.


\section{Summary}
\label{chap_summary}
We performed high angular resolution observations in the 1.3 mm lines and the dust continuum emission using the SMA, of the UC H\textsc{ii} region G33.92+0.11. 
In addition, we observed the NH$_{3}$ (J, K) = (1,1), (2,2) and (3,3) lines with higher velocity resolution ($\le$0.6 km\,s$^{-1}$) using the VLA D--array. 
A total of 19 spectral lines are selected for detailed inspection and analysis. 
We compare these new observations with observations of another UC H\textsc{ii} region G10.6-0.4, which has comparable molecular mass, size, bolometric luminosity, and is located at a comparable distance. 
Each of these two targets may represent respectively, examples of highly inclined and face--on, rotationally flattened massive molecular clumps. 
Our main conclusions are summarized as follows: 

\begin{itemize}
\item[1.] From $\ge$1 pc to $\le$0.1 pc scale, the molecular gas in G33.92+0.11 has a morphological configuration resembling a \textit{Hub--Filament System}. The UC H\textsc{ii} region is embedded in the center of the \textit{hub} (i.e. G33.92+0.11 A). The molecular gas appears to be heated by the UC H\textsc{ii} region and the embedded OB stars. Observations of NH$_{3}$ transitions show higher rotational temperature ($\gtrsim$35 K) close to the center of G33.92+0.11 A.  Observations of CH$_{3}$CN J=12--11 transitions show the rotational temperature of 49.3 K at the center of G33.92+0.11 A.
\item[2.] The brightest component of the UC H\textsc{ii} region has an elongated shape. We marginally resolve the blueshifted and redshfited motion of the ionized gas in this elongated region. The velocity scale of the motion is $\sim\pm$10 km\,s$^{-1}$, which can be explained by the thermal expansion of the ionized gas, or a collimated, low velocity ionized jet. 
\item[3.] Numerous satellite dense cores are embedded in the resolved molecular filaments. In the center of the \textit{hub} G33.92+0.11 A, a pair of centralized massive cores show several times higher molecular mass than the rest of the satellite dense cores. The centralized massive cores account for 11\% of the total molecular mass in the G33.92+0.11 A region (including arm--S1, S2). Together with the satellite dense cores, the resolved compact cores account for 24.3 \% of the total molecular mass in the G33.92+0.11 A region.
\item[4.] We suggest (hierarchical) fragmentation is a plausible mechanism to resolve the angular momentum issue of the core/(proto)stellar accretion. 
\item[5.] Cores embedded in the G33.92+0.11 A region appear to be chemically young as suggested by the low upper limit of the SO$_{2}$ abundance measured toward this source. This is in contrast with the bright SO$_{2}$ emission found toward the target G10.6-0.4 (Liu et al. 2010a). The abundance X(SO$_{2}$) is constrained to be less than 3$\times$10$^{-9}$.
\item[6.] An overall north--south (or northeast--southwest) velocity gradient is seen in molecular gas in G33.92+0.11. At a parsec scale radius, the velocity gradient is about 0.96 km\,s$^{-1}$\,pc$^{-1}$. A higher velocity gradient of 4.6 km\,s$^{-1}$\,pc$^{-1}$ may be seen in the central $\pm$10$''$ (i.e. $\pm$0.344 pc) area in G33.92+0.11 A. The velocity profile is asymmetrical in the central 0.6 pc area in G33.92+0.11 A, and show higher blueshifted velocity than the redshifted velocity. The intrinsic spatial and velocity asymmetries appear to be essential in such strongly self--gravitating OB cluster forming molecular clumps.  The asymmetry of the velocity profile in the target G10.6-0.4 is reported in Liu et al. (2010a). 
\item[7.] We resolved zones that show chemical diversities. The molecular line emission in the central 0.1--0.2 region in G33.92+0.11 A can be characterized by a hot core chemistry. We do not see signatures of strong shocks in this central region. In diameters of 0.2--0.5 pc in G33.92+0.11 A, several places show strong SiO and OCS emission, which are consistent with the shock enrichment of these species. In more extended areas, stratifications of the 8 $\mu$m emission, the $^{13}$CO/C$^{18}$O emission, the SO emission, and the $^{13}$CS emission resembles a PDR chemistry induced by the illumination from the central OB cluster. Follow up observations on this target may improve our understanding of the chemical evolution in the OB cluster forming massive molecular envelope. 
\end{itemize}

We emphasize that our main focus is the formation of an overall OB cluster. 
Studies of accretion of individual OB stars cannot be addressed because of the still insufficient angular resolution. 
ALMA observations in the future will provide much better angular/velocity resolution and sensitivity, to systematically survey other similar targets, or fainter objects with lower masses or lower temperatures.

\acknowledgments
This work is based in part on observations made with the \textit{Spitzer Space Telescope}, which is operated by the Jet Propulsion Laboratory, California Institute of Technology. 
We thank the GLIMPSE team (PI: E. Churchwell) for making the IRAC images available to the community.
We thank the SMA staff for making our SMA observations possible.
Liu thanks Glen Petipas, Daniel Espada, Ken Harbour Young, and SMA operators for their assistance with the filler observations.
Liu thanks Roberto Galv{\'a}n-Madrid for useful and encouraging discussions. 
Liu thanks Thomas Peters for useful comments while organizing this project. 
Liu thanks Xing Lu for help developing the codes for data analysis.
Liu thanks ASIAA for supporting the relevant thesis researches. 
%

{\it Facilities:} \facility{SMA, VLA, Spitzer Space Telescope}



\begin{table*}[h]\footnotesize
\begin{tabular}{  | p{2.5cm} | ccccccc | }\hline\hline
Tracks  & 1 & 2 & 3 & 4 & 5 & 6 & 7 \\
Observing Dates  		&  2009Nov02  & 2010Mar27  &  2011Mar16  &  2011Aug13  &  2011Sep18   &  2011Sep19   &   2011Sep20    \\\hline
Receiver (GHz)        	&  230	&  230  &  230  &   230   &   230  &  230   &   230  \\
Array Configuration	&  compact  &  subcompact  &  subcompact  &  extended  & very extended & very extended  & very extended   \\
Observing Mode        &  Line & Line  &   Line &   Line &   Line &   Line &   Line    \\
USB Frequency (GHz) & 228.6--232.6    & 229--233 &  228.8--232.8 & 228.8--232.8 & 228.8--232.8 & 228.8--232.8 &  228.8--232.8      \\
LSB Frequency  (GHz) & 216.6--220.6    & 217--221 &  216.8--220.8 & 216.8--220.8 & 216.8--220.8 & 216.8--220.8 &  216.8--220.8      \\
uv range ($k\lambda$) &  16--87 & 5.5--31 & 4--52  & 25--175  &  40--390  & 40--390  &  40--390  \\
$\tau_{\mbox{\tiny{225 GHz}}} $  & 0.05 &  0.05  & 0.2  & 0.2  & 0.25  & 0.2  &  0.2   \\
Number of Antennas &  7  &  7  &  7  &  7  &  8  &  8  &  8  \\
Time on Target Loop (hr) &      2     & 0.4   & 8  &  8 & 6 & 6  & 6  \\
Mode & Single & Single & Mosaic  &  Mosaic  & Single  & Single  & Single  \\
Remark & filler & filler & regular & regular & filler & filler & filler \\
Flux Calibrator 	& Titan			&	Titan			&	Saturn		&	Callisto		&	Callisto	&	Callisto	&	Callisto	 \\
Passband Calibrator	& 3C273	&	3C279		&	3C279		&	3C454.3	&	3C84		&	3C84		&	3C84		\\\hline
\end{tabular}
\caption{\normalsize{Summary of SMA observations. The spectral line mode has velocity resolution of 1.2\,km\,s$^{-1}$. The velocity resolution of continuum mode observations is a fact of 4 degraded. We list the time on target loop instead of time on source to provide the idea of uv sampling. In each track, we observed 1830+063 every 10--15 minutes for gain calibration. The observing efficiency is about 70\%. The single pointing observations target on R.A.= 18$^{\mbox{h}}$52$^{\mbox{m}}$50$^{\mbox{s}}$.272 and Decl.= $+$00$^{\circ}$55$'$29$''$.604. Mosaic observations iterate though the pointings of \textbf{Main:} R.A.= 18$^{\mbox{h}}$52$^{\mbox{m}}$50$^{\mbox{s}}$.272 and Decl.= $+$00$^{\circ}$55$'$29$''$.604, \textbf{N:} R.A.= 18$^{\mbox{h}}$52$^{\mbox{m}}$50$^{\mbox{s}}$.272 and Decl.= $+$00$^{\circ}$55$'$57$''$.104, \textbf{S:} R.A.= 18$^{\mbox{h}}$52$^{\mbox{m}}$50$^{\mbox{s}}$.272 and Decl.= $+$00$^{\circ}$55$'$02$''$.104, \textbf{NE:} R.A.= 18$^{\mbox{h}}$52$^{\mbox{m}}$51$^{\mbox{s}}$.860 and Decl.= $+$00$^{\circ}$55$'$43$''$.354, \textbf{NW:} R.A.= 18$^{\mbox{h}}$52$^{\mbox{m}}$48$^{\mbox{s}}$.684 and Decl.= $+$00$^{\circ}$55$'$45$''$.354, \textbf{SE:} R.A.= 18$^{\mbox{h}}$52$^{\mbox{m}}$51$^{\mbox{s}}$.860 and Decl.= $+$00$^{\circ}$55$'$15$''$.854, \textbf{SW:} R.A.= 18$^{\mbox{h}}$52$^{\mbox{m}}$48$^{\mbox{s}}$.684 and Decl.= $+$00$^{\circ}$55$'$15$''$.854.}}
\label{table_tracks}
\end{table*}

\vspace{3cm}

\begin{table}[h]
\begin{tabular}{lccccccc}\hline\hline
Species & Line   &   Rest Frame Frequency &  $E_{u}$/k &  Log$_{10}$$A_{ij}$   &  Observing Tracks  & uv range & Detection \\
           &           &      (GHz)                               &    (K)            &             &                            &        ($k\lambda$)\\\hline
H$_{2}$S & 2$_{2,0}$--2$_{1,1}$  &  216.71043  & 84.0  & -4.32  &  1  &  16--87		& Y \\
SiO & 5--4            &   217.10998  & 31.3  & -3.28  &  1--7 				  &   4--390	& Y \\
O$^{13}$CS &18--17  & 218.19900  & 99.5  & -4.52 & 1--7 			&  4--150		& N  \\
OCS &18--17  &  218.90336  &  99.8 & -4.52 & 2--7 						&  4--100		& Y  \\
C$^{18}$O &  2--1  &   219.56036  & 15.8   & -6.22  &  1--7			&  4--180 		& Y	\\
SO & 5$_{6}$--4$_{5}$        &  219.94944   & 35.0  & -3.87  &  1--7 	&  4--100		& Y	\\
$^{13}$CO & 2--1  &   220.39868  &  15.9 & -6.21  &  1--7 				& 	4--180		& Y	\\
CH$_{3}$CN & 12$_{0}$--11$_{0}$ & 220.74724  & 68.9 & -3.04 &	2--7		& 4--100	& Y \\
CH$_{3}$CN & 12$_{1}$--11$_{1}$ & 220.74299  & 76.0 & -3.04 & 2--7		& 4--100	& Y\\
CH$_{3}$CN & 12$_{2}$--11$_{2}$ & 220.73027  & 97.4 & -3.05 & 2--7 		& 4--100	& Y\\
CH$_{3}$CN & 12$_{3}$--11$_{3}$ & 220.70908  & 133.2 & -3.07 & 2--7 	& 4--100	& Y\\
CH$_{3}$CN & 12$_{4}$--11$_{4}$ & 220.67932  & 183.2 & -3.09 &  1--7	& 4--100	& N\\
CH$_{3}$CN & 12$_{5}$--11$_{5}$ & 220.64112  & 247.4 & -3.12 &  1--7	& 4--100	& N\\
CH$_{3}$CN & 12$_{6}$--11$_{6}$ & 220.59449  & 325.9 & -3.16 &  1--7	& 4--100	& N\\
CH$_{3}$CN & 12$_{7}$--11$_{7}$ & 220.53931  & 418.6 & -3.21 &  1--7	& 4--100	& N\\
SO$_{2}$ & 11$_{5,7}$--12$_{4,8}$ & 229.34763 &  122.0 & -4.72 & 1--7	& 4--100	& Y \\
$^{12}$CO & 2--1  &   230.53800  &  16.6 &	-6.16   &  1--7 & 4--180 						& Y \\
OCS & 19--18      &   231.06098   & 111.9  & -4.45  &  1--7		& 4--100					& Y \\
$^{13}$CS & 5--4   &  231.22100   & 33.3  &  -3.60 &  1--7		& 4--150					& Y \\
N$_{2}$D$^{+}$ & 3--2 & 231.32166 & 22.2 & -2.67	& 1--7		& 4--150					& N \\ 
H$\alpha$ & 30$\alpha$         &   231.90093  & --  & --  &  1--7	& 4--390					& Y \\\hline
NH$_{3}$ & (1,1)		&		23.69450			&	23.3	&	-6.78	&	VLA	2003Feb14	& 2.6--7.6	&  Y\\
NH$_{3}$ & (2,2)	    &		23.72264			&	64.4	&	-6.65	&	VLA	2003Feb14	& 2.6--7.6	&  Y\\
NH$_{3}$ & (3,3)   	&		23.87013			&	123.5	&	-6.59	&	VLA	2003May13	& 2.1--8.2	& Y\\\hline
\end{tabular}
\caption{Selected spectral lines in discussions. $E_{u}$ is the upper level energy, and $A_{ij}$ is the Einstein's $A$ coefficient.}
\label{table_line}
\end{table}

\begin{table}[h]
\begin{tabular}{ccccccc}\hline\hline
Clump/Core			&			$S_{\mbox{1.3mm}}$		&  $S_{\mbox{ff}}$ 	&		Mass$^{\beta=1}$		& Mass$^{\beta=1.5}$	 & Mass$^{\beta=2}$	 &  ($S^{Ai}_{1.3mm}$ - $S^{Ai}_{ff}$)/($S^{A}_{1.3mm}$ - S$^{A}_{ff}$) \\
						&				(mJy)						&				(mJy)	    &		($M_{\odot}$) &		($M_{\odot}$) &		($M_{\odot}$) 				& ($\%$)\\\hline
A					&					2800												&    670					&		1600	&	3000	&	5800		& ---			\\
A1					&		330		&	180	& 84	&	160	&	310	& 		7.2	\\
A2					&		200		&	45	& 87	&	170	&	320	& 7.5		\\
A3					&		58			&	---	& 33	&	63	&	120	& 2.8		\\
A4					&		55			&	---	& 31	&	60	&	120	&	2.6		\\
A5					&		54			&	---	& 30	&	59	&	110	&	2.6		\\
A6					&		31			&	---	& 17	&	34	&	65		&	1.5	\\
A7					&		28			&	---	& 	16	&	30		&	59	 & 1.3			\\
A8					&		14			&	---	& 	7.9	&	15	&	29		& 0.67				\\
A9					&		50			&	---	& 28	&	54	&	110		& 2.4			\\
A10				&		58			&	---	& 33	&	63	&	120		& 2.8			\\
A11				&		23			&	---	& 13	&	25	&	48		& 1.1			\\\hline
B					&		320		&	---	& 180	&	350	&	670	& ---	\\
C					&		12			&	---	&	6.8	&	13		&	25		& ---	\\
D					&		250		&	---	&	140	&	270		&	520		& ---	\\\hline
\end{tabular}
\caption{Estimates of the molecular mass based on the 1.3 mm dust continuum emission. The $S_{\mbox{1.3mm}}$ and $S_{\mbox{ff}}$ are the continuum fluxes at 1.3 mm, and the free--free contribution of the continuum flux at 1.3 mm. 
We estimate the molecular mass of the compact (hot) molecular cores A1--11 assuming the averaged gas excitation temperature of 45K. We estimate the molecular mass of the clumps G33.92+0.11 A--D assuming the averaged gas excitation temperature of 35K. In the G33.92+0.11 A region, the compact cores A1--11 contribute to $\sim$32.5\% of the total of the 1.3 mm thermal dust continuum emission, which corresponds to 24.3\% of the overall mass. Cores A1 and A2 contribute to $\sim$14.7\% of 1.3 mm thermal dust emission in the G33.92+0.11 A region, which corresponds to 11\% of the overall mass.}
\label{table_cores}
\end{table}

\clearpage

\begin{table}[h]
\begin{tabular}{cccccc}\hline\hline
Outflow Source  & $\alpha$   				& $\delta$				  & Mass  & Momentum & Energy \\
                        &     (J2000)			&		 (J2000)			& ($M_{\odot}$) & ($M_{\odot}$\,km\,s$^{-1}$) & (10$^{44}$ erg) \\\hline
A1			&	18$^{\mbox{h}}$52$^{\mbox{m}}$50$^{\mbox{s}}$.135		&	$+$00$^{\circ}$55$'$31$''$.0	&	2.3	& 12  	& 20\\
A2			&	18$^{\mbox{h}}$52$^{\mbox{m}}$50$^{\mbox{s}}$.468		&	$+$00$^{\circ}$55$'$27$''$.2  &	0.053	& 1.1 & 3.0\\
A5			&	18$^{\mbox{h}}$52$^{\mbox{m}}$50$^{\mbox{s}}$.468		&	$+$00$^{\circ}$55$'$34$''$.4  &	0.31	&  3.1	& 6.3\\
A10		&	18$^{\mbox{h}}$52$^{\mbox{m}}$50$^{\mbox{s}}$.041		&	$+$00$^{\circ}$55$'$18$''$.0	&	0.18	&  1.4	& 4.2\\
C 			&	18$^{\mbox{h}}$52$^{\mbox{m}}$51$^{\mbox{s}}$.135		&	$+$00$^{\circ}$55$'$29$''$.0	&	1.2	&  4.6	& 6.0 \\\hline
\end{tabular}
\caption{The derived $^{12}$CO 2--1 parameters for the high velocity gas (Section \ref{chap_hv}). The momentum and energy are for the (line--of--sight) projected components.}
\label{table_outflow}
\end{table}

\vspace{2cm}

\begin{table}[h]\scriptsize
\begin{tabular}{c|cccc|ccc|ccc|c}\hline\hline
         	&	\multicolumn{4}{c|}{NH$_{3}$ (1,1)}				& \multicolumn{3}{c|}{NH$_{3}$ (2,2)}  & \multicolumn{3}{c|}{NH$_{3}$ (3,3)} & ($A_{11}\sigma_{11}\xi$)/($3A_{33}\sigma_{33}$)\\\hline
Region	& $A_{11}$	&	$v_{0,11}$	& $\sigma_{11}$	& (satellite/main)$\times$3 ($\xi$)	& $A_{22}$	&	$v_{0,22}$	& $\sigma_{22}$			 & $A_{33}$	&	$v_{0,33}$	& $\sigma_{33}$	& \\
			&	(K)		&	(km\,s$^{-1}$)	& (km\,s$^{-1}$)	&		& (K)		&	(km\,s$^{-1}$)	& (km\,s$^{-1}$) &  (K)		&	(km\,s$^{-1}$)	& (km\,s$^{-1}$) &\\\hline
NW	&	2.30	&	106.0	&	0.94	&	1.00	&	1.47	&	106.2	&	0.80	& 0.85	&	106.0	&	1.23	& 0.69 \\
		&	2.62	&	108.5	&	1.07	&	1.02	&	1.60	&	108.6	&	0.80	&	1.93	&	108.5	&	1.20  & 0.41 \\\hline
SW	&	1.38	&	106.1	&	0.96	&	1.24	&	1.29	& 	106.0	&	1.06	&	1.29	&	106.0	&	1.02 	& 0.42 \\	
		&	5.96	&	107.9	&	0.86	&	1.27	&	3.93	&	108.0	&	0.90	&	3.56	&	108.0	&	0.99	& 0.62	\\\hline
NE		&	4.7	&	107.6	&	0.85	&	1.03	&	3.14	&	107.7	&	1.05	&	2.09	&	107.8	&	1.49	& 0.47 \\
SE		&	3.11	&	106.3	&	1.27	&	1.03	&	2.20	&	106.6	&	1.24	&	1.99	&	106.9	&	1.17	&	0.61\\
arm--S	&	3.11	&	107.2	&	0.78	&	1.30	&	1.66	&	107.2	&	0.67	&	0.78	&	107.1	&	0.78	&	1.70	\\
B		&	3.85	&	107.8	&	0.84	&	1.28	&	2.07	&	107.7	&	0.83	&	1.12	&	108.0	&	0.97	& 1.27 \\\hline
\end{tabular}
\caption{Gaussian components to characterize the NH$_{3}$ spectra in Figure \ref{fig_nh3spectra}. The Gaussian profile has the form of ($A_{JK}\,e^{-\frac{(x-v_{0,JK})^{2}}{2\sigma_{JK}^{2}}}$). For the fittings of the NH$_{3}$ (1,1) transitions, we assume the satellite hyperfine lines have the same $\sigma$ and $v_{0}$ with the main hyperfine line, and fit the main to satellite lines amplitude ratio. 
In optically thin cases, the main to satellite hyperfine line ratio of the (J,K)=(1,1) transitions is about 3.33. 
The main to satellite hyperfine line ratio becomes much higher for higher (J,K) levels. 
The full width at half maximum (FWHM) of the Gaussians can be obtained by FWHM$\sim$2.35$\times$$\sigma$. The last column ($A_{11}\sigma_{11}\xi$)/($3A_{33}\sigma_{33}$) is the ratio of the (J,K)=(1,1) satellite hyperfine line flux to the (J,K)=(3,3) main hyperfine line flux, which characterize the gas excitation temperature.}
\label{table_nh3gau}
\end{table}

\vspace{2cm}

\begin{table}[h]\scriptsize
\begin{tabular}{c|ccc|ccc|ccc}\hline\hline
         	&	\multicolumn{3}{c|}{OCS 19--18}				& \multicolumn{3}{c|}{SO 5$_{6}$--4$_{5}$}  & \multicolumn{3}{c}{SiO 5--4} \\\hline
Region	& $A_{OCS}$	&	$v_{0,OCS}$	& $\sigma_{OCS}$	& $A_{SO}$	&	$v_{0,SO}$	& $\sigma_{SO}$			 & $A_{SiO}$	&	$v_{0,SiO}$	& $\sigma_{SiO}$	\\
			&	(K)		&	(km\,s$^{-1}$)	& (km\,s$^{-1}$)			& (K)		&	(km\,s$^{-1}$)	& (km\,s$^{-1}$) &  (K)		&	(km\,s$^{-1}$)	& (km\,s$^{-1}$) \\\hline
Core A3	&	0.28	&	106.8	&	1.5	&	3.8	&	107.1	&	1.4	&	---	&	---	&	---	\\	\hline
Core A5	&	0.40	&	109.4	&	1.0	&	4.8	&	108.2	&	1.5	&	0.74	&	109.0	&	1.35	\\	
         	&	0.16	&	106.3	&	1.2	&		&		&		&	0.22	&	105.0	&	2.5	\\\hline
Core A8	&	0.30	&	107.5	&	2.6	&	3.2	&	108.1	&	1.9 	&	0.75	&	107.0	&	1.9	\\	
         	&	     	&	         	&	     	&	     	&	      	&	     	&	0.61	&	110.0	&	8.5	\\\hline			
\end{tabular}
\caption{Gaussian components to characterize the SO, OCS and SiO spectra in Figure \ref{fig_highvelspec}. The Gaussian profile has the form of ($A_{species}\,e^{-\frac{(x-v_{0,species})^{2}}{2\sigma_{species}^{2}}}$).
The full width at half maximum (FWHM) of the Gaussians can be obtained by FWHM$\sim$2.35$\times$$\sigma$.}
\label{table_highvel}
\end{table}

\clearpage

\clearpage

\clearpage

\clearpage

\clearpage

\clearpage

\clearpage

\clearpage

\clearpage

\clearpage

\clearpage

\clearpage

\clearpage

\clearpage

\end{document}